\documentclass[12pt]{article}
\usepackage[table,dvipsnames]{xcolor}
\usepackage[most]{tcolorbox}
\usepackage[utf8]{inputenc}	
\usepackage[titletoc,title]{appendix}
\usepackage{calc,amsmath,amsthm,amsfonts,amssymb,amscd,url,doi,enumitem,mathrsfs,bm}
\usepackage{multirow,booktabs,multicol,setspace,cancel}
\usepackage{fullpage,lastpage,listings,enumerate}
\usepackage{wrapfig,empheq,framed,tabularx,tikz,feynmp-auto,graphicx,caption,subcaption,mwe}
\usepackage[retainorgcmds]{IEEEtrantools}
\usepackage[margin=3cm]{geometry}
\usepackage{footnote}

\usetikzlibrary{positioning}
\newlist{todolist}{itemize}{1}
\setlist[todolist]{label=$\square$}

\begin{document} 
\begin{fmffile}{notes_fm}



\pagenumbering{gobble}

\begin{center}
{ University of Heidelberg \\ \vspace{0.2cm} Department of Physics and Astronomy \\ \vspace{0.2cm}  Institute for Theoretical Physics } \\

\vspace{1.3cm}
\textsc{Master Thesis in Physics}

\vspace {1.2cm}
\hrule
\vspace{0.2cm}
\Large{\textbf{A Global Analysis of the \\ \vspace{0.1cm} Standard Model Effective Field Theory \\ \vspace{0.1cm} in the Production and Decay Channels \\ \vspace{0.1cm} of a Single Top Quark }} \vspace{0.5cm}

\hrule

\vspace{1.0cm}
\normalsize{
\textit{submitted by \hspace{11cm} Supervisors} \\ \vspace{0.2cm}
Rhea Penelope Moutafis \hspace{8.3cm} Tilman Plehn\textsuperscript{1} \\ \vspace{0.2cm}
born in Egling, Germany \hspace{8.5cm} Dirk Zerwas\textsuperscript{2} } \\ \vspace{1cm}

\textit{\textsuperscript{1}{Institute for Theoretical Physics, Heidelberg, Germany}} \\ \vspace{0.2cm}
\textit{\textsuperscript{2}{Laboratoire de l'Acc\'el\'erateur Lin\'eaire, Orsay, France}} \\ \vspace{1.7cm}

This master thesis has been carried out by Rhea Moutafis \\ \vspace{0.2cm}
at the Institute for Theoretical Physics \\ \vspace{0.2cm} and the Laboratoire de l'Acc\'el\'erateur Lin\'eaire \\ \vspace{0.2cm}
under the supervision of Tilman Plehn and Dirk Zerwas. \\ \vspace{1.7cm}

April 30, 2019

\end{center}

\newpage
\thispagestyle{plain} 
\mbox{}
\clearpage

\vspace*{0.3cm}
\textbf{ A Global Analysis of the Standard Model Effective Field Theory in the Production and Decay Channels of a Single Top Quark }

A global analysis of the Standard Model Effective Field Theory (SMEFT) with \textsc{SFitter} is performed using measurements of single top quark production and top quark decay processes from ATLAS and CMS at energies of $\sqrt{s} =$ 7, 8 and 13~TeV.
NLO QCD corrections are included in the theoretical predictions of all involved processes.
Correlations among theoretical and experimental uncertainties are accounted for using \textsc{DataPrep}, which has been developed by the author.
Using the dataset prepared by this software, constraints for the relevant degrees of freedom are derived. 
Almost all constraints are more than three times more stringent than the values of the literature.
This thesis sets the technical and conceptual framework for a global fit of the entire top quark and Higgs boson sectors, which the author is involved in.
The project marks an important milestone towards a truly global analysis constraining physics beyond the Standard Model in a model-independent way.

\vspace{3cm}
\textbf{ Eine Globale Analyse der Effektiven Feldtheorie des Standardmodells in den Produktions- und Zerfallskan\"alen eines Einzelnen Top-Quarks }

Eine globale Analyse der effektiven Feldtheorie des Standardmodells (SMEFT) mit \textsc{SFitter} wird unter Verwendung von Messungen von Zerfalls- und Produktionsprozessen eines einzelnen Top-Quarks von ATLAS und CMS bei Energien von $\sqrt{s} =$ 7, 8 und 13 TeV ausgef\"uhrt.
NLO QCD-Korrekturen werden in den theoretischen Vorhersagen aller einbezogenen Prozesse verwendet.
Korrelationen unter theoretischen und experimentellen Unsicherheiten werden durch die Verwendung von \textsc{DataPrep}, einer von der Autorin entwickelten Software, ber\"ucksichtigt.
Mit den so pr\"aparierten Daten werden Schranken f\"ur die relevanten Freiheitsgrade ermittelt.
Fast alle Schranken sind mehr als dreimal so stringent wie die Literaturwerte.
Diese Arbeit setzt den technischen und konzeptuellen Rahmen f\"ur einen globalen Fit der gesamten Top Quark- und Higgs Boson-Sektoren, an dem der Autor ma{\ss}geblich beteiligt ist.
Das Projekt ist ein wichtiger Meilenstein auf dem Weg zu einer wirklich globalen Analyse, die Physik jenseits des Standardmodells mit einer modellunabh\"angigen Methode beschr\"ankt.

\newpage
\thispagestyle{plain} 
\mbox{}
\clearpage 


\setcounter{tocdepth}{1}
\tableofcontents
\clearpage

\newpage
\thispagestyle{plain} 
\mbox{}
\clearpage


\newpage
\pagenumbering{arabic}
\section{Introduction}

The Standard Model (SM) has been extremely successful so far, the last major triumph being the discovery of the Higgs boson in 2012~\cite{higgs,discovery}.
However, there are various issues with the SM, for example the hierarchy problem and the strong CP problem.
These, among other issues, are reasons to search for physics beyond the SM (BSM).

The top quark plays a special role in most BSM scenarios. 
Studies of the top quark sector allow another perspective on physics at the electroweak scale because of the strong coupling between the top quark and the Higgs boson.
Thus, if the Higgs boson is the key to new physics, BSM effects should also be visible in the top quark sector.

The LHC could deliver evidence for BSM physics by direct production of new particles.
Should new particles be too heavy to be directly produced at the LHC, they could still leave imprints in the cross sections and kinematic distributions of the SM particles via interferences or virtual effects.
These effects can be described with a Standard Model Effective Theory (SMEFT)~\cite{Weinberg:1979sa}-\cite{Grzadkowski:2010es}. 
In this framework, the effects of BSM dynamics at high energy scales are parametrized in terms of higher-dimensional operators which respect gauge and Lorentz symmetries and are built up from the SM fields.
This is an excellent way to constrain BSM physics in a model-independent way.

One difficulty with SMEFT is the sheer number of operators.
Even assuming conservation of baryon and lepton number~\cite{Grzadkowski:2010es}, one ends up with $N_\text{op}=59$ operators at dimension six for flavor universality, and over 2000 without flavor assumptions.
This means that a large parameter space needs to be investigated.
In this study, only those operators are included that are relevant for the production and decay of a single top quark.

The goal of this work is to constrain the Wilson coefficients corresponding to these dimension-six operators using \textsc{SFitter}.
In this framework, Monte Carlo (MC) toys are generated to construct the probability distribution in the space of the Wilson coefficients.
Full next-to-leading order (NLO) simulations are used for almost all theoretical predictions.
There are only a couple of exceptions:
for the predictions of the cross sections of single top quark production in association with a $W$ or $Z$ boson, NLO $K$-factors are used.
Also, for some SMEFT contributions at order $\mathcal{O}(\Lambda^{-4})$ to the production in association with a $Z$ boson, predictions at leading order (LO) are used.

SMEFT has been applied to the Higgs sector multiple times~\cite{Ellis:2014jta}-\cite{Englert:2015hrx}.
SMEFT analyses in the top quark sector also already exist in the literature, for example by the \textsc{TopFitter}~\cite{Buckley:2016cfg}-\cite{Buckley:2015lku} and the \textsc{SMEFiT}~\cite{Hartland:2019bjb} collaborations.
However, a global fit of the top quark sector has never been conducted in conjunction with measurements involving a Higgs boson, as will be the goal of the publication following this thesis.
In addition, correlations between the systematic uncertainties have never been studied in such great detail as in this thesis.

The outline of this thesis is as follows:
In section~\ref{sec_SMEFT}, the SMEFT framework in the top quark sector and and the relevant dimension-six operators are introduced.
In section~\ref{sec_data}, the influence of the SMEFT operators on the production and decay processes of a single top quark at the LHC are described, as well as the measurements and the theoretical predictions of the SM and SMEFT cross-sections.
In section~\ref{sec_methodology}, the methodology of \textsc{SFitter}, specifically tailored to the measurements covered in this analysis, is presented.
In section~\ref{sec_results} the results are evaluated.
In section~\ref{sec_conclusion}, the results are summarized and new ideas are proposed to generalize the analysis of this thesis to other measurements. 

\clearpage


\section{SMEFT for the top quark sector}
\label{sec_SMEFT}

In this section, the SMEFT framework is introduced.
After that, the operators that are relevant for single top production and decay processes are presented.
Finally, NLO QCD effects on the theoretical predictions are discussed.

\subsection{Introduction to SMEFT}
\label{subsec:SMEFT:intro}

The effects of new heavy BSM particles with a typical mass scale of $M\simeq \Lambda$ can be parametrized for $E\ll \Lambda$ as a power expansion
\begin{equation}
\mathcal{L}_\text{SMEFT} = \mathcal{L}_\text{SM} +\ \sum_i^{N_\text{d6}} \frac{c_i}{\Lambda^2}\ \mathcal{O}_i^{(6)} +\ \sum_j^{N_\text{d8}} \frac{b_j}{\Lambda^4}\ \mathcal{O}_j^{(8)} + ...,
\end{equation}
where $\mathcal{L}_\text{SM}$ is the SM Lagrangian, and $\{ \mathcal{O}_i^{(6)} \}$ and $\{ \mathcal{O}_j^{(8)} \}$ stand for the elements of the operator basis of mass dimension $d=6$ and $d=8$, respectively.
The dimensions $d=5$ and $d=7$ are not considered here as they violate lepton or baryon number conservation~\cite{Degrande:2012wf,Kobach:2016ami}.
The Warsaw basis~\cite{Grzadkowski:2010es} is chosen, and operators with mass dimensions $d\geq 8$ are regarded as negligible.

The effect of the dimension-six SMEFT operators on cross-sections, differential distributions, or other observables, can be written as
\begin{equation}
\sigma_\text{SMEFT} = \sigma_\text{SM} +\ \sum_i^{N_\text{d6}} \sigma_i \frac{c_i}{\Lambda^2} +\ \sum_{i,j}^{N_\text{d6}} \tilde{\sigma}_{ij} \frac{c_i c_j}{\Lambda^4} + ...,
\label{eq_SMEFT_sigma}
\end{equation}
where $\sigma_\text{SM}$ denotes the SM prediction.
As this study is limited to CP conserving operators, the Wilson coefficients $c_i$ are real.
The second term arises from SMEFT operators interfering with the SM operators.
The last term arises from SMEFT operators interacting with each other.

The approach of this study includes quadratic dimension-six terms, but not linear dimension-eight terms, which also are of order $\mathcal{O}(\Lambda^{-4})$.
While the linear dimension-eight terms are usually suppressed in a valid EFT approach, there are several reasons to include quadratic dimension-six terms:
\begin{itemize}
\item In BSM models with relatively large couplings the quadratic dimension-six terms can become dominant without implying that the EFT becomes invalid, see for example~\cite{Contino:2016jqw}-\cite{Biekoetter:2014jwa}.
\item The linear dimension-six terms are often suppressed. 
In some cases, a dimension-six operator does not interfere with a SM operator at all because it has a different helicity and color structure or a different CP parity than the SM operator.
\item For some operators, the quadratic term is dependent on the energy of the process, while the linear one is not. One would therefore expect the quadratic terms to be dominant at high energies.
\end{itemize}

One problem with the SMEFT is the huge amount of operators.
Without further assumptions than those stated above, we get $N_\text{d6} = 2499$ operators~\cite{DAmbrosio:2002vsn}.
This would mean having to deal with a huge operator space with possibly many flat directions.

To reduce the number of operators to a feasible amount, the strategy recommended by the LHC Top Quark Working Group~\cite{AguilarSaavedra:2018nen} is adopted.
This includes the assumption that the flavor structure of the Wilson coefficients is diagonal, and that the Yukawa couplings are nonzero only for the top and bottom quarks.
Furthermore, only CP-even operators are considered, and only those are used which induce modifications in the interactions of the top quark with other SM fields.
With these assumptions, one is left with seven operators for single top quark production and decay.

In general, the Wilson coefficients run with the scale and thus depend on the typical momentum transfer of the process.
This dependency is not included as the focus of this analysis is on processes where $E\lesssim \Lambda$, $E$ being the typical energy scale of the process.
In addition, including NLO QCD effects also reduce the scale dependence, making it less significant.

Certain operators can induce a growth of the cross sections through two mechanisms: 
One the one hand, if an operator involves higher dimension Lorentz structures with additional derivatives or four-fermion interactions, energy-growth arises.
On the other hand, unitarity cancellations that take place in the SM amplitudes can be spoiled through operators even if they do not contain new structures~\cite{Degrande:2018fog}.
This is an important feature of SMEFT because measurements at higher energies will be increasingly sensitive to the values of certain Wilson coefficients.

There are limits to this feature as the realm of the validity of the SMEFT is only in the region where $E \ll \Lambda$, where the value of $\Lambda$ is not known.
The recommendation~\cite{AguilarSaavedra:2018nen} is to impose a kinematic cut $E_\text{cut}$ such that the condition
\begin{equation}
E<E_\text{cut} < \Lambda
\end{equation}
is always guaranteed. 
The assumption in this study is that $\Lambda=1$ TeV.
As all the upper limits of kinematic distributions in the current dataset are well below 1 TeV, the results of this analysis are the same for a large range of $E_\text{cut}$.

\clearpage


\subsection{Relevant Operators}

In the scope of this study, only operators that actually contain a top quark are considered.
A brief discussion of those operators which do not contain a top quark but could nevertheless be relevant is given in section~\ref{subsec:openot}.
A complete list of all dimension-six operators can be found in appendix~\ref{app:ope}.

All dimension-six operators are built from the same objects: fermion fields of dimension 3/2, field strength tensors of dimension two, Higgs doublets $\varphi$ of dimension one, and covariant derivatives of dimension one.
If an operator is built from fermion fields, these are involved in the induced interaction.
Field strength tensors result in either a $W$ or $Z$ boson or a photon being involved.
A Higgs boson can be involved in the induced interaction. 
However, if spontaneous symmetry breaking takes place, the Higgs field only contributes with its vacuum expectation value and is thus not actually involved in the induced interaction.
Finally, if an operator includes a covariant derivative of the general form $D_\mu = \partial_\mu + ig \frac{\tau^I}{2} W_\mu^I + \frac{ig'}{2} B_\mu$, the resulting interaction can but does not need to involve a vector boson or a photon.

Table~\ref{tab:SMEFT:params} summarizes the Wilson coefficients that contain a top quark and are constrained by production and decay processes of a single top quark.
The Wilson coefficients used in the fit are often not the ones derived directly from the Warsaw-basis operators, but rather linear combinations of these.
This method is adopted in order to eliminate flat directions.
Consider for example a cross section depending on the operators $A$ and $B$ such that
\begin{equation}
\sigma_\text{SMEFT} = \sigma_\text{SM} +\ \sigma_A \frac{c_A}{\Lambda^2} +\ \sigma_B \frac{c_B}{\Lambda^2},
\label{eq:ope:sigma}
\end{equation}
where $c_A, c_B$ are the corresponding Wilson coefficients.
Further assume that the Wilson coefficients have the same sign and are similar in size, and that there are no contributions at order $\mathcal{O}(\Lambda^{-4})$.
In this case, it is not clear which operator, $A$ or $B$, is constrained by this cross section measurement.
However, if instead of the operators $A$ and $B$ one used $A+B$ and $A-B$, then we end up with one Wilson coefficient that $\sigma_\text{SMEFT}$ is very sensitive to, and one to which it is not.
The latter could be constrained by a measurement of a different physical process. 
With this method, one therefore gets rid of two correlated operators and gets two more distinct ones.

\begin{table}[h]
  \centering
  \begin{tabular}{|llll|}
  \hline
  \textsc{SFitter} & Wilson coeff. & Operators & Interactions \\  
  \hline
  \texttt{D6qq38} & $c_{Qq}^{3,8}$ & $\mathcal{O}_{qq}^{1(i33i)} - \mathcal{O}_{qq}^{3(i33i)} $ & $ttqq$ \\
  \texttt{D6qq31} & $c_{Qq}^{3,1}$ & $\mathcal{O}_{qq}^{3(ii33)} + \frac{1}{6}( \mathcal{O}_{qq}^{1(i33i)} - \mathcal{O}_{qq}^{3(i33i)} )$ & $ttqq$ \\
  \texttt{D6tg} & $c_{tG}$ & $\text{Re}\{\mathcal{O}_{uG}^{(33)} \} $ & $ttg(H), ttgg(H)$ \\  
  \texttt{D6tw} & $c_{tW}$ & $\text{Re}\{\mathcal{O}_{uW}^{(33)} \} $ & $tbW(H), ttZ(H), tt\gamma(H)$ \\  
  \texttt{D6bw} & $c_{bW}$ & $\text{Re}\{\mathcal{O}_{dW}^{(33)} \} $ & $tbW(H)$ \\  
  \texttt{D6phiphi} & $c_{\varphi tb}$ & $\text{Re}\{\mathcal{O}_{\varphi ud}^{(33)} \} $ & $tbW(H)(H)$ \\  
  \texttt{D6phiq3} & $c_{\varphi Q}^3$ & $\mathcal{O}_{\varphi q}^{3(33)} $ & $tt\gamma (H)(H), ttZ(H)(H),$\\  
   &  & & $tbW(H)(H), ttH(H)$\\  
  \texttt{D6tz} & $c_{tZ}$ & $\text{Re}\{ -s_W \mathcal{O}_{uB}^{(33)} + c_W \mathcal{O}_{uW}^{(33)} \} $ & $ttZ(H)$ \\
  \texttt{D6phiqm} & $c_{\varphi Q}^-$ & $ \mathcal{O}_{\varphi q}^{1(33)} - \mathcal{O}_{\varphi q}^{3(33)} $ & $tt\gamma (H)(H), ttZ(H)(H),$\\
     &  & & $ttH(H)$\\  
  \texttt{D6phit} & $c_{\varphi t}$ & $\mathcal{O}_{u\varphi}^{(33)}$ & $ttH(H)(H)$\\
  \hline
  \end{tabular}
  \caption{Operators used in this analysis.
  The relevant Wilson coefficients are listed alongside their \textsc{SFitter} code, the operators they arise from, and the interactions these operators induce. The operators are listed in equations~\ref{eq-SMEFT-ops1} and~\ref{eq-SMEFT-ops2}. A capital $Q$ in the name of the Wilson coefficient indicates that the corresponding operators involve a quark of the third generation, as opposed to a quark of any generation when the name contains a lowercase $q$. The sine and cosine of the weak mixing angle are denoted as $s_W$ and $c_W$, respectively. An $(H)$ denotes an optionally included Higgs boson, i.e. $ttg(H)$ refers to $ttg$ and $ttgH$. The Wilson coefficients $c_i$ are normalized to $\Lambda^{-2}$ in \textsc{SFitter}.
  }
  \label{tab:SMEFT:params}
\end{table}

As an example, $c_{Qq}^{3,8}$ is constituted from $\mathcal{O}_{qq}^{1(i33i)}$ and $\mathcal{O}_{qq}^{3(i33i)}$. 
Using equation~\ref{eq-SMEFT-ops1}, one can compute the operators as 
\begin{equation}
\begin{aligned}
\mathcal{O}_{qq}^{1(i33i)} - \mathcal{O}_{qq}^{3(i33i)} &= 
2(\bar{q}_u \gamma^\mu t)(\bar{b} \gamma_\mu q_d)
+2(\bar{q}_d \gamma^\mu b)(\bar{t} \gamma_\mu q_u) \\
&-2(\bar{q}_u \gamma^\mu b)(\bar{b} \gamma_\mu q_u)
-2(\bar{q}_d \gamma^\mu t)(\bar{t} \gamma_\mu q_d),
\end{aligned}
\label{eq:SMEFT:qq38}
\end{equation}
where $q_u$ and $q_d$ are left-handed up- and down-type quarks of the $i$-th generation, respectively.
One can see from this that the color current forms an octet in interactions such as $q_i q_3 \rightarrow q_i q_3$, just like the corresponding SM interaction which is mediated by a $W$ boson. 
Even though the color flow is similar, these operators induce a new structure because four-quark interactions without a mediator are not present in the SM.
Additionally, equation~\ref{eq:SMEFT:qq38} contributes to interactions like $q_i q_i \rightarrow q_3 q_3$ via the Fiertz identity. 
In these interactions, we have a color singlet, as opposed to an octet in the SM.

The operators that $c_{Qq}^{3,1}$ arises from include the previously discussed operator pair, but also have an additional term arising from $\mathcal{O}_{qq}^{3(ii33)}$. 
Its contribution reads
\begin{equation}
\begin{aligned}
\mathcal{O}_{qq}^{3(ii33)} &= 
2(\bar{q}_u \gamma^\mu q_d)(\bar{b} \gamma_\mu t)
+2(\bar{q}_d \gamma^\mu q_u)(\bar{t} \gamma_\mu b) \\
&+(\bar{q}_u \gamma^\mu q_u)(\bar{t} \gamma_\mu t)
+(\bar{q}_u \gamma^\mu q_u)(\bar{b} \gamma_\mu b) \\
&+(\bar{q}_d \gamma^\mu q_d)(\bar{t} \gamma_\mu t)
+(\bar{q}_d \gamma^\mu q_d)(\bar{b} \gamma_\mu b).
\end{aligned}
\label{eq:SMEFT:qq31}
\end{equation}
In contrast to the findings from equation~\ref{eq:SMEFT:qq38}, we have a color octet in $q_i q_i \rightarrow q_3 q_3$ interactions, and a singlet in $q_i q_3 \rightarrow q_i q_3$ interactions.
This contribution therefore adds additional color structure to the four-quark interactions that are already induced by $\mathcal{O}_{qq}^{1(i33i)} - \mathcal{O}_{qq}^{3(i33i)}$.

While the operators discussed so far induce new interactions, others can also result in modifications of SM interactions.
This is the case with $\mathcal{O}_{uG}^{(33)}$ as it can modify the $ttg$ vertex.
In addition, it can induce new vertices such as $ttgH, ttgg, ttggH$ and $ttggHH$.
Like the operators previously discussed, this operator does not contain a covariant derivative, and is thus independent of the momenta of the involved particles.

Similarly, $\mathcal{O}_{uW}^{(33)}$ and $\mathcal{O}_{dW}^{(33)}$ modify the SM vertices $tbW$ and $ttZ$. 
Both operators do not contain derivatives either.
The main difference between the two operators is that $\mathcal{O}_{uW}^{(33)}$ induces interactions of $ttZ(H)$ and $tt\gamma(H)$, which $\mathcal{O}_{dW}^{(33)}$ does not.
In addition, $\mathcal{O}_{dW}^{(33)}$ cannot contribute in an interference term as there is no coupling between a right handed anti-bottom- and a right handed top quark in the SM.
This is not the case for $\mathcal{O}_{uW}^{(33)}$, which does interfere with the SM.

Similarly to $\mathcal{O}_{dW}^{(33)}$, $\mathcal{O}_{\varphi ud}^{(33)}$ contains a right handed anti-bottom- and top quark coupling and thus cannot interfere with the SM.
That is why there are only quadratic SMEFT contributions from this operator.
In addition, it contains a covariant derivative, implying that its contribution is dependent on the momenta of the involved particles.
This is important insofar as its contribution grows with the energy of the physical process.

The operator $\mathcal{O}_{\varphi q}^{3(33)}$ also contains a covariant derivative and its contribution therefore grows with the process energy.
It modifies couplings of gauge bosons to quarks, and induces new vertices where an additional Higgs boson is involved.
Linear SMEFT terms are possible from this operator.

The Wilson coefficient $c_{tZ}$ is especially sensitive to the $ttZ(H)$ vertex.
This can be understood by considering the effect of electroweak symmetry breaking on the $W_3$ and $B$ bosons,
\begin{equation}
\begin{bmatrix}
\gamma \\
Z^0
\end{bmatrix}
=
\begin{bmatrix}
c_W & s_W \\
-s_W & c_W
\end{bmatrix}
\begin{bmatrix}
B \\
W_3
\end{bmatrix}.
\label{eq:SMEFT:ewsb}
\end{equation}
With the linear combination involving the electroweak mixing angles $-s_W$ and $c_W$, one gets exactly the contribution involving a $Z$ boson.

The Wilson coefficient $c_{\varphi Q}^-$ differs from $c_{\varphi Q}^3$ mainly through the contribution from $\mathcal{O}_{\varphi q}^{1(33)}$.
Because of the subtraction of covariant derivatives containing gauge bosons and the Pauli matrices in $\mathcal{O}_{\varphi q}^{1(33)}$, any vertices involving $W$ bosons drop out. 
That is why this coefficient is, similarly to $c_{tZ}$, particularly important for processes that involve a $Z$ boson.

Finally, $c_{\varphi t}$ modifies the $ttH$ vertex, and induces new $ttHH$ and $ttHHH$ interactions.
It contains no derivative, meaning that there is no growing contribution at higher momenta of the involved particles.

\clearpage


\subsection{Operators not involving top quarks}
\label{subsec:openot}

As stated before, only use operators are used that involve at least one top quark and thus assume that all other operators are well constrained by other measurements.
In this section, a brief overview is given over operators that could be relevant for the single top quark sector but do not involve a top quark, and how these operators are constrained.

Operators involving a modification of the electroweak gauge-boson couplings to light fermions are relevant for the interpretation of the single top and $tZ$ measurements.
In the Warsaw basis, and under the assumed flavor structure, these are $\mathcal{O}_{\phi q}^{(1)},\ \mathcal{O}_{\phi q}^{(3)},\ \mathcal{O}_{\phi u},$ $\mathcal{O}_{\phi d},\ \mathcal{O}_{\phi l}^{(3)},\ \mathcal{O}_{\phi l}^{(1)},\ \mathcal{O}_{\phi e},\ \mathcal{O}_{ll}^{(3)},\ \mathcal{O}_{\phi WB}$, and $ \mathcal{O}_{\phi D}$.
Among these 10 operators, 8 degrees of freedom are well constrained by electroweak observables~\cite{Falkowski:2014tna}.
The other two directions can only be constrained by diboson production processes~\cite{Alonso:2013hga,Grojean:2006nn,Brivio:2017bnu}.
They can be parametrized as~\cite{Hagiwara:1993ck}
\begin{equation}
\begin{aligned}
& \mathcal{O}_{HW} = ( D^\mu \varphi )^\dagger \tau_I ( D^\nu \varphi ) W_{\mu\nu}^I, \\
& \mathcal{O}_{HB} = ( D^\mu \varphi )^\dagger ( D^\nu \varphi ) B_{\mu\nu}. \\
\end{aligned}
\end{equation}

Together with the basis operator $\mathcal{O}_W$, they form the set of operators that modify the triple-gauge-boson couplings (TGC). 
These couplings would enter the $tZ$ process, but they are well constrained from diboson production at LEP2.

One aspect is that processes like the single top quark production in association with a $Z$ boson (called $tZ$ production for short) could enhance the sensitivity to anomalous TGC. 
The reason is that diagrams of $tZ$ production in the SM have large cancellations among each other as required by unitarity.
These cancellations are spoiled by anomalous TGC, leading to enhanced cross sections at large energies.
As found in~\cite{Degrande:2018fog}, this effect is present but small compared to the sensitivities from diboson production.
In this thesis, these operators in the associated production of top quarks are neglected.

In addition, $tZ$ production, with a non-resonant $Z$ boson, and the helicity fractions of a $W$ boson from top decay could be affected by two-lepton-two-quark operators, such as $( \bar{t} \gamma^\mu t) ( \bar{e} \gamma_\mu e )$ and  $( \bar{Q} \gamma^\mu \tau^I Q)( \bar{l} \gamma_\mu \tau^I l )$.
The problem is that including these operators implies a reinterpretation of the experimental measurements, for example of the extra- polation from the fiducial to the total phase space.
These operators are therefore not covered in this analysis and left for future work.

In this study, I only use operators that contain a top quark.
It needs to be emphasized, however, that this is not a good approximation in general and thus needs to be checked for each sector of physical processes.
If they are not constrained well enough by other analyses, operators not containing the particle produced in the process do need to be included.

\clearpage


\subsection{NLO QCD effects}
\label{subsec:SMEFT:NLO}

Using predictions at NLO QCD level make sense given the high precision of available top quark measurements.
There are several reasons why NLO QCD corrections to SMEFT effects are necessary, including the following:
\begin{itemize}
\item QCD corrections to total rates can be large, especially for processes that are proportional to $\alpha_s$ at Born level. Additionally, NLO QCD corrections reduce the theoretical uncertainties from scale variations, thus leading to a potential improvement of the bounds on the SMEFT Wilson coefficients.
\item QCD corrections can distort the distributions of key observables. Given that differential distributions are included in the fit, not including QCD corrections to certain differential distributions could lead to incorrect conclusions on the nature of BSM physics.
\item Given the high precision of recent experimental measurements, the SM predictions of the differential kinematic differential distributions are implemented at NNLO QCD. These corrections are found using $K$-factors on the NLO predictions.
\end{itemize}

For the decay processes of a top quark, the SM prediction at NLO QCD is available in the form of analytical results~\cite{Zhang:2014rja}.
For the predictions of almost all other processes, a fully automated MC simulation at NLO QCD is used for the SM prediction and SMEFT effects. 
The only exception are the predictions for the quadratic SMEFT terms for the single top quark production in association with a $Z$ boson, where the predictions at LO are used as those at NLO are not available to the author at the time of publishing this thesis.
The theoretical uncertainties of the SMEFT effects are neglected in the fit as they are small compared to the other uncertainties.

\clearpage

%
%
%
%
%
%
%
%
%
%
%
%
%
%
%
%
%
%
%
%


\section{Measurements and predictions}
\label{sec_data}

In the last section, SMEFT and the dimension-six operators that are relevant for single top quark production and decay were introduced.
The aim of this section is to gain more intuition about the sensitivity of the Wilson coefficients to the different physical processes. 
\subsection{Physical Processes}


\subsubsection{Single Top Quark Production in $s$- and $t$-channel}
\label{subsubsec:data:ts}

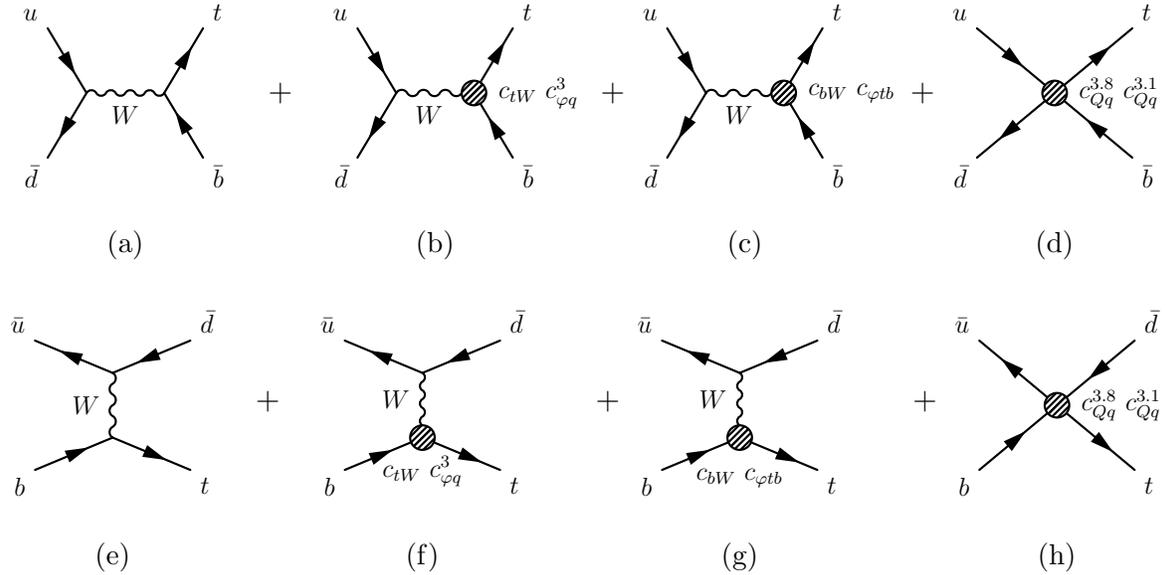
\begin{figure}[h]
\centering
\vspace{3mm}
\begin{subfigure}{.18\textwidth}
  \centering
  \vspace{0.5cm}
  \resizebox{\textwidth}{!}{
  \begin{fmfgraph*}(90,60)
  \fmfleft{i1,i2}
  \fmfright{f1,f2}
  
  \fmflabel{$\bar{d}$}{i1}
  \fmflabel{$u$}{i2}
  \fmflabel{$\bar{b}$}{f1}
  \fmflabel{$t$}{f2}
  
  \fmf{fermion}{i2,v1,i1}
  \fmf{fermion}{f1,v2,f2}
  \fmf{photon,label=$W$}{v1,v2}
  
  \end{fmfgraph*}
  }
  \vspace{2mm}
  \caption{}
\end{subfigure}%
\hspace{5mm}%
\begin{subfigure}{.02\textwidth}%
\vspace{-9mm}
+
\end{subfigure}%
\hspace{5mm}%
\begin{subfigure}{.18\textwidth}
  \centering
  \vspace{0.5cm}
  \resizebox{\textwidth}{!}{
  \begin{fmfgraph*}(90,60)
  \fmfleft{i1,i2}
  \fmfright{f1,f2}
  
  \fmflabel{$\bar{d}$}{i1}
  \fmflabel{$u$}{i2}
  \fmflabel{$\bar{b}$}{f1}
  \fmflabel{$t$}{f2}
  
  \fmf{fermion}{i2,v1,i1}
  \fmf{fermion}{f1,v2,f2}
  \fmf{photon,label=$W$}{v1,v2}
  
  \fmfblob{4mm}{v2}
  \fmfv{label.dist=4mm, label=$c_{tW}\ c_{\varphi q}^3$}{v2}
  
  \end{fmfgraph*}
  }
  \vspace{2mm}
  \caption{}
\end{subfigure}%
\hspace{8mm}%
\begin{subfigure}{.02\textwidth}%
\vspace{-9mm}
+
\end{subfigure}%
\hspace{2mm}%
\begin{subfigure}{.18\textwidth}
  \centering
  \vspace{0.5cm}
  \resizebox{\textwidth}{!}{
  \begin{fmfgraph*}(90,60)
  \fmfleft{i1,i2}
  \fmfright{f1,f2}
  
  \fmflabel{$\bar{d}$}{i1}
  \fmflabel{$u$}{i2}
  \fmflabel{$\bar{b}$}{f1}
  \fmflabel{$t$}{f2}
  
  \fmf{fermion}{i2,v1,i1}
  \fmf{fermion}{f1,v2,f2}
  \fmf{photon,label=$W$}{v1,v2}

  \fmfblob{4mm}{v2}
  \fmfv{label.dist=4mm, label=$c_{bW}\ c_{\varphi tb}$}{v2}  
  
  \end{fmfgraph*}
  }
  \vspace{2mm}
  \caption{}
\end{subfigure}%
\hspace{8mm}%
\begin{subfigure}{.02\textwidth}%
\vspace{-9mm}
+
\end{subfigure}%
\hspace{2mm}%
\begin{subfigure}{.18\textwidth}
  \centering
  \vspace{0.5cm}
  \resizebox{\textwidth}{!}{
  \begin{fmfgraph*}(90,60)
  \fmfleft{i1,i2}
  \fmfright{f1,f2}
  
  \fmflabel{$\bar{d}$}{i1}
  \fmflabel{$u$}{i2}
  \fmflabel{$\bar{b}$}{f1}
  \fmflabel{$t$}{f2}
  
  \fmf{phantom}{i2,v1,f1}  
  \fmffreeze
  \fmf{fermion}{i2,v1,f2}
  \fmf{fermion}{f1,v1,i1}

  \fmfblob{4mm}{v1}
  \fmfv{label.dist=4mm, label=$c_{Qq}^{3.8}\ c_{Qq}^{3.1}$}{v1}  
    
  \end{fmfgraph*}
  }
  \vspace{2mm}
  \caption{}
\end{subfigure} 
\begin{subfigure}{0.18\textwidth}
  \centering
  \vspace{10mm}
  \resizebox{\textwidth}{!}{
  \begin{fmfgraph*}(90,60)
  \fmfleft{i1,i2}
  \fmfright{f1,f2}
  
  \fmflabel{$b$}{i1}
  \fmflabel{$\bar{u}$}{i2}
  \fmflabel{$t$}{f1}
  \fmflabel{$\bar{d}$}{f2}
  
  \fmf{fermion}{i1,v1,f1}
  \fmf{fermion}{f2,v2,i2}
  \fmf{photon,label=$W$}{v1,v2}
  
  \end{fmfgraph*}
  }
  \vspace{2mm}
  \caption{}
\end{subfigure}%
\hspace{5mm}%
\begin{subfigure}{.02\textwidth}%
\vspace{-5mm}
+
\end{subfigure}%
\hspace{5mm}%
\begin{subfigure}{0.18\textwidth}
  \centering
  \vspace{10mm}
  \resizebox{\textwidth}{!}{
  \begin{fmfgraph*}(90,60)
  \fmfleft{i1,i2}
  \fmfright{f1,f2}
  
  \fmflabel{$b$}{i1}
  \fmflabel{$\bar{u}$}{i2}
  \fmflabel{$t$}{f1}
  \fmflabel{$\bar{d}$}{f2}
  
  \fmf{fermion}{i1,v1,f1}
  \fmf{fermion}{f2,v2,i2}
  \fmf{photon,label=$W$}{v1,v2}
  
  \fmfblob{4mm}{v1}
  \fmfv{label.dist=3mm, label=$c_{tW}\ c_{\varphi q}^3$}{v1}
  
  \end{fmfgraph*}
  }
  \vspace{2mm}
  \caption{}
\end{subfigure}%
\hspace{9mm}%
\begin{subfigure}{.02\textwidth}%
\vspace{-5mm}
+
\end{subfigure}%
\hspace{2mm}%
\begin{subfigure}{0.18\textwidth}
  \centering
  \vspace{10mm}
  \resizebox{\textwidth}{!}{
  \begin{fmfgraph*}(90,60)
  \fmfleft{i1,i2}
  \fmfright{f1,f2}
  
  \fmflabel{$b$}{i1}
  \fmflabel{$\bar{u}$}{i2}
  \fmflabel{$t$}{f1}
  \fmflabel{$\bar{d}$}{f2}
  
  \fmf{fermion}{i1,v1,f1}
  \fmf{fermion}{f2,v2,i2}
  \fmf{photon,label=$W$}{v1,v2}
  
  \fmfblob{4mm}{v1}
  \fmfv{label.dist=5mm, label=$c_{bW}\ c_{\varphi tb}$}{v1}
  
  \end{fmfgraph*}
  }
  \vspace{2mm}
  \caption{}
\end{subfigure}%
\hspace{9mm}%
\begin{subfigure}{.02\textwidth}%
\vspace{-5mm}
+
\end{subfigure}%
\hspace{2mm}%
\begin{subfigure}{0.18\textwidth}
  \centering
  \vspace{10mm}
  \resizebox{\textwidth}{!}{
  \begin{fmfgraph*}(90,60)
  \fmfleft{i1,i2}
  \fmfright{f1,f2}
  
  \fmflabel{$b$}{i1}
  \fmflabel{$\bar{u}$}{i2}
  \fmflabel{$t$}{f1}
  \fmflabel{$\bar{d}$}{f2}

  \fmf{phantom}{i2,v1,f1}  
  \fmffreeze
  \fmf{fermion}{f2,v1,i2}
  \fmf{fermion}{i1,v1,f1}

  \fmfblob{4mm}{v1}
  \fmfv{label.dist=4mm, label=$c_{Qq}^{3.8}\ c_{Qq}^{3.1}$}{v1}  
    
  \end{fmfgraph*}
  }
  \vspace{2mm}
  \caption{}
\end{subfigure} 
\vspace{3mm}
\caption{Example Feynman diagrams for the single top quark $s$- and $t$-channel production. Subfigures (a-d) are the $s$-channel, (e-h) the $t$-channel diagrams. Subfigures (a) and (e) show the SM amplitude, (b) and (f) the corrections from $c_{tW}$ and $c_{\phi q}^3$, (c) and (g) those from $c_{bW}$ and $c_{\varphi tb}$, and (d) and (h) the four-fermion interaction from $c_{Qq}^{3,8}$ and $c_{Qq}^{3,1}$.
The contributions of $c_{tW}$ and $c_{\phi q}^3$, and $c_{bW}$ and $c_{\varphi tb}$ are shown separately to emphasize the different structures of the new interactions.
}
\vspace{5mm}
\label{img:data:feynst}
\end{figure}

Figure~\ref{img:data:feynst} displays example Feynman diagrams of the SM and the additional contributions by the dimension-six operators.
The modifications and new interactions induced by these operators have been marked with the Wilson coefficients for conciseness.

The different structures of the operators result in different sensitivities to a given process.
As an example, the SM cross section of $s$-channel single top production with the contributions of $c_{\varphi q}^3$, $c_{tW}$ and $c_{Qq}^{3,1}$ at order $\mathcal{O}(\Lambda^{-2})$ reads~\cite{Zhang:2010dr}
\begin{equation}
\begin{aligned}
\sigma_{u\bar{d}\rightarrow t\bar{b}} & = \left( 1 + \frac{2c_{\varphi q}^3 v^2}{\Lambda^2} \right) \frac{g^4 (s-m_t^2)^2(2s+m_t^2)}{384\pi s^2 (s-m_W^2)^2} \\
& +\ c_{tW}\ \frac{g^2m_t m_W (s-m_t^2)^2} {8\sqrt{2} \pi \Lambda^2 s(s-m_W^2)^2} \\
& +\ c_{Qq}^{3,1}\ \frac{g^2(s-m_t^2)^2(2s+m_t^2)} {48\pi \Lambda^2 s^2 (s-m_W^2)},
\end{aligned}
\label{eq:data:Xs}
\end{equation}
where $g$ is the electroweak coupling constant, and $s$ and $t$ are the usual Mandelstam variables: $s = (p_t - p_b)^2$ for $s$- and $t$-channel, $t= (p_u - p_b)^2$ for $s$-channel and $t= (p_u - p_t)^2$ for $t$-channel processes.
One can see from this expression that the influence of $c_{\varphi q}^3$ simply results in a modification of the SM cross section as it has exactly the same structure.
Even though the contribution induced by $\mathcal{O}_{\varphi q}^3$ grows with the momentum of the involved particles, a kinematic distribution of the $s$-channel cross section will not increase the sensitivity to $c_{\varphi q}^3$ as its contribution represents a constant offset from the SM term.
In contrast to this, $c_{tW}$ and $c_{Qq}^{3,1}$ do not arise from operators that contain derivatives like the former, but show a different dependence on the center-of-mass energy $\sqrt{s}$ than the SM term.

The cross section of $t$-channel single top quark production with the same effective contributions has a similar structure,
\begin{equation}
\begin{aligned}
\sigma_{ub\rightarrow dt} & = \left( 1 + \frac{2c_{\varphi q}^3 v^2}{\Lambda^2} \right) \frac{g^4 (s-m_t^2)^2}{64\pi  s\ m_W^2 (s-m_t^2+m_W^2)} \\
& -\ c_{tW}\ \frac{g^2m_t m_W \left(s-m_t^2 - (s - m_t^2 +m_W^2)\text{log} \frac{s-m_t^2+m_W^2} {m_W^2} \right)} {4\sqrt{2} \pi \Lambda^2 s(s-m_t^2+m_W^2)} \\
& -\ c_{Qq}^{3,1}\ \frac{g^2(s-m_t^2) \text{log} \frac{s-m_t^2+m_W^2} {m_W^2} } {8\pi \Lambda^2s}.
\end{aligned}
\label{eq:data:Xt}
\end{equation}
What is important is that the contributions from $c_{tW}$ and $c_{Qq}^{3,1}$ have a different sign and a different dependency on the center-of-mass energy than in $s$-channel production, see equation~\ref{eq:data:Xs}.
That is why it makes sense to use the $s$- and $t$-channel measurements separately in the fit.

For the $t$-channel single top quark production, kinematic distributions are available. 
These can be used to better constrain operators that have a different dependency on the center-of-mass energy than the SM contribution, like $c_{tW}$ and $c_{Qq}^{3,1}$.

\begin{figure}[h!]
\centering
\vspace{3mm}
\begin{subfigure}{.48\textwidth}
  \centering
  \includegraphics[width=\textwidth]{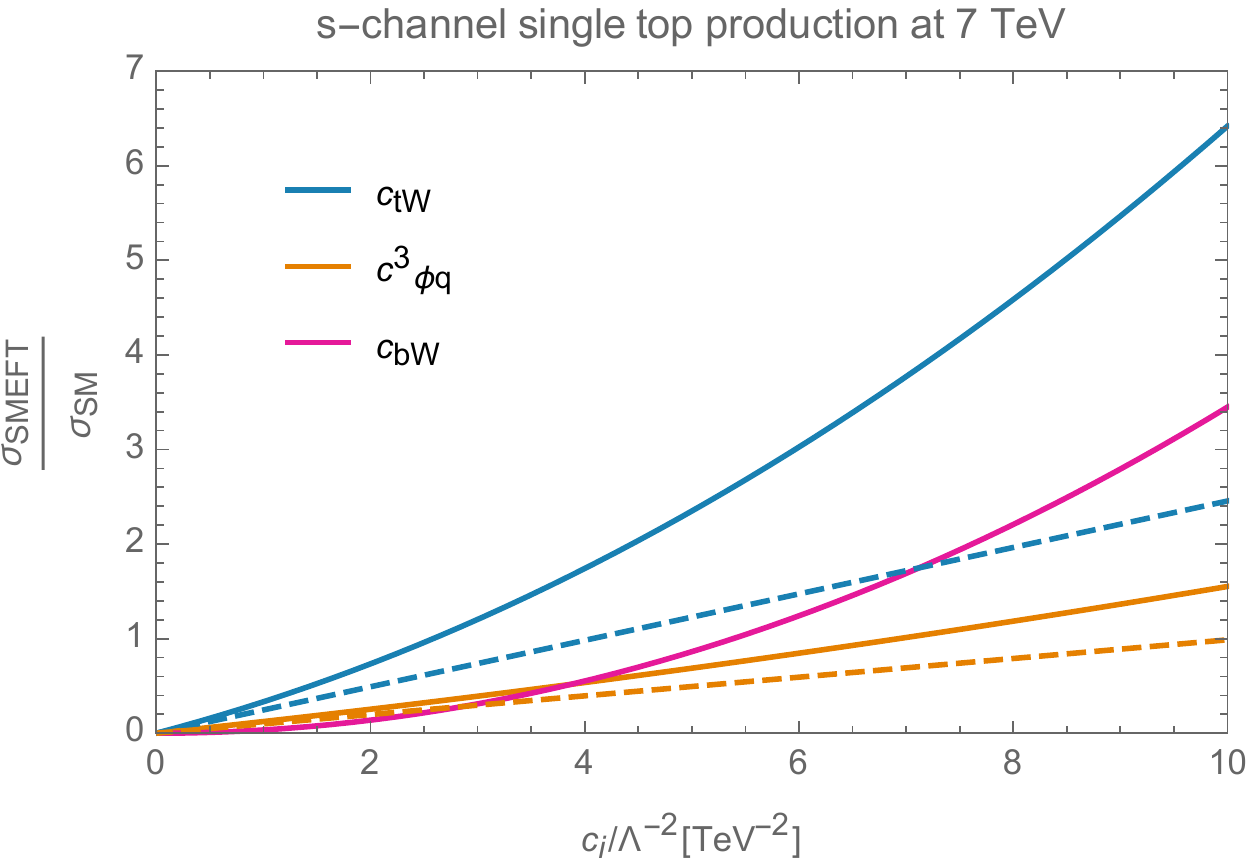}
  \caption{}
\end{subfigure}%
\hspace{.04\textwidth}%
\begin{subfigure}{.48\textwidth}
  \centering
  \includegraphics[width=\textwidth]{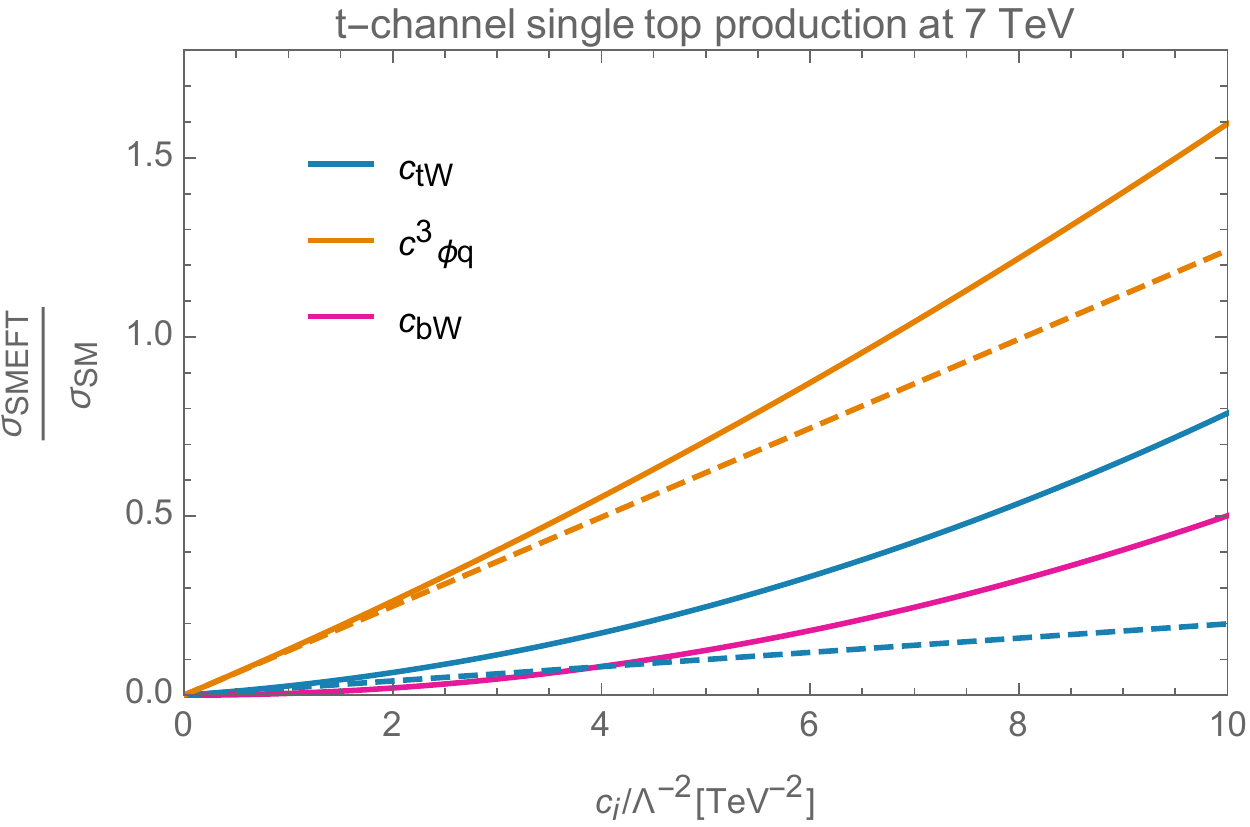}
  \caption{}
\end{subfigure}
\begin{subfigure}{.48\textwidth}
  \centering
  \includegraphics[width=\textwidth]{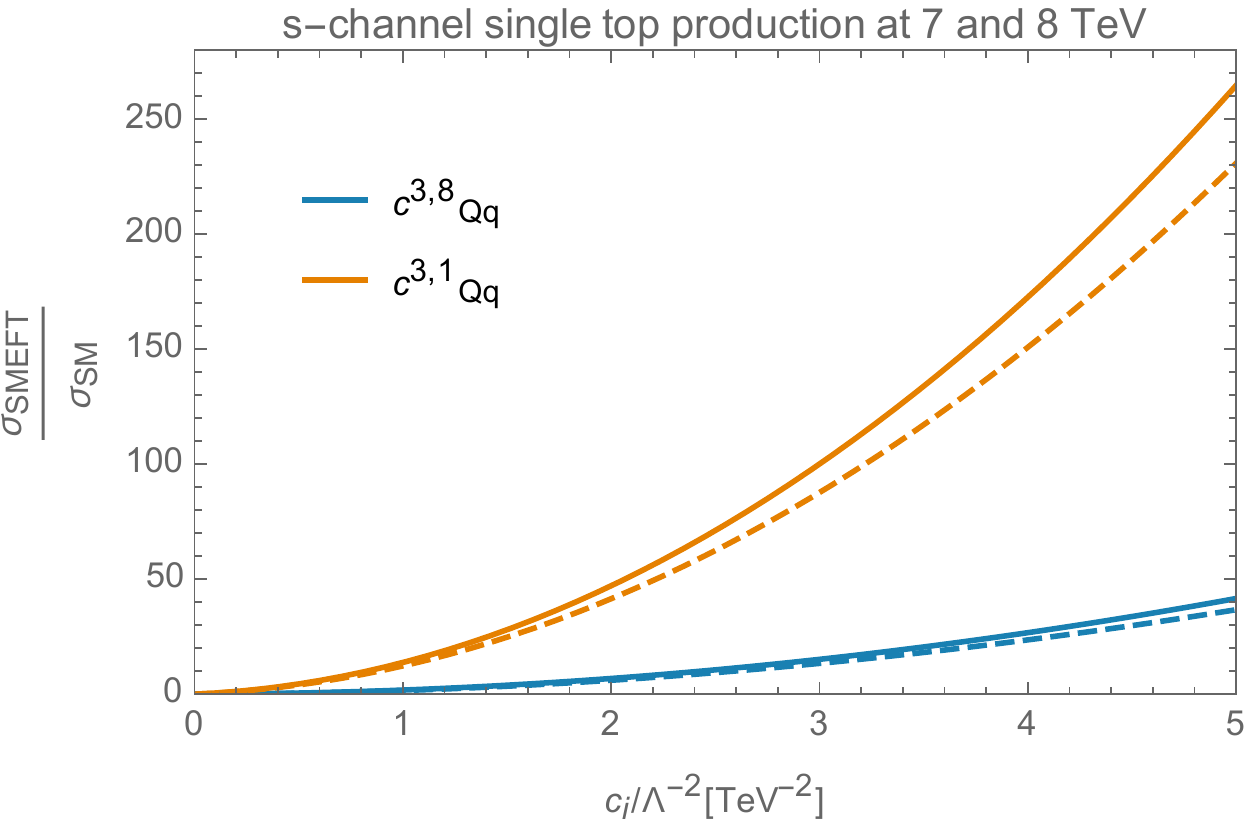}
  \caption{}
\end{subfigure}%
\hspace{.04\textwidth}%
\begin{subfigure}{.48\textwidth}
  \centering
  \includegraphics[width=\textwidth]{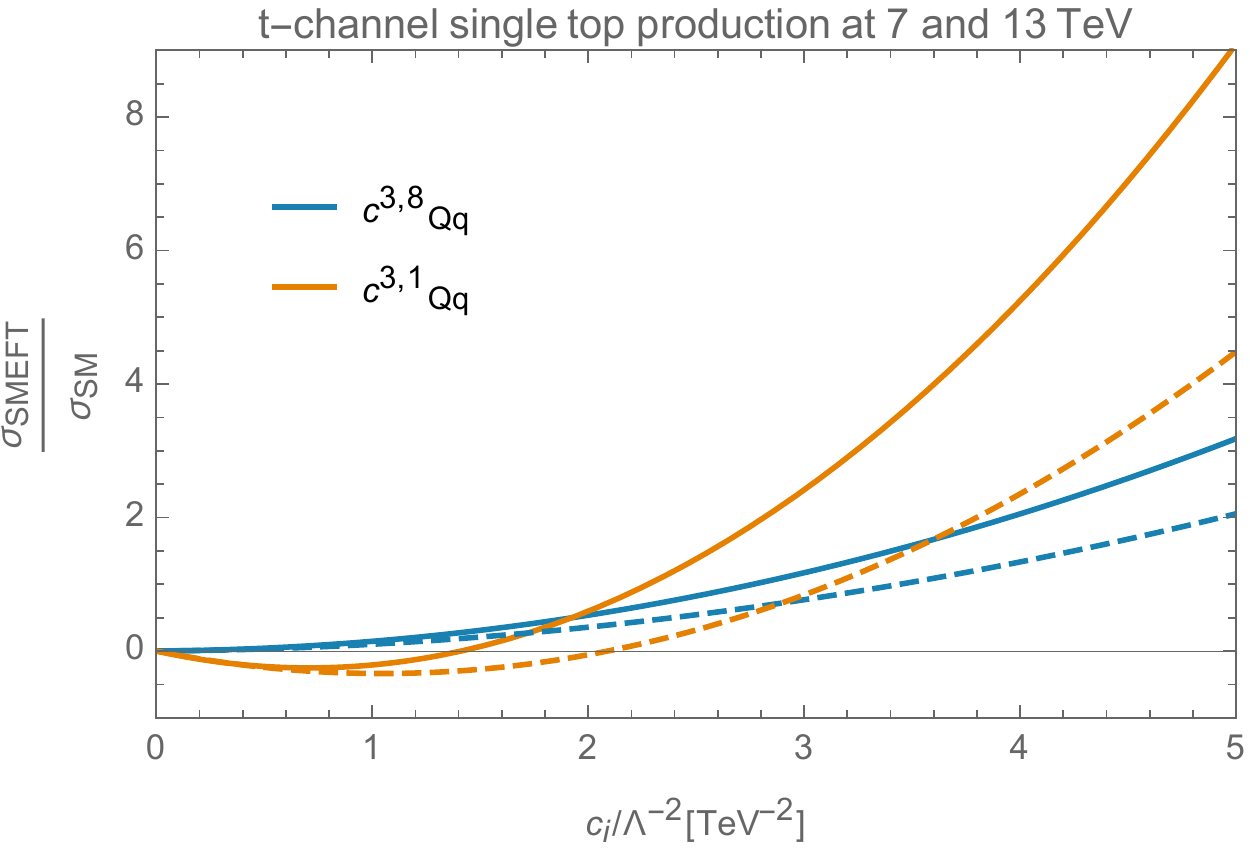}
  \caption{}
\end{subfigure}
\caption{Ratii of the cross sections of $s$- and $t$-channel single top quark production with the contributions of different operators, as a function of the corresponding Wilson coefficient.
  Subfigure (a): corrections of $c_{tW}$, $c_{\varphi q}^3$ and $c_{bW}$ in $s$-channel production at 7 TeV. The dashed lines mark the contributions at order $\mathcal{O}(\Lambda^{-2})$, the solid lines those at order $\mathcal{O}(\Lambda^{-4})$. 
  The contribution of $c_{bW}$ at order $\mathcal{O}(\Lambda^{-2})$ is zero.
  Subfigure (b): same as (a) for $t$-channel production.
  Subfigure (c): corrections of $c_{Qq}^{3,8}$ and $c_{Qq}^{3,1}$ in $s$-channel production at 7 TeV (dashed) and 8 TeV (solid line) at order $\mathcal{O}(\Lambda^{-4})$.
  Subfigure (d): corrections of $c_{Qq}^{3,8}$ and $c_{Qq}^{3,1}$ in $t$-channel production at 7 TeV (dashed) and 13 TeV (solid line) at order $\mathcal{O}(\Lambda^{-4})$.
  }
\label{img:data:st713}
\end{figure}

Figure~\ref{img:data:st713} shows the ratii of the SMEFT cross sections involving various operators to the SM cross sections.
One can clearly see that most operators are a lot more sensitive to $s$-channel- than to $t$-channel production.
For example, at $c_{Qq}^{3,1}=5$, the $t$-channel cross section at 7 TeV increases by a factor of 4.5, while the $s$-channel cross section at the same energy increases by a factor of 230.

One interesting aspect is also that $c_{Qq}^{3,1}$ contributes negative values to the $t$-channel cross section in lower ranges but positive ones to $s$-channel cross sections, as already shown in equations~\ref{eq:data:Xs} and~\ref{eq:data:Xt}.
In contrast to that, the contribution of $c_{tW}$ to the $t$-channel cross section remains positive.
One can see that in the second line in equation~\ref{eq:data:Xt}, which is positive for all $s$.

One could conclude that the dimension-six operators are way more sensitive to $s$-channel- than $t$-channel single top quark production. 
In contrast to the $s$-channel, however, for the $t$-channel production kinematic distributions of the differential cross section as a function of the transverse momentum of the top quark or its rapidity are available.
These could enhance the sensitivities of operators that are dependent on these variables, e.g. $c_{tW}$ and $c_{Qq}^{3,1}$ (equation~\ref{eq:data:Xt}).



\subsubsection{Single Top Quark Production in Association with a Vector Boson}

\begin{figure}[h]
\centering
\vspace{3mm}
\begin{subfigure}{.18\textwidth}
  \centering
  \vspace{0.5cm}
  \resizebox{\textwidth}{!}{
  \begin{fmfgraph*}(90,60)
  \fmfleft{i1,i2}
  \fmfright{f1,f2}
  
  \fmflabel{$b$}{i1}
  \fmflabel{$g$}{i2}
  \fmflabel{$W$}{f1}
  \fmflabel{$t$}{f2}
  
  \fmf{fermion}{i1,v1}
  \fmf{fermion, label=$t$}{v1,v2}
  \fmf{fermion}{v2,f2}
  \fmf{gluon}{v2,i2}
  \fmf{photon}{f1,v1}
  
  \end{fmfgraph*}
  }
  \vspace{2mm}
  \caption{}
\end{subfigure}%
\hspace{5mm}%
\begin{subfigure}{.02\textwidth}%
\vspace{-9mm}
+
\end{subfigure}%
\hspace{5mm}%
\begin{subfigure}{.18\textwidth}
  \centering
  \vspace{0.5cm}
  \resizebox{\textwidth}{!}{
  \begin{fmfgraph*}(90,60)
  \fmfleft{i1,i2}
  \fmfright{f1,f2}
  
  \fmflabel{$b$}{i1}
  \fmflabel{$g$}{i2}
  \fmflabel{$W$}{f1}
  \fmflabel{$t$}{f2}
  
  \fmf{gluon}{v1,i2}
  \fmf{fermion}{i1,v1}
  \fmf{fermion, label=$b$}{v1,v2}
  \fmf{fermion}{v2,f2}
  \fmf{photon}{v2,f1}
  
  \end{fmfgraph*}
  }
  \vspace{2mm}
  \caption{}
\end{subfigure}%
\hspace{5mm}%
\begin{subfigure}{.02\textwidth}%
\vspace{-9mm}
+
\end{subfigure}%
\hspace{5mm}%
\begin{subfigure}{.18\textwidth}
  \centering
  \vspace{0.5cm}
  \resizebox{\textwidth}{!}{
  \begin{fmfgraph*}(90,60)
  \fmfleft{i1,i2}
  \fmfright{f1,f2}
  
  \fmflabel{$b$}{i1}
  \fmflabel{$H$}{i2}
  \fmflabel{$W$}{f1}
  \fmflabel{$t$}{f2}
  
  \fmf{fermion}{i1,v1}
  \fmf{fermion, label=$t$}{v1,v2}
  \fmf{fermion}{v2,f2}
  \fmf{dashes}{v2,i2}
  \fmf{photon}{f1,v1} 
  
  \end{fmfgraph*}
  }
  \vspace{2mm}
  \caption{}
\end{subfigure}%
\hspace{5mm}%
\begin{subfigure}{.02\textwidth}%
\vspace{-9mm}
+
\end{subfigure}%
\hspace{5mm}%
\begin{subfigure}{.18\textwidth}
  \centering
  \vspace{0.5cm}
  \resizebox{\textwidth}{!}{
  \begin{fmfgraph*}(90,60)
  \fmfleft{i1,i2}
  \fmfright{f1,f2}
  
  \fmflabel{$b$}{i1}
  \fmflabel{$H$}{i2}
  \fmflabel{$W$}{f1}
  \fmflabel{$t$}{f2}
  
  \fmf{dashes}{v1,i2}
  \fmf{fermion}{i1,v1}
  \fmf{fermion, label=$b$}{v1,v2}
  \fmf{fermion}{v2,f2}
  \fmf{photon}{v2,f1}
    
  \end{fmfgraph*}
  }
  \vspace{2mm}
  \caption{}
\end{subfigure} 
\begin{subfigure}{.02\textwidth}%
\vspace{-5mm}
+
\end{subfigure}%
\hspace{4mm}%
\begin{subfigure}{0.18\textwidth}
  \centering
  \vspace{10mm}
  \resizebox{\textwidth}{!}{
  \begin{fmfgraph*}(90,60)
  \fmfleft{i1,i2}
  \fmfright{f1,f2}
  
  \fmflabel{$b$}{i1}
  \fmflabel{$g$}{i2}
  \fmflabel{$W$}{f1}
  \fmflabel{$t$}{f2}
  
  \fmf{fermion}{i1,v1}
  \fmf{fermion, label=$t$}{v1,v2}
  \fmf{fermion}{v2,f2}
  \fmf{gluon}{v2,i2}
  \fmf{photon}{f1,v1}
  
  \fmfblob{4mm}{v2}
  \fmfv{label.dist=4mm, label=$c_{tG}$}{v2}
  
  \end{fmfgraph*}
  }
  \vspace{2mm}
  \caption{}
\end{subfigure}%
\hspace{4mm}%
\begin{subfigure}{.02\textwidth}%
\vspace{-5mm}
+
\end{subfigure}%
\hspace{4mm}%
\begin{subfigure}{0.18\textwidth}
  \centering
  \vspace{10mm}
  \resizebox{\textwidth}{!}{
  \begin{fmfgraph*}(90,60)
  \fmfleft{i1,i2}
  \fmfright{f1,f2}
  
  \fmflabel{$b$}{i1}
  \fmflabel{$g$}{i2}
  \fmflabel{$W$}{f1}
  \fmflabel{$t$}{f2}
  
  \fmf{fermion}{i1,v1}
  \fmf{fermion, label=$t$}{v1,v2}
  \fmf{fermion}{v2,f2}
  \fmf{gluon}{v2,i2}
  \fmf{photon}{f1,v1}
  
  \fmfblob{4mm}{v1}
  \fmfv{label.dist=3mm, label=$c_{tW}\ c_{\varphi q}^3$}{v1}
  
  \end{fmfgraph*}
  }
  \vspace{2mm}
  \caption{}
\end{subfigure}%
\hspace{4mm}%
\begin{subfigure}{.02\textwidth}%
\vspace{-5mm}
+
\end{subfigure}%
\hspace{4mm}%
\begin{subfigure}{0.18\textwidth}
  \centering
  \vspace{10mm}
  \resizebox{\textwidth}{!}{
  \begin{fmfgraph*}(90,60)
  \fmfleft{i1,i2}
  \fmfright{f1,f2}
  
  \fmflabel{$b$}{i1}
  \fmflabel{$H$}{i2}
  \fmflabel{$W$}{f1}
  \fmflabel{$t$}{f2}
  
  \fmf{fermion}{i1,v1}
  \fmf{fermion, label=$t$}{v1,v2}
  \fmf{fermion}{v2,f2}
  \fmf{dashes}{v2,i2}
  \fmf{photon}{f1,v1}
  
  \fmfblob{4mm}{v2}
  \fmfv{label.dist=3mm, label=$c_{\varphi q}^3$}{v2}
  
  \end{fmfgraph*}
  }
  \vspace{2mm}
  \caption{}
\end{subfigure}%
\hspace{4mm}%
\begin{subfigure}{.02\textwidth}%
\vspace{-5mm}
+
\end{subfigure}%
\hspace{4mm}%
\begin{subfigure}{0.18\textwidth}
  \centering
  \vspace{10mm}
  \resizebox{\textwidth}{!}{
  \begin{fmfgraph*}(90,60)
  \fmfleft{i1,i2}
  \fmfright{f1,f2}
  
  \fmflabel{$b$}{i1}
  \fmflabel{$H$}{i2}
  \fmflabel{$W$}{f1}
  \fmflabel{$t$}{f2}
  
  \fmf{fermion}{i1,v1}
  \fmf{fermion, label=$t$}{v1,v2}
  \fmf{fermion}{v2,f2}
  \fmf{dashes}{v2,i2}
  \fmf{photon}{f1,v1}
  
  \fmfblob{4mm}{v1}
  \fmfv{label.dist=4mm, label=$c_{bW}\ c_{\varphi tb}$}{v1}
    
  \end{fmfgraph*}
  }
  \vspace{2mm}
  \caption{}
\end{subfigure} 
\vspace{3mm}
\caption{Example Feynman diagrams for the single top quark production in association with a $W$ boson. Subfigures (a-d) represent the SM contributions, figures (a-h) contributions induced from dimension-six operators.
Subfigure (e) shows corrections from $c_{tG}$, (f) those from $c_{tW}$ and $c_{\varphi q}^3$, (g) another one from $c_{\varphi q}^3$, and (h) those from $c_{bW}$ and $c_{\varphi tb}$.
The contributions of $c_{tW}$ and $c_{\phi q}^3$, and $c_{bW}$ and $c_{\varphi tb}$ are shown separately to emphasize the different structures of the new interactions.
}
\vspace{5mm}
\label{img:data:feyntW}
\end{figure}
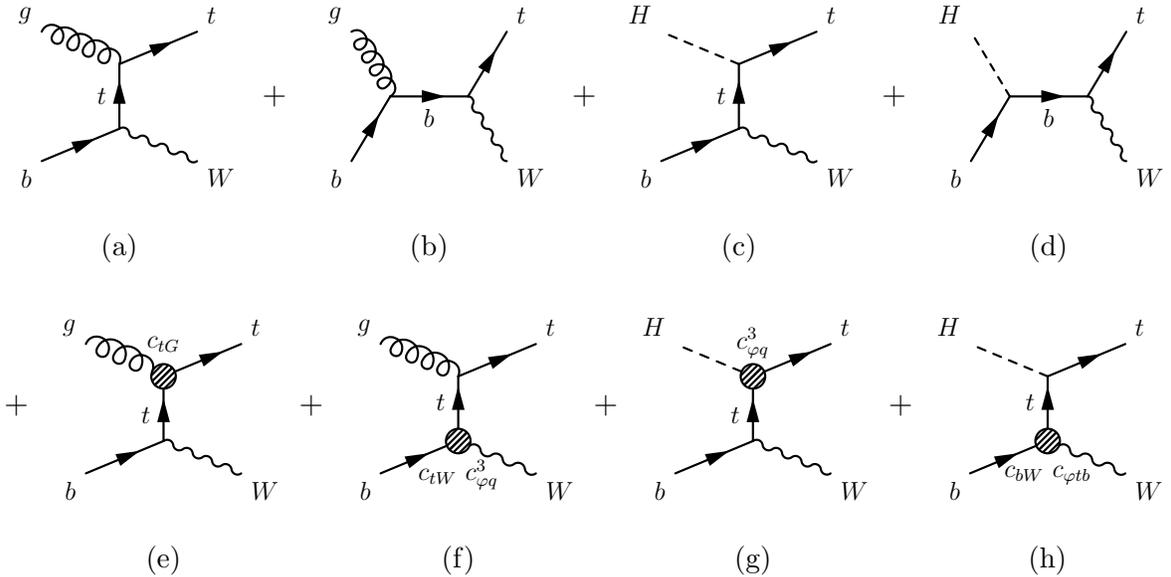

Figure~\ref{img:data:feyntW} shows example Feynman diagrams for the single top quark production in association with a $W$ boson. 
While the rate of the process $gb\rightarrow tW$ is much larger than that of $Hg\rightarrow tW$, it is displayed to emphasize the connection between Higgs and top quark physics.
One can also see from subfigure (b) that $c_{\varphi q}^3$ modifies the $ttH$ vertex, making this an interesting production channel.
In subfigure (h), the Higgs boson could be exchanged by a gluon.
As already mentioned, the total rate of the process involving a gluon is a lot higher, but the involvement of a Higgs boson is a valid option.

The fact that the production of a single top quark in association with a $W$ boson does not constrain the four-quark operators is clear - the process involves less than four quarks, even with next-to-leading-order loop corrections.
Like the single top quark production in $s$- and $t$-channel, this process constrains the operators that correspond to $c_{tW}$, $c_{\phi q}^3$, $c_{bW}$ and $c_{\varphi tb}$ because it has a $Wtb$ vertex.
In addition, this process constrains $c_{tG}$ because of its $ttg$ vertex.

The SM cross section of this process with the contributions of $c_{\varphi q}^3$, $c_{tW}$ and $c_{tG}$ at order $\mathcal{O}(\Lambda^{-2})$ reads~\cite{Zhang:2010dr}
\begin{equation}
\begin{aligned}
\sigma_{gb\rightarrow Wt} & = \left( 1 + \frac{2c_{\varphi q}^3 v^2}{\Lambda^2} \right) \frac{g^2 g_s^2 }{384 s^3 m_W^2} \bigg( -((3m_t^2-2m_W^2)s + 7(m_t^2-m_W^2)(m_t^2+2m_W^2))\lambda^{1/2} \\
& +\ 2(m_t^2 + 2m_W^2)(s^2 + 2(m_t^2-m_W^2)s + 2(m_t^2-m_W^2)^2)\ \text{log}\ \frac{s + m_t^2 -m_W^2+\lambda^{1/2}} {s + m_t^2 -m_W^2-\lambda^{1/2}}  \bigg) \\
& -\ c_{tW} \frac{g_s^2m_t m_W } {24\sqrt{2} \Lambda^2 s^3} \bigg( (s + 21(m_t^2 -m_W^2))\lambda^{1/2}\\
& +\ 2(s^2 -6(m_t^2 - m_W^2)s - 6(m_t^2-m_W^2)^2)\ \text{log}\ \frac{s + m_t^2 -m_W^2+\lambda^{1/2}} {s + m_t^2 -m_W^2-\lambda^{1/2}}  \bigg) \\
& +\ c_{tG} \frac{g^2g_s  vm_t} {24\sqrt{2} \Lambda^2 s^2} \left(  2s\ \text{log}\ \frac{s + m_t^2 -m_W^2+\lambda^{1/2}} {s + m_t^2 -m_W^2-\lambda^{1/2}} + \lambda^{1/2} \right),
\end{aligned}
\label{eq:data:XtW}
\end{equation}
where $\lambda = s^2+m_t^4+m_W^4 - 2sm_t^2 - 2sm_W^2 -2m_t^2m_W^2$.
Similarly to the $s$- and $t$-channel production of a single top quark, $c_{\varphi q}^3$ represents a constant offset, see equations~\ref{eq:data:Xs} and~\ref{eq:data:Xt}.
The contribution of $c_{tW}$ has the same sign as in the $t$-channel production.
In addition, $c_{tG}$ shows a dependence on the center-of-mass energy $\sqrt{s}$ even though it does not contain a covariant derivative.

\begin{figure}[h!]
\centering
\vspace{3mm}
\begin{subfigure}{.48\textwidth}
  \centering
  \includegraphics[width=\textwidth]{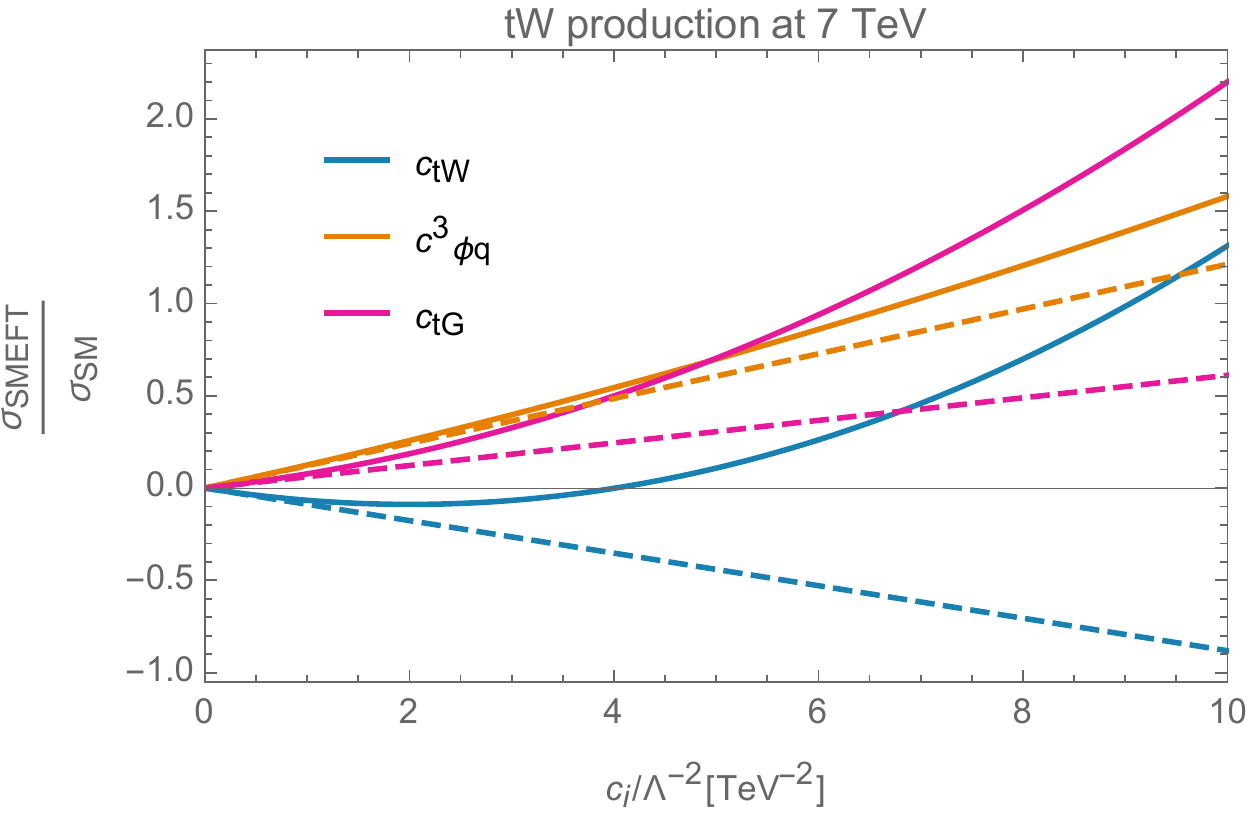}
  \caption{}
\end{subfigure}%
\hspace{.04\textwidth}%
\begin{subfigure}{.48\textwidth}
  \centering
  \includegraphics[width=\textwidth]{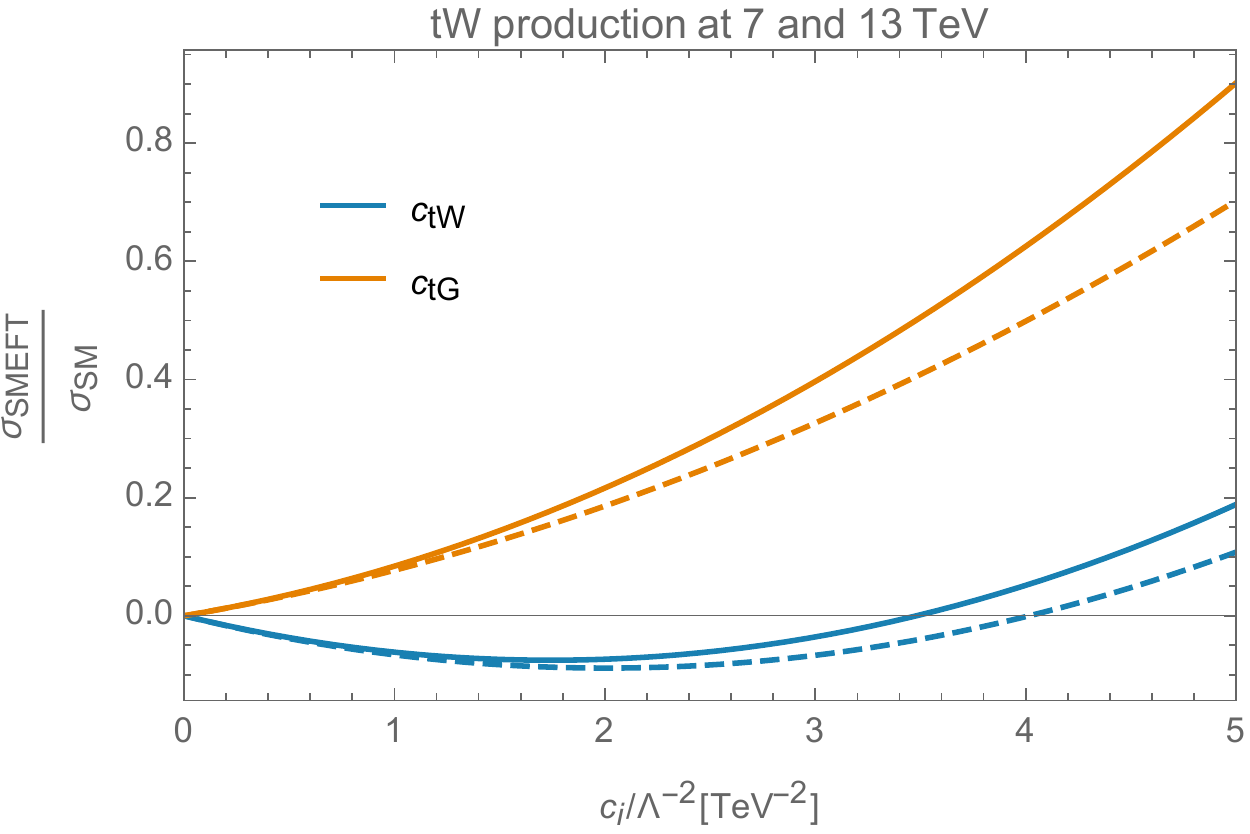}
  \caption{}
\end{subfigure}
\caption{Ratii of the cross sections of single top quark production in association with a $W$ boson with the contributions of different operators, as a function of the corresponding Wilson coefficient.
  Subfigure (a): corrections of $c_{tW}$, $c_{\varphi q}^3$ and $c_{tG}$ at 7 TeV. The dashed lines mark the contributions at order $\mathcal{O}(\Lambda^{-2})$, the solid lines those at order $\mathcal{O}(\Lambda^{-4})$. 
  Subfigure (b): corrections of $c_{tW}$ and $c_{tG}$ at 7 TeV (dashed) and 13 TeV (solid line) at order $\mathcal{O}(\Lambda^{-4})$.
  }
\label{img:data:tW713}
\end{figure}

Figure~\ref{img:data:tW713} shows the ratii of the SMEFT cross sections to the SM cross sections.
The corrections by $c_{bW}$ and $c_{\varphi tb}$ are not shown in subfigure (a) as they are rather small.
One can clearly see the importance of the corrections at order $\mathcal{O}(\Lambda^{-4})$.
For example, the $c_{tW}$ corrections are negative at order $\mathcal{O}(\Lambda^{-2})$, but positive at order $\mathcal{O}(\Lambda^{-4})$ and values greater than 4.

Subfigure (b) illustrates the effect of SMEFT energy growth. 
The Wilson coefficient $c_{\varphi q}^3$ is unaffected by different energies, as expected.
One can see, however, that the contribution of $c_{tG}$ to the SMEFT cross section at 13 TeV is about 30 percent larger than that at 7 TeV.
Also, the contribution of $c_{tW}$ practically doubles in the same energy range, even though its contribution at 13 TeV is about five times smaller than that of $c_{tG}$.


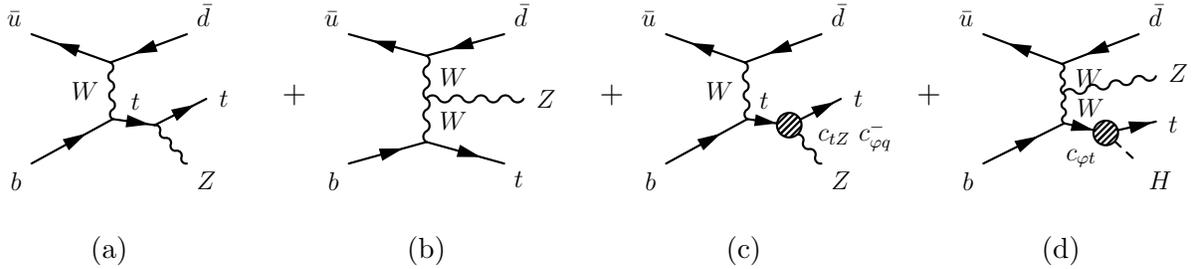
\begin{figure}[h]
\centering
\vspace{3mm}
\begin{subfigure}{0.18\textwidth}
  \centering
  \vspace{10mm}
  \resizebox{\textwidth}{!}{
  \begin{fmfgraph*}(90,60)
  \fmfleft{i1,i2}
  \fmfright{f0,f1,f2}
  
  \fmflabel{$b$}{i1}
  \fmflabel{$\bar{u}$}{i2}
  \fmflabel{$t$}{f1}
  \fmflabel{$\bar{d}$}{f2}
  \fmflabel{$Z$}{f0}
  
  \fmf{fermion}{i1,v1}
  \fmf{fermion, label=$t$,tension=2}{v1,v3}
  \fmf{fermion}{v3,f1}
  \fmf{fermion}{f2,v2,i2}
  \fmf{photon,label=$W$}{v1,v2}
  \fmf{photon}{v3,f0}
  
  \end{fmfgraph*}
  }
  \vspace{2mm}
  \caption{}
\end{subfigure}%
\hspace{9mm}%
\begin{subfigure}{.02\textwidth}%
\vspace{-5mm}
+
\end{subfigure}%
\hspace{2mm}%
\begin{subfigure}{0.18\textwidth}
  \centering
  \vspace{10mm}
  \resizebox{\textwidth}{!}{
  \begin{fmfgraph*}(90,60)
  \fmfleft{i1,i2}
  \fmfright{f1,f2,f3}
  
  \fmflabel{$b$}{i1}
  \fmflabel{$\bar{u}$}{i2}
  \fmflabel{$t$}{f1}
  \fmflabel{$\bar{d}$}{f3}
  \fmflabel{$Z$}{f2}
  
  \fmf{fermion}{i1,v1,f1}
  \fmf{fermion}{f3,v2,i2}
  \fmf{photon,label=$W$}{v1,v3}
  \fmf{photon,label=$W$}{v3,v2}
  \fmffreeze
  \fmf{photon}{v3,f2}
  
  \end{fmfgraph*}
  }
  \vspace{2mm}
  \caption{}
\end{subfigure}%
\hspace{9mm}%
\begin{subfigure}{.02\textwidth}%
\vspace{-5mm}
+
\end{subfigure}%
\hspace{2mm}%
\begin{subfigure}{0.18\textwidth}
  \centering
  \vspace{10mm}
  \resizebox{\textwidth}{!}{
  \begin{fmfgraph*}(90,60)
  \fmfleft{i1,i2}
  \fmfright{f0,f1,f2}
  
  \fmflabel{$b$}{i1}
  \fmflabel{$\bar{u}$}{i2}
  \fmflabel{$t$}{f1}
  \fmflabel{$\bar{d}$}{f2}
  \fmflabel{$Z$}{f0}
  
  \fmf{fermion}{i1,v1}
  \fmf{fermion, label=$t$,tension=2}{v1,v3}
  \fmf{fermion}{v3,f1}
  \fmf{fermion}{f2,v2,i2}
  \fmf{photon,label=$W$}{v1,v2}
  \fmf{photon}{v3,f0}
  
  \fmfblob{4mm}{v3}
  \fmfv{label.dist=5mm, label.angle=-20, label=$c_{tZ}\ c_{\varphi q}^-$}{v3}
  
  \end{fmfgraph*}
  }
  \vspace{2mm}
  \caption{}
\end{subfigure}%
\hspace{9mm}%
\begin{subfigure}{.02\textwidth}%
\vspace{-5mm}
+
\end{subfigure}%
\hspace{2mm}%
\begin{subfigure}{0.18\textwidth}
  \centering
  \vspace{10mm}
  \resizebox{\textwidth}{!}{
  \begin{fmfgraph*}(90,60)
  \fmfleft{i1,i2}
  \fmfright{f0,f1,f2,f3}
  
  \fmflabel{$b$}{i1}
  \fmflabel{$\bar{u}$}{i2}
  \fmflabel{$H$}{f0}
  \fmflabel{$t$}{f1}
  \fmflabel{$Z$}{f2}
  \fmflabel{$\bar{d}$}{f3}
  
  \fmf{fermion}{i1,v1}
  \fmf{fermion, tension=2}{v1,v3}
  \fmf{fermion}{v3,f1}
  \fmf{fermion}{f3,v2,i2}
  \fmf{dashes}{v3,f0}
  \fmf{phantom}{v1,v2}
  \fmffreeze
  \fmf{photon,label=$W$}{v1,v4}
  \fmf{photon,label=$W$}{v4,v2}
  \fmffreeze
  \fmf{photon}{v4,f2}

  \fmfblob{4mm}{v3}
  \fmfv{label.dist=3mm, label.angle=-120, label=$c_{\varphi t}$}{v3}  
    
  \end{fmfgraph*}
  }
  \vspace{2mm}
  \caption{}
\end{subfigure} 
\vspace{3mm}
\caption{Example Feynman diagrams for the production of a single top quark in association with a $Z$ boson. Subfigures (a-b) show SM contributions, subfigure (c) shows contributions from $c_{tZ}$ and $c_{\varphi q}^-$, and (d) contributions from $c_{\varphi t}$.
  }
\vspace{5mm}
\label{img:data:feyntZ}
\end{figure}

\begin{figure}[h]
\centering
\vspace{3mm}
\begin{subfigure}{.48\textwidth}
  \centering
  \includegraphics[width=\textwidth]{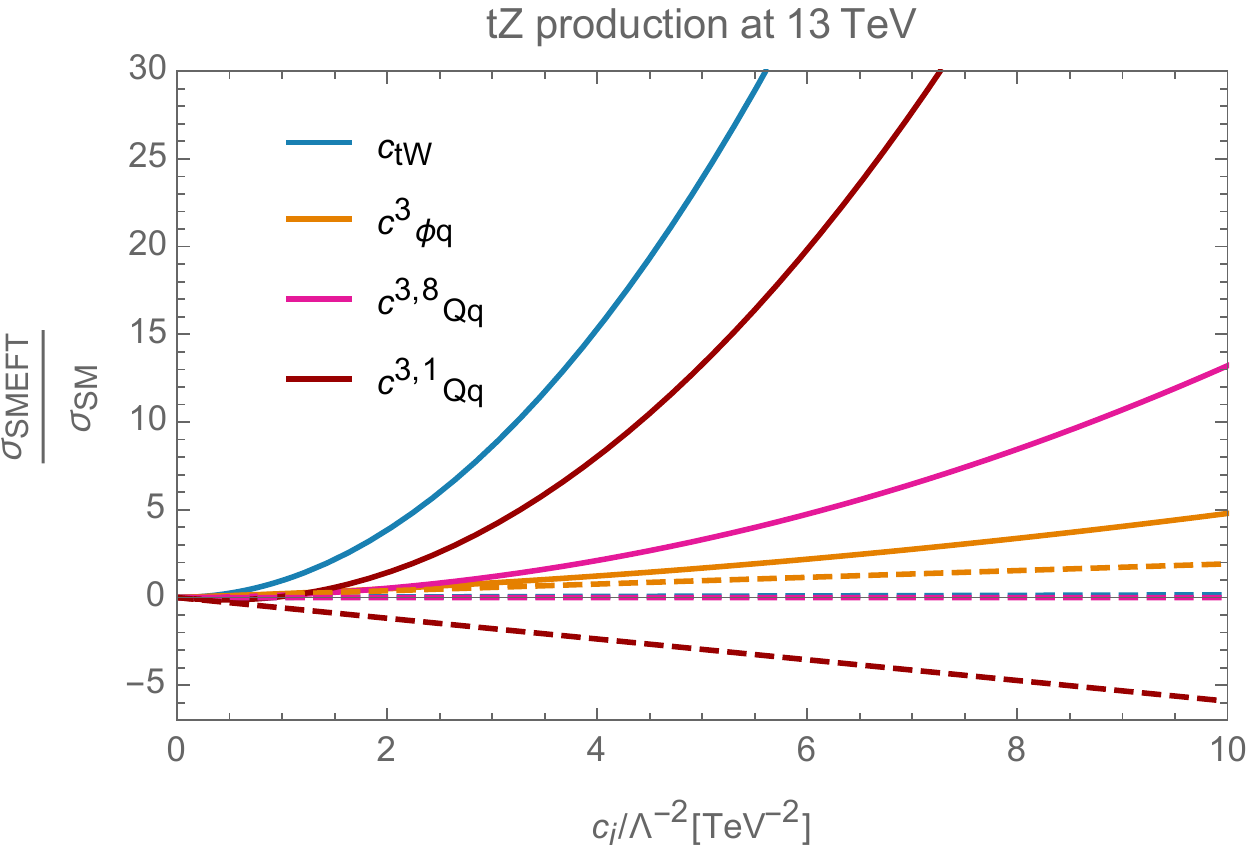}
  \caption{}
\end{subfigure}%
\hspace{.04\textwidth}%
\begin{subfigure}{.48\textwidth}
  \centering
  \includegraphics[width=\textwidth]{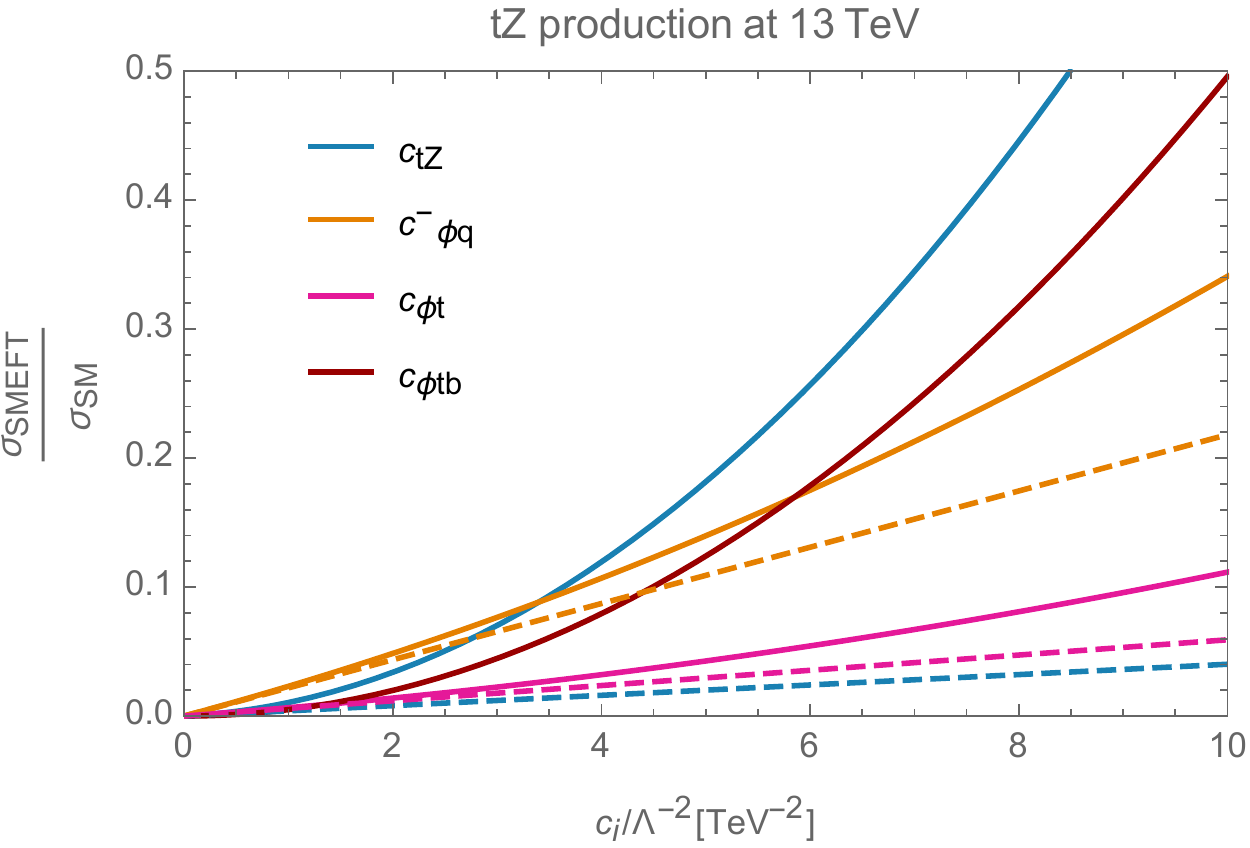}
  \caption{}
\end{subfigure}
\caption{Ratii of the cross sections of single top quark production in association with a $Z$ boson with the contributions of different operators, as a function of the corresponding Wilson coefficient.
  Subfigure (a): corrections of $c_{tW}$, $c_{\varphi q}^3$, $c_{Qq}^{3,8}$ and $c_{Qq}^{3,1}$ at 13 TeV. The dashed lines mark the contributions at order $\mathcal{O}(\Lambda^{-2})$, the solid lines those at order $\mathcal{O}(\Lambda^{-4})$. 
  The contributions of $c_{tW}$ and $c_{Qq}^{3,8}$ at order $\mathcal{O}(\Lambda^{-2})$ are close to zero.
  Subfigure (b): corrections of $c_{tZ}$, $c_{\varphi q}^-$, $c_{\varphi t}$ and $c_{\varphi tb}$ at 13 TeV.
  }
\label{img:data:tZ13}
\end{figure}

Figure~\ref{img:data:feyntZ} shows example diagrams of single top quark production in association with a $Z$ boson. 
In contrast to the production in association with a $W$ boson, some four-quark operators can have an impact on the EFT cross section, for example by leaving away the $W$ boson mediator in subfigure (a).
Consider first the operators that give rise to $c_{Qq}^{3,8}$, as stated in equation~\ref{eq:SMEFT:qq38}.
These involve a $t\bar{b}$, $b\bar{t}$, or $t\bar{t}$ quark pair. 
However, the process in subfigure (a) only involves a $tb$ quark pair.
Therefore, these operators can only play a role in loop corrections to this process.

The operator $\mathcal{O}_{qq}^{3(ii33)}$, see equation~\ref{eq:SMEFT:qq31}, does not induce any interaction even at NLO.
The reason is that the color flow induced by this operator cannot take place even in NLO corrections. 
While the corresponding Wilson coefficient $c_{Qq}^{3,1}$ does correspond to the operators of equation~\ref{eq:SMEFT:qq38} also, their contribution to the SMEFT cross section is small.
In this study, this contribution is therefore neglected.

In subfigure (a), the process resembles the $t$-channel production of a single top quark, where a $Z$ boson is radiated off.
This is analogously possible with the $s$-channel production.
One can therefore conclude that the single top quark production in association with a $Z$ boson also constrains $c_{tW}$, $c_{\phi q}^3$, $c_{bW}$ and $c_{\varphi tb}$.

In addition to the aforementioned Wilson coefficients, this process constrains $c_{tZ}$ and $c_{\varphi q}^-$ because of the possible $ttZ$ vertex, see subfigure (c).
Also, the possible $ttH$ vertex in subfigure (d) implies that $c_{\varphi t}$ is sensitive to this process.
In principle, this Wilson coefficient is also sensitive to the single top quark production in association with a $W$ boson, see the $ttH$ vertex in figure~\ref{img:data:feyntZ}(c).
However, the production in association with a $W$ boson where a gluon is involved is much more likely.
This contribution by involving gluons is not given in the production in association with a $Z$ boson. 
That is why the contribution of $c_{\varphi t}$ to the cross section of the production in association with a $W$ boson is neglected in this study.

Figure~\ref{img:data:tZ13} shows the contributions of various operators to the ratio of the SMEFT to the SM cross section of single top quark production in association with a $Z$ boson.
Energy growth behavior has not been taken into account as the only available measurements so far are at 13 TeV.
Subfigure (a) shows the contributions of operators that also play a role in single top quark $s$- and $t$-channel production. 
One interesting fact is that the linear term of $c_{tW}$ is small, which is not the case in $s$- and $t$-channel production. 
That is because $c_{tW}$ affects the $ttZ$-vertex, as opposed to the $Wtb$-vertex previously.
The difference in symmetries of the SM $ttZ$-vertex to those of the induced one by dimension-six operators leads to the linear suppression. 

Comparing these contributions with figure~\ref{img:data:st713}, one finds that most contributions lie between those to the $s$- and $t$-channels at 13 TeV.
The only exception is $c_{\varphi q}^3$: at a value of 10, the SMEFT cross section of the $s$-channel production increases by a factor 0.7 and that of the $t$-channel production by about 1.55 (not shown in the figures). 
In the production in association with a $Z$ boson (called $tZ$ production for short), however, it increases by a factor of about five.
The reason lies in the fact that $c_{\varphi q}^3$ has the same structure as the SM prediction in the $s$- and $t$-channel single top quark production.
This contribution is the same as for $tZ$ production.
In contrast, the SM cross section for $tZ$ production is a lot smaller than that of the $s$- and $t$-channel processes.
Therefore, the relative contribution of $c_{\varphi q}^3$ on the SMEFT cross section of $tZ$ production is a lot higher.
This highlights the potential of the single top quark production in association with a $Z$ boson to better constrain $c_{\varphi q}^3$.

Subfigure~\ref{img:data:st713}(b) shows the contributions of $c_{\varphi tb}$, $c_{tZ}$, $c_{\varphi q}^-$ and $c_{\varphi t}$.
One can immediately see that the corrections are a lot smaller than those in subfigure (a).
Three of the Wilson coefficients are not constrained by any other process of this study, their corrections are small, and only few measurements of $tZ$ production are available so far.
That is why these three coefficients are dropped from this analysis.
One should keep in mind, however, that it does make sense to include them in a fit where top quark pair production in association with a $Z$ boson is also studied, as this process also constrains them.



\subsubsection{Top Quark Decay}

\begin{figure}[h]
\centering
\vspace{3mm}
\begin{subfigure}{0.18\textwidth}
  \centering
  \vspace{10mm}
  \resizebox{\textwidth}{!}{
  \begin{fmfgraph*}(90,60)
  \fmfleft{i1,i2}
  \fmfright{f1,f2}
  
  \fmflabel{$t$}{i1}
  \fmflabel{$\bar{\nu}$}{i2}
  \fmflabel{$b$}{f1}
  \fmflabel{$e^+$}{f2}
  
  \fmf{fermion}{i1,v1,f1}
  \fmf{fermion}{f2,v2,i2}
  \fmf{photon,label=$W$}{v1,v2}
  
  \end{fmfgraph*}
  }
  \vspace{2mm}
  \caption{}
\end{subfigure}%
\hspace{5mm}%
\begin{subfigure}{.02\textwidth}%
\vspace{-5mm}
+
\end{subfigure}%
\hspace{5mm}%
\begin{subfigure}{0.18\textwidth}
  \centering
  \vspace{10mm}
  \resizebox{\textwidth}{!}{
  \begin{fmfgraph*}(90,60)
  \fmfleft{i1,i2}
  \fmfright{f1,f2}
  
  \fmflabel{$t$}{i1}
  \fmflabel{$\bar{\nu}$}{i2}
  \fmflabel{$b$}{f1}
  \fmflabel{$e^+$}{f2}
  
  \fmf{fermion}{i1,v1,f1}
  \fmf{fermion}{f2,v2,i2}
  \fmf{photon,label=$W$}{v1,v2}
  
  \fmfblob{4mm}{v1}
  \fmfv{label.dist=3mm, label=$c_{tW}\ c_{\varphi q}^3$}{v1}
  
  \end{fmfgraph*}
  }
  \vspace{2mm}
  \caption{}
\end{subfigure}%
\hspace{5mm}%
\begin{subfigure}{.02\textwidth}%
\vspace{-5mm}
+
\end{subfigure}%
\hspace{5mm}%
\begin{subfigure}{0.18\textwidth}
  \centering
  \vspace{10mm}
  \resizebox{\textwidth}{!}{
  \begin{fmfgraph*}(90,60)
  \fmfleft{i1,i2}
  \fmfright{f1,f2}
  
  \fmflabel{$t$}{i1}
  \fmflabel{$\bar{\nu}$}{i2}
  \fmflabel{$b$}{f1}
  \fmflabel{$e^+$}{f2}
  
  \fmf{fermion}{i1,v1,f1}
  \fmf{fermion}{f2,v2,i2}
  \fmf{photon,label=$W$}{v1,v2}
  
  \fmfblob{4mm}{v1}
  \fmfv{label.dist=5mm, label=$c_{bW}\ c_{\varphi tb}$}{v1}
  
  \end{fmfgraph*}
  }
  \vspace{2mm}
  \caption{}
\end{subfigure}%
\hspace{3mm}%
\begin{subfigure}{.02\textwidth}%
\vspace{-5mm}
+
\end{subfigure}%
\hspace{8mm}%
\begin{subfigure}{0.18\textwidth}
  \centering
  \vspace{10mm}
  \resizebox{\textwidth}{!}{
  \begin{fmfgraph*}(90,60)
  \fmfleft{i1,i2,i3}
  \fmfright{f1,f2}
  
  \fmflabel{$g$}{i1}
  \fmflabel{$t$}{i2}
  \fmflabel{$\bar{\nu}$}{i3}
  \fmflabel{$b$}{f1}
  \fmflabel{$e^+$}{f2}
  
  \fmf{fermion}{i2,v2}
  \fmf{fermion,tension=1.5}{v2,v1}
  \fmf{fermion}{v1,f1}
  \fmf{gluon}{i1,v2}
  \fmf{fermion}{f2,v3,i3}
  \fmf{photon,label=$W$}{v1,v3}

  \fmfblob{4mm}{v2}
  \fmfv{label.dist=4mm, label.angle=-30, label=$c_{tG}$}{v2}  
    
  \end{fmfgraph*}
  }
  \vspace{2mm}
  \caption{}
\end{subfigure} 
\vspace{3mm}
\caption{Example Feynman diagrams for top quark decay. 
  Subfigure (a) shows the SM diagram, (b) the corrections from $c_{tW}$ and $c_{\phi q}^3$, (c) those from $c_{bW}$ and $c_{\varphi tb}$, and (d) that from $c_{tG}$.
The contributions of $c_{tW}$ and $c_{\phi q}^3$, and $c_{bW}$ and $c_{\varphi tb}$ are shown separately to emphasize the different structures of the new interactions.
}
\vspace{5mm}
\label{img:data:feynWhel}
\end{figure}
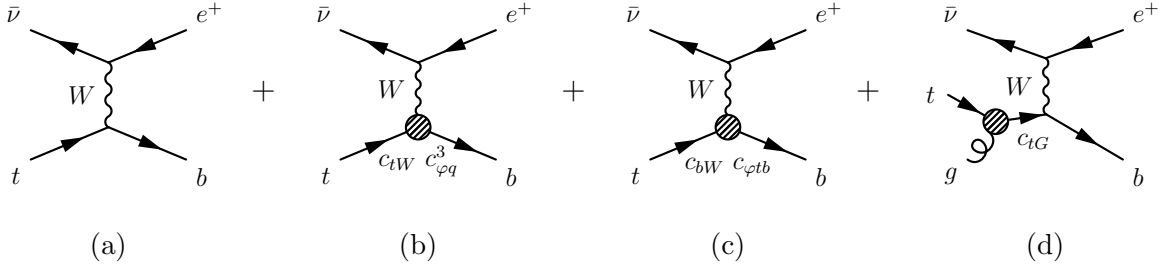

Figure~\ref{img:data:feynWhel} shows example diagrams of top decay, where the $W$ boson decays leptonically.
Obviously, there are no contributions by four-quark operators.
One can also see the corrections from $c_{tW}$, $c_{\phi q}^3$, $c_{bW}$ and $c_{\varphi tb}$ at the $Wtb$ vertex.
In addition, $c_{tG}$ is constrained via gluon radiation.
In this fit, the helicity fractions of the $W$ bosons are implemented.
As these are measured with respect to the rest frame of the top quark, a discussion of energy-growth behavior is not applicable.

The helicity fractions are linked to the differential decay width of the top quark via~\cite{AguilarSaavedra:2006fy}
\begin{equation}
\frac{1}{\Gamma} \frac{\text{d}\Gamma}{\text{d cos}\ \theta} = \frac{3}{8}\ (1 + \text{cos}\ \theta)^2 F_R + \frac{3}{8}\ (1 - \text{cos}\ \theta)^2 F_R + \frac{3}{4}\ \text{sin}^2 \theta\ F_0,
\label{eq:data:GammaFWpol}
\end{equation}
where $F_0$ is the longitudinal and $F_L$ and $F_R$ are the left- and right-handed transverse helicity fractions, respectively. 
The angle between the 3-momentum of the $W$ boson in the $t$ rest frame and that of the charged lepton in the $W$ boson rest frame is referred to as $\theta$.
The differential decay rate with SMEFT contributions of $c_{\varphi q}^3$ and $c_{tW}$ up to order $\mathcal{O}(\Lambda^{-2})$ is, with the narrow width approximation for the $W$ boson~\cite{Zhang:2010dr}
\begin{equation}
\begin{aligned}
\frac{\text{d}\Gamma}{\text{d cos}\ \theta} & = \left( 1 + \frac{2\ c_{\phi q}^3 v^2}{\Lambda^2} \right) \frac{g^4}{4096 \pi^2 m_t^2m_W \Gamma_W} \Big( m_t^2 + m_W^2 + (m_t^2-m_W^2)\ \text{cos}\ \theta \Big) (1-\text{cos}\ \theta) \\
& +\ \frac{c_{tW} g^2}{128\sqrt{2} \pi^2 \Lambda^2 m_t^2\Gamma_W}\ m_W^2 (m_t^2 -m_W^2)^2 (1 -\text{cos}\ \theta),
\end{aligned}
\label{eq:data:GammaWpol}
\end{equation}
where $\Gamma_W$ is the width of the $W$ boson in the SM.
Similarly to the findings in the previous subsections, the contribution $c_{\varphi q}^3$ to the differential decay width represents a constant offset of the SM.
The contribution of $c_{tW}$, in contrast, has a different structure than the SM.

Evaluating $\frac{1}{\Gamma} \frac{\text{d}\Gamma}{\text{d cos}\ \theta}$, one finds that the term containing $c_{\varphi q}^3$ drops out.
This indicates that there is no contribution of $c_{\varphi q}^3$ at order $\mathcal{O}(\Lambda^{-2})$.
The helicity fractions at this order read
\begin{equation}
\begin{aligned}
F_0 &= \frac{m_t^2}{m_t^2+ 2m_W^2} - \frac{4\sqrt{2}\ c_{tW} v^2 }{\Lambda^2} \frac{m_t m_W (m_t^2 -m_W^2)}{(m_t^2+2m_W^2)^2}, \\
F_L &= \frac{2m_W^2}{m_t^2+ 2m_W^2} + \frac{4\sqrt{2}\ c_{tW} v^2 }{\Lambda^2} \frac{m_t m_W (m_t^2 -m_W^2)}{(m_t^2+2m_W^2)^2}, \\
F_R &= 0.
\end{aligned}
\label{eq:data:FWpol}
\end{equation}
One interesting feature is that the sign of $c_{tW}$ switches between $F_0$ and $F_L$.
This suggests that implementing $F_0$ and $F_L$ in the fit would constrain $c_{tW}$ especially well.
The other Wilson coefficients that are shown in figure~\ref{img:data:feynWhel} can only be constrained by the helicity fractions if contributions of order $\mathcal{O}(\Lambda^{-4})$ are included in the fit.

\begin{figure}[t!]
\centering
\vspace{3mm}
\begin{subfigure}{.48\textwidth}
  \centering
  \includegraphics[width=\textwidth]{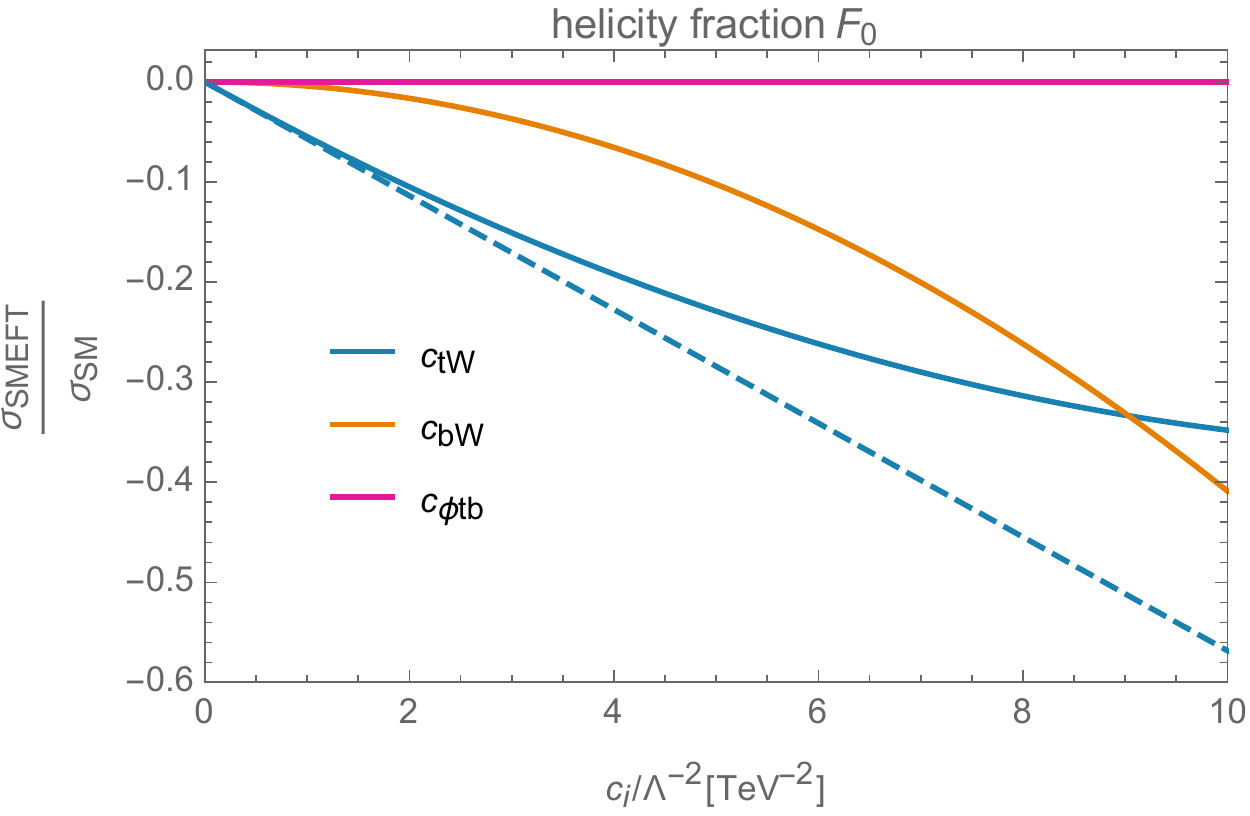}
  \caption{}
\end{subfigure}%
\hspace{.04\textwidth}%
\begin{subfigure}{.48\textwidth}
  \centering
  \includegraphics[width=\textwidth]{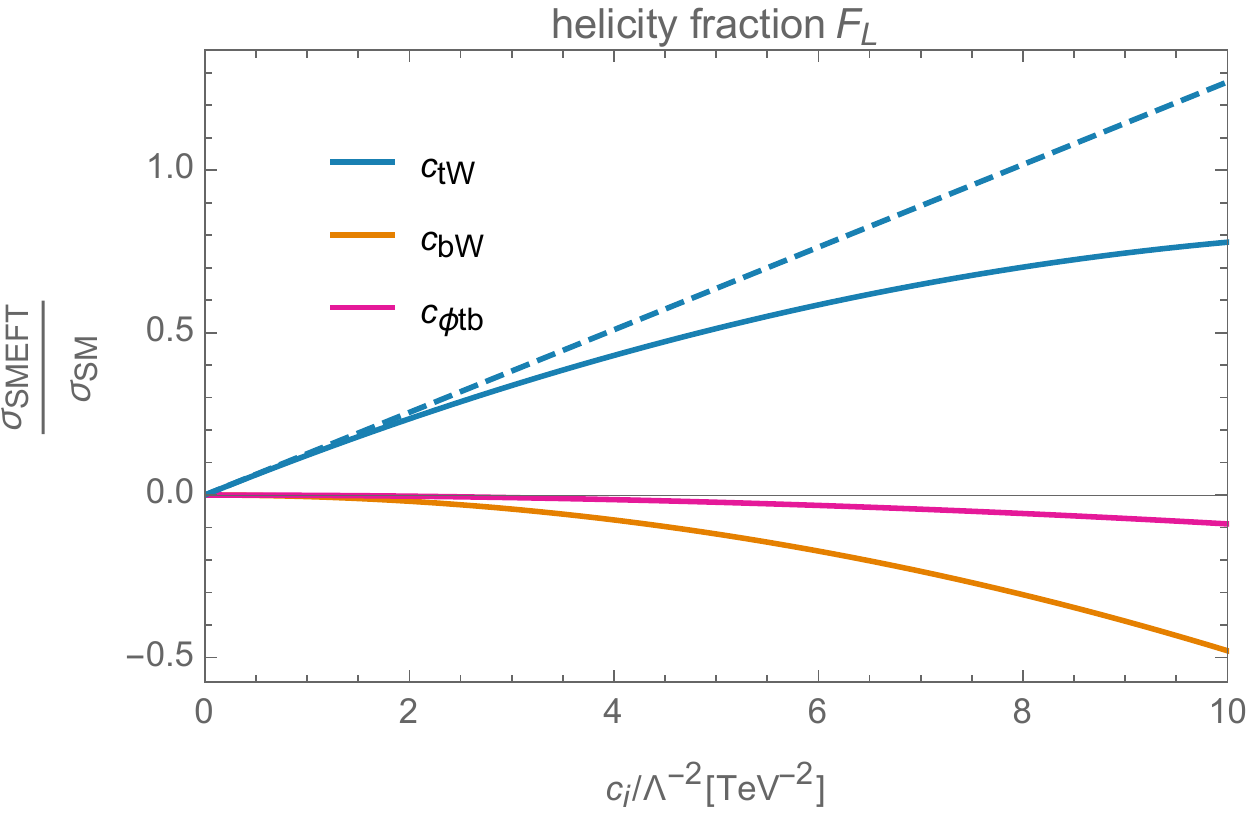}
  \caption{}
\end{subfigure}
\caption{Ratii of helicity fractions of the $W$ boson from top decay with the contributions of the operators giving rise to $c_{tW}$, $c_{bW}$ and $c_{\varphi tb}$, to their SM prediction.
  Subfigure (a): helicity fraction $F_0$. 
  Subfigure (b): helicity fraction $F_L$.
  The contributions of $c_{tW}$ and $c_{bW}$ at order $\mathcal{O}(\Lambda^{-2})$ are zero.
  }
\label{img:data:Whel}
\end{figure}

This is also illustrated in figure~\ref{img:data:Whel}, where only $c_{tW}$ has nonzero correction terms at order $\mathcal{O}(\Lambda^{-2})$.
The contributions of $c_{tG}$ are not included in the figures as they are small - it is better constrained through single top quark production in association with a $W$ boson.
The relative contributions of dimension-six operators to $F_R$ are not shown as this makes little sense with its prediction being almost zero.

Comparing the contributions to $F_0$, subfigure (a), and $F_L$, subfigure (b), one sees a sign flip for $c_{tW}$, as observed in equation~\ref{eq:data:FWpol}. 
This confirms that it makes sense to include $F_0$ and $F_L$ separately in the fit.
One also sees that $c_{\varphi tb}$ is more sensitive to $F_L$, even if the impact is pretty small.
The impact of $c_{bW}$ is comparable for both observables.

\clearpage


\subsection{Sensitivities}

\begin{table}[h]
  \centering
  \vspace{0.4cm}
  \begin{tabular}{|lccccc|}
  \hline
  Wilson coeff. & $s$-channel & $t$-channel & $tW$ & $tZ$ & $t$ decay \\
  \hline
  $c_{Qq}^{3,8}$ & $\checkmark$ & $\checkmark$ &  & $[\checkmark]$ & \\
  $c_{Qq}^{3,1}$ & $\checkmark$ & $\checkmark$ &  &  & \\
  $c_{tG}$ & & & $\checkmark$ &  & $[\checkmark]$ \\
  $c_{tW}$ & $\checkmark$& $\checkmark$ & $\checkmark$ & $\checkmark$ & $\checkmark$ \\
  $c_{bW}$ & $(\checkmark)$ & $(\checkmark)$ & $(\checkmark)$ & $(\checkmark)$ & $(\checkmark)$\\
  $c_{\varphi tb}$ & $(\checkmark)$ & $(\checkmark)$ & $(\checkmark)$ & $(\checkmark)$ & $(\checkmark)$\\
  $c_{\varphi q}^3$ & $\checkmark$ & $\checkmark$ & $\checkmark$ & $\checkmark$ & $(\checkmark)$\\
  \hline
  \end{tabular}
  \caption{List of Wilson coefficients that are relevant for this study. 
  The check marks indicate that a given process constrains the corresponding operator.
  A check mark in square brackets indicates that a given process constrains the corresponding operators, but only at NLO.
  A check mark in round brackets indicates that a given process constrains the corresponding operators, but only at $\mathcal{O}(\Lambda^{-4})$.
  }
  \label{tab:data:sensitivity}
\end{table}

Table~\ref{tab:data:sensitivity} summarizes the findings of the previous subsections.
Even though this study only covers single top quark production and decay, the importance of a global fit becomes clear from the table as many operators are constrained by multiple processes.
For example, while the Wilson coefficients $c_{tZ}$, $c_{\varphi q}^-$ and $c_{\varphi t}$ are sensitive to single top quark production in association with a $Z$ boson, they are not included because their sensitivity is small and there is only one available measurement to constrain these.

One can see from this table that various operators only contribute at $\mathcal{O}(\Lambda^{-4})$ for reasons previously discussed.
The impact of the $\mathcal{O}(\Lambda^{-4})$ of dimension-six operators is investigated in detail later on.
At this point, however, it is clear that taking higher orders into account is essential to constrain Wilson coefficients such as $c_{bW}$ and $c_{\varphi tb}$.

For the fit, the same theoretical predictions of the cross sections, differential cross sections and helicity fractions are used as by the SMEFiT collaboration~\cite{Hartland:2019bjb}.
The predictions for the single top quark t-channel production are at NNLO QCD level as they include NNLO $K$-factors.
Those for the single top quark production in the s-channel and in association with a $W$ boson have been calculated at NLO QCD level.
For the $W$-helicity in top quark decay processes, analytical results at NLO QCD have been used.

\clearpage


\subsection{Measurements}

The previously presented operators are constrained by measurements from the LHC at $\sqrt{s}=7$, 8 and 13 TeV.
These measurements supersede those of Tevatron. 
Table~\ref{tab:data:measurements} shows the dataset used in the present analysis.

The table lists measurements of inclusive cross sections, differential distributions in the $t$-channel single top production and helicity fractions. 
As explained in section~\ref{subsubsec:data:ts}, the differential distributions could constrain certain operators stricter. 
This is shown not to have a sizable effect in the final results, as will be shown in section~\ref{subsubsec:results:Snd}.
 
The not-normalized differential distributions are used in the final fit and the corresponding measurements of inclusive cross sections are omitted to avoid over-counting.
Where only the normalized differential distribution is given, the not-normalized one is reconstructed by multiplying the bins of the normalized distribution with the measurement of the corresponding total cross section.
This multiplication was carried out under the assumption that the uncertainties are not correlated.
The experimental uncertainties of the bins of the not-normalized distributions were then obtained by quadratical addition of the uncertainties of the inclusive cross section and the bins of the normalized distributions.
This is a fairly conservative approach and a good approximation in this case.
The accuracy of this method has been checked using the normalized and not normalized distributions of~\cite{Aad:2014fwa} and~\cite{Aaboud:2017pdi}.
Another cross check was performed by determining that the deviation of the sum of the reconstructed not-normalized bins from the sum of their predictions is indeed the same as the deviation of the measured total cross section from its prediction.

\begin{table}[p!]
  \centering
  \vspace{0.5cm}
  \begin{tabular}{|llllll|}
  \hline
  process & experiment & energy [TeV] & observable & $N_\text{dat}$ & reference \\
  \hline
  $s$-channel & CMS & 7 & $\sigma_\text{tot} (t + \bar{t})$ & 1 & \cite{Khachatryan:2016ewo} \\
   & CMS & 8 & $\sigma_\text{tot} (t + \bar{t})$ & 1 & \cite{Khachatryan:2016ewo} \\
   & ATLAS & 8 & $\sigma_\text{tot} (t + \bar{t})$ & 1 & \cite{Aad:2015upn} \\
  \hline
  $t$-channel & ATLAS & 7 & $\sigma_\text{tot} (t + \bar{t})$ & 1 & \cite{Aad:2012ux} \\
   & CMS & 7 & $\sigma_\text{tot} (t + \bar{t})$ & 1 & \cite{Chatrchyan:2012ep} \\
   & ATLAS & 8 & $\sigma_\text{tot} (t)$ & 1 & \cite{Aaboud:2017pdi} \\
   & & & $\sigma_\text{tot} (\bar{t})$ & 1 & \\
   & CMS & 8 & $\sigma_\text{tot} (t)$ & 1 & \cite{Khachatryan:2014iya} \\
   & & & $\sigma_\text{tot} (\bar{t})$ & 1 &  \\
   & ATLAS & 13 & $\sigma_\text{tot} (t)$ & 1 & \cite{Aaboud:2016ymp} \\
   & & & $\sigma_\text{tot} (\bar{t})$ & 1 &  \\
   & CMS & 13 & $\sigma_\text{tot} (t)$ & 1 & \cite{Sirunyan:2016cdg} \\
   & & & $ \sigma_\text{tot} (\bar{t})$ & 1 & \\
  \hline
  $t$-channel distributions & ATLAS & 7 & $\text{d} \sigma_\text{tot} (t) / \text{d} y_t $ & 4 & \cite{Aad:2014fwa} \\
   &  &  & $\text{d} \sigma_\text{tot} (\bar{t}) / \text{d} y_t $ & 4 & \\  
   & ATLAS & 8 & $\text{d} \sigma_\text{tot} (t) / \text{d} y_t $ & 4 & \cite{Aaboud:2017pdi} \\
   &  &  & $\text{d} \sigma_\text{tot} (\bar{t}) / \text{d} y_t $ & 4 & \\
   & CMS & 13 & $\text{d} \sigma / \text{d} | y^{t+\bar{t}} |  $ & 4 &\cite{CMS:2016xnv} \\
  \hline
  $tW$ & ATLAS & 7 & $ \sigma_\text{tot} (tW) $ & 1 & \cite{Aad:2012xca} \\  
   & CMS & 7 & $ \sigma_\text{tot} (tW) $ & 1 & \cite{Chatrchyan:2012zca} \\  
   & ATLAS & 8 & $ \sigma_\text{tot} (tW) $ & 1 & \cite{Aad:2015eto} \\
   & CMS & 8 & $ \sigma_\text{tot} (tW)  $ & 1 & \cite{Chatrchyan:2014tua} \\
   & ATLAS & 13 & $ \sigma_\text{tot} (tW) $ & 1 & \cite{Aaboud:2016lpj} \\
   & CMS & 13 & $ \sigma_\text{tot} (tW)  $ & 1 & \cite{Sirunyan:2018lcp} \\
  \hline
   $tZ$ & ATLAS & 13 & $ \sigma_\text{tot} (tZq)  $ & 1 & \cite{Aaboud:2017ylb} \\
  \hline
  $t$ decay & ATLAS & 7 & $F_0$ & 1 & \cite{Aad:2012ky} \\
   &  &  & $F_L$ & 1 & \\
   & CMS & 7 & $F_0$ & 1 & \cite{Chatrchyan:2013jna} \\
   &  &  & $F_L$ & 1 & \\
   & ATLAS & 8 & $F_0$ & 1 & \cite{Aaboud:2016hsq} \\
   &  &  & $F_L$ & 1 & \\
   & CMS & 8 & $F_0$ & 1 & \cite{Khachatryan:2016fky} \\
   &  &  & $F_L$ & 1 & \\
  \hline
  \end{tabular}
  \caption{Measurements used in the fit presented here. One can see the physical process, the experiment, the center-of-mass energy $\sqrt{s}$ in TeV, the observable that was measured, the number of measurements, the number of data points $N_\text{dat}$, and the corresponding publication.
  Measurements of the same process at the same energy but different experiments are averaged.
  }
  \label{tab:data:measurements}
\end{table}

All experimental analyses concerning $t$-channel distributions list distributions over the momentum of the top quark and its rapidity. 
To avoid over-counting, only the distributions is included which has the better potential to constrain the Wilson coefficients, i.e. which has smaller uncertainties and to which the coefficients are more sensitive.
Where $t$-channel distributions both of ATLAS and CMS are available, only one distribution is chosen because the theoretical uncertainties of the two distributions are correlated.
Not taking these correlations into account would lead to a distortion of the log-likelihood, as described in section~\ref{subsec:meth:theocorr}.
That is why the distribution in~\cite{CMS:2014ika} remains unused.

In the measurements of the inclusive cross section in $s$- and $t$-channel production, the ratio $R = \sigma (tq) / \sigma(\bar{t}q)$ is also often given. 
Similarly to the normalized distributions, this ratio appears to be more precise because of the partial cancellation of systematic uncertainties.
However, in the ratio the information about the total cross section is lost.
That is why implementing the ratio instead of the inclusive cross sections brings no advantages for the fit.

In the cases where the same process at the same energy was measured both by ATLAS and CMS, the fit results would contain unphysical effects if they were simply implemented as independent measurements.
The reason is that the theoretical uncertainties of the two measurements are fully correlated. 
To respect this in the fit, one must either introduce nuisance operators or use the weighted average of the two measurements.
Nuisance operators would be implemented as supplementary operators and modify the prediction of the fit.
This is equivalent to using the weighted average of two measurements, but the problem is that we would need ten additional operators in the fit - one for the $s$-channel measurements at 8 TeV, one each for the $tW$-measurements at 7, 8 and 13 TeV, and six for the measurements of the top decay as the latter are independent of the center-of-mass energy.
This would result in 17 fit parameters for 38 measurements.
In contrast, averaging the measurements results in seven parameters for 28 measurements because the number of fit parameters remains unchanged.
This is a major improvement because instead of roughly two measurements per parameter, the fit now has four.
The procedure of averaging measurements is described in detail in section~\ref{subsec:meth:theocorr}.

As the helicity fractions from top quark decay processes are independent of the energy, we get only two entries for the fit after averaging.
As the helicity fractions are correlated via $F_0 + F_L + F_R =1$, only two observables from each analysis have been used.
$F_0$ and $F_L$ are chosen as in most analyses $F_R$ is the least promising observable.
In fact, there is another measurement of the helicity fraction available~\cite{Aad:2015yem}, but it only contains $f_1 = F_L + F_R$.
Including it would thus have made the averaging difficult.

Another measurement of the top quark production in association with a $Z$ boson at 13 TeV is available from CMS~\cite{Sirunyan:2017nbr}. 
This has not been used, however, as the publication only gives the fiducial cross section and thus averaging the measurements would be rather difficult.

One could also consider implementing measurements of other variables, for example the polarization asymmetry $A_{P\pm}$, the spin correlations variable $A_{\Delta\phi}$, and the $A_{c_1c_2}$ and $A_{\text{cos}\phi}$ asymmetries~\cite{ATLAS:2018rgl},\cite{Khachatryan:2016xws}.
However, this is left for future work.


\section{Methodology of SFitter}
\label{sec_methodology}

In this section, the types of uncertainties of a measurement are explained, and how Monte Carlo toys are constructed using these in \textsc{SFitter}.
Then, the log-likelihood and the the impact of correlations among theoretical and systematic uncertainties on it is studied.


\subsection{Types of uncertainties}

The \textsc{SFitter} framework aims at finding the best set of operators for a given dataset.
This is equivalent to maximizing the likelihood of a set of model operators.
This, in turn, is equal to the probability of measuring these results given that the fitted set of $n$ parameters $\{ c^{(k)}\}$, $k=1,\dots,n$, to $m$ measurements $\{x_\text{meas}^{(i)}$, $i=1, \dots, m$, is the true one~\cite{Plehn:2009nd},
\begin{equation}
\mathcal{L}\big(\{ c_\text{mod}^{(k)}\} | \{ x_\text{meas}^{(i)}\} \big) = P\big(\{ x_\text{meas}^{(i)}\} | \{ c_\text{mod}^{(k)}\} \big).
\label{eq:meth:unc1}
\end{equation}

What we are really interested in is the probability of the fitted model given the current measurements.
Using Bayes' theorem, one could write equation~\ref{eq:meth:unc1} as
\begin{equation}
\begin{aligned}
P\big(\{ c_\text{mod}^{(k)}\} | \{ x_\text{meas}^{(i)}\} \big) &= P\big(\{ x_\text{meas}^{(i)}\} | \{ c_\text{mod}^{(k)}\} \big)\ \frac{ P\big(\{ c_\text{mod}^{(k)}\} \big) } { P\big(\{ x_\text{meas}^{(i)}\} \big) } \\
&= \mathcal{L}\big(\{ c_\text{mod}^{(k)}\} | \{ x_\text{meas}^{(i)}\} \big) \ \frac{ P\big(\{ c_\text{mod}^{(k)}\} \big) } { P\big(\{ x_\text{meas}^{(i)}\} \big) } .
\end{aligned}
\label{eq:meth:unc2}
\end{equation}
One problem that arises with the Bayesian approach is how to determine the prior $P\big(\{ c_\text{mod}^{(k)}\} \big)$.
As it is a statement about the model or the model parameter choice, there is no way it can be determined from experiment.
An approximation would be to use the SM as a prior, provided that the BSM effects are small.

One can learn from the Bayesian approach how to deal with different types of uncertainties.
In the current framework, we have theoretical, statistical and systematic uncertainties.
One could introduce nuisance parameters $r$ for each uncertainty, which ensures that the measurements can be smeared around their central value according to the uncertainties.
This method is adopted in the Monte Carlo (MC) toy method, see the section below.

Another way to maximize the likelihood function is with the Frequentist approach.
Here, the number of parameters is reduced to non-flat directions.
This is in principle ideal for SMEFT fits, where there is typically a large number of flat directions.
The problem is that then the normalization of the likelihood function as a probability measure is not justified any more.
One therefore uses the profile likelihood, which can be derived from the likelihood function by maximizing it for each fit parameter $c^{(k)}$. 

In the current framework, Gaussian statistical and systematic uncertainties are used.
Theoretical uncertainties from the SM predictions are used, but the theoretical uncertainties on the predictions of $\sigma_i$ and $\tilde{\sigma}_{ij}$ from eq.~\ref{eq_SMEFT_sigma} are neglected.
A more conservative approach would be to use SM-like uncertainties on $\sigma_i$ and $\tilde{\sigma}_{ij}$, but this is left for future work.
The theoretical uncertainties are flat because it is impossible to define a best value in this case.
The profile likelihood of the combination of a Gaussian with a box uncertainty goes as follows:
Assume a Gaussian uncertainty distribution with width $\sigma$ and a box uncertainty with width $y_\text{max} - y_\text{min}$.
These two uncertainties shall not be correlated.
Then the maximum of the profile likelihood is
\begin{equation}
\begin{aligned}
\mathcal{L}(y) &= \max_x \Theta(y_\text{max} -x)\ \Theta(x -y_\text{min})\  e^{-(x-y)^2/(2\sigma^2)} \\
& = \max_{x\in [y_\text{min}, y_\text{max}]} e^{-(x-y)^2/(2\sigma^2)} \\
&= \begin{cases}
e^{-(y-y_\text{min})^2/(2\sigma^2)} & y<y_\text{min} \\
1 & y \in [y_\text{min}, y_\text{max}] \\
e^{-(y-y_\text{max})^2/(2\sigma^2)} & y>y_\text{max}.
\end{cases}
\end{aligned}
\label{eq:meth:unc3}
\end{equation}

This construction of the likelihood is called \text{RFit} scheme~\cite{Hocker:2001xe} and is used in the current framework.


\subsection{Monte Carlo Toys}

The MC toy method is a useful approach in particle physics phenomenology to propagate experimental uncertainties from the input dataset to the constraints of the operators~\cite{Hartland:2019bjb}.
The advantage of this method is that it does not require any assumptions about the underlying probability distribution of the fitted parameters.
What makes it particularly interesting for SMEFT is that it is suitable for parameter spaces that are large and have many flat directions.

The idea is to reconstruct the experimental input using $k=1,\dots,N_\text{toy}$ toys of the original measurements.
In this study, $N_\text{toy}=10,000$ toys are used, which is more than enough to minimize any significant statistical fluctuations.
These toys $O_i^{(\text{toy})}$ are constructed from the nominal values $O_i^{(\text{exp})}$ of the $i=1,\dots,N_\text{meas}$ measurements and its uncertainties as follows:
\begin{equation}
O_i^{(\text{toy})(k)} = O_i^{(\text{exp})} 
+ \sum_{\alpha=1}^{N_\text{theo}} r_{i,\alpha}^{(k)}\ \sigma_{i,\alpha}^{(\text{theo})}
+ \sum_{\beta=1}^{N_\text{stat}} r_{i,\beta}^{(k)}\ \sigma_{i,\beta}^{(\text{stat})}
+ \sum_{\gamma=1}^{N_\text{syst}} r_{i,\gamma}^{(k)}\ \sigma_{i,\gamma}^{(\text{syst})}.
\label{eq:meth:toys1}
\end{equation}
Here, $\sigma_{i,\alpha}^{(\text{theo})}$, $\sigma_{i,\beta}^{(\text{stat})}$ and $\sigma_{i,\gamma}^{(\text{syst})}$ denote the theoretical, statistical and systematic uncertainties, respectively.
$N_\text{theo}$, $N_\text{stat}$, $N_\text{stat}$ are the corresponding numbers of uncertainties.
The coefficients $r_{i,\alpha}^{(k)}$ are random numbers from a flat distribution between minus one and one, while $r_{i,\beta}^{(k)}$ and $r_{i,\gamma}^{(k)}$ are from a univariate Gaussian distribution.

We assume that two theoretical uncertainties $\sigma_{i,\alpha}^{(\text{theo})}$ and $\sigma_{i',\alpha}^{(\text{theo})}$ are not correlated unless they arise from theoretical predictions of measurements of the same process at the same energy.
In the latter case, the two measurements are averaged, which is explained in detail in subsection~\ref{subsec:meth:theocorr}.
Statistical uncertainties are not correlated by nature.
Correlations between two systematic uncertainties from measurements $i$ and $i'$ with indices $\gamma$ and $\gamma'$ are implemented by setting $r_{i,\gamma}^{(k)}=r_{i',\gamma'}^{(k)}$.
This is further explained in section~\ref{subsec:meth:statcorr}.


\subsection{Log-Likelihood}
\label{subsec:meth:loglik}

The goal of \textsc{SFitter} is to find the set of theoretical predictions $O_i^{(\text{pred})}(\{ c_i^{(k)} \})$ that consitute the best fit to toys $O_i^{(\text{toy})(k)}$.
Thereby, $\{ c_i^{(k)} \}$ is the set of Wilson coefficients that the SMEFT predictions depend on.
These predictions are found by minimizing the log-likelihood $\chi^2=-2\mathcal{L}$ with respect to $\{ c_i^{(k)} \}$.
For a fit of the $k$-th MC toy from $N_\text{meas}$ measurements, this is equivalent to
\begin{equation}
(\chi^2)^{(k)} = \min_{\{ c_i^{(k)} \}}\ \sum_{i,j=1}^{N_\text{meas}} \left( O_i^{(\text{pred})}(\{ c_l^{(k)}\}) - O_i^{(\text{toy})(k)} \right) (\text{cov}^{(k)})_{ij}^{-1} \left( O_j^{(\text{pred})}(\{ c_l^{(k)}\}) - O_j^{(\text{toy})(k)} \right),
\label{eq:meth:chi1}
\end{equation}
where the covariance matrix is defined via
\begin{equation}
(\text{cov}^{(k)})_{ij} = \left(  
\sum_{\alpha=1}^{N_\text{theo}} \big( \sigma_{i,\alpha}^{\text{(theo)}} \big)^2 
+ \sum_{\beta=1}^{N_\text{stat}} \big( \sigma_{i,\beta}^{\text{(stat)}} \big)^2 
\right) \delta_{ij} 
+ \sum_{\gamma=1}^{N_\text{syst}}\ \sigma_{i,\gamma}^{\text{(syst)}}\ \sigma_{j,\gamma}^{\text{(sys)}}\ O_i^{\text{(toy)(k)}} O_j^{\text{(toy)(k)}}.
\label{eq:meth:chi3}
\end{equation}
The first two terms go with $\delta_{ij}$ because the correlations between theoretical uncertainties are taken care of by averaging, and statistical uncertainties are uncorrelated.

\begin{figure}[h]
\centering
  \includegraphics[width=0.9\textwidth]{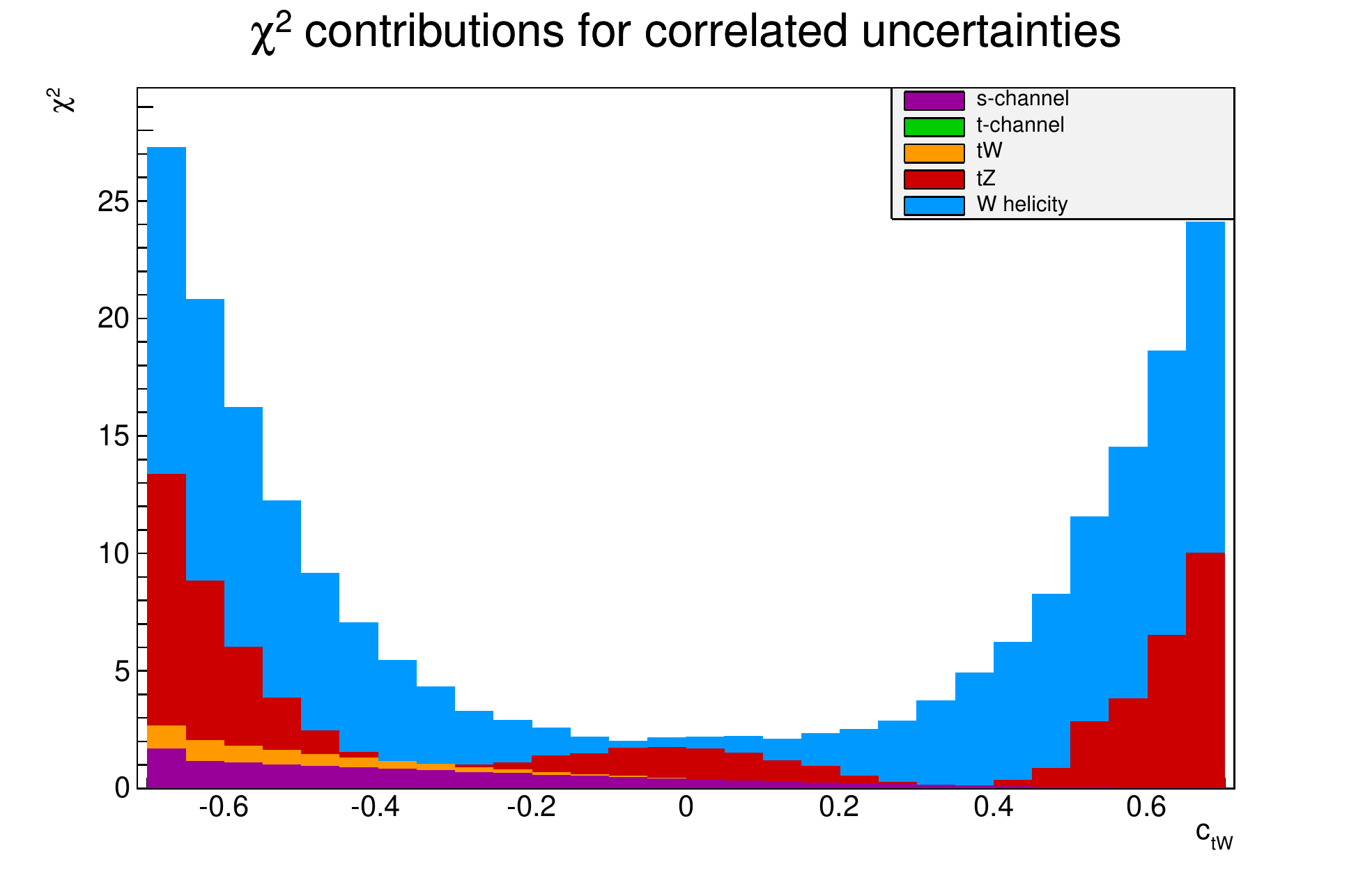}
  \vspace{-3mm}
  \caption{Contributions of measurements of the studied physical processes to the log-likelihood $\chi^2$ as a function of the Wilson coefficient $c_{tW}$.
  All measurements of table~\ref{tab:data:measurements} apart from the $t$-channel distributions are used.
  The nominal value is used, without employing any toys, and correlations between theoretical and systematic uncertainties are taken into account.
  }
  \label{img:meth:chi2}
\end{figure}

Figure~\ref{img:meth:chi2} shows the contributions of various measurements to the log-likelihood in a single-parameter fit of $c_{tW}$.
The measurements are from table~\ref{tab:data:measurements}, where only the measurements of the total cross sections of the $t$-channel single top quark production were used and the corresponding distributions are omitted.
Instead of toys, the nominal value of each measurement is used. 

The contributions to the log-likelihood are found by using the grid function in \textsc{SFitter}.
In this mode, \textsc{SFitter} calculates the log-likelihood of each measurement at equidistant points in the space of $c_{tW}$.
One can clearly see that the contributions of the single top quark production in $s$-channel and in association with a $W$ boson fall linearly.
The reason is that in both channels, the quadratic term of $c_{tW}$ is subleading.
The $s$-channel measurements are lower than their SM prediction, and the linear term $\sigma_{tW}$ (with $\sigma_{tW}$ like in equation~\ref{eq_SMEFT_sigma}) is positive.
The reverse is true for the production in association with a $W$ boson.

The measurements of the $t$-channel production do not contribute to the log-likelihood at all. 
Not only are they in excellent agreement with the SM, but the bands of the theoretical uncertainties are so big that the SMEFT predictions remain inside them for the entire depicted range of $c_{tW}$.
In contrast, the one measurement of the production in association with a $Z$ boson has a large contribution to the log-likelihood.
The shape of this distribution with the two local minima at around $c_{tW}\simeq\pm0.3$ reflects the fact that the quadratic term is leading in this case.
The reason is that the symmetries in the SM $ttZ$ vertex are different from those in 

Finally, the measurements of the helicity fractions have only a small contribution around $c_{tW}=0$, but increase rapidly for higher / lower values.
This shows that $c_{tW}$ most sensitive to measurements of the helicity fractions and single top quark production in association with a $Z$ boson.
In total, one obtains a best fit point of $c_{tW}\simeq -0.1$ from this analysis.


\subsection{Correlated theoretical uncertainties}
\label{subsec:meth:theocorr}

As previously mentioned, two theoretical uncertainties are assumed to be uncorrelated unless they arise from the theoretical prediction of the same observable of the same process at the same energy, i.e. two predictions are identical.
In the dataset that was used for figure~\ref{img:meth:chi2}, this occurs many times. 
The measurements of $s$-channel production at 8 TeV are averaged, all those of $t$-channel production and production in association with a $W$ boson. 
As the helicity fractions of the $W$ boson are independent of the center-of-mass energy, the measurements of $F_0$ and $F_L$ are also averaged across energies.

\begin{figure}[b!]
\centering
  \includegraphics[width=0.9\textwidth]{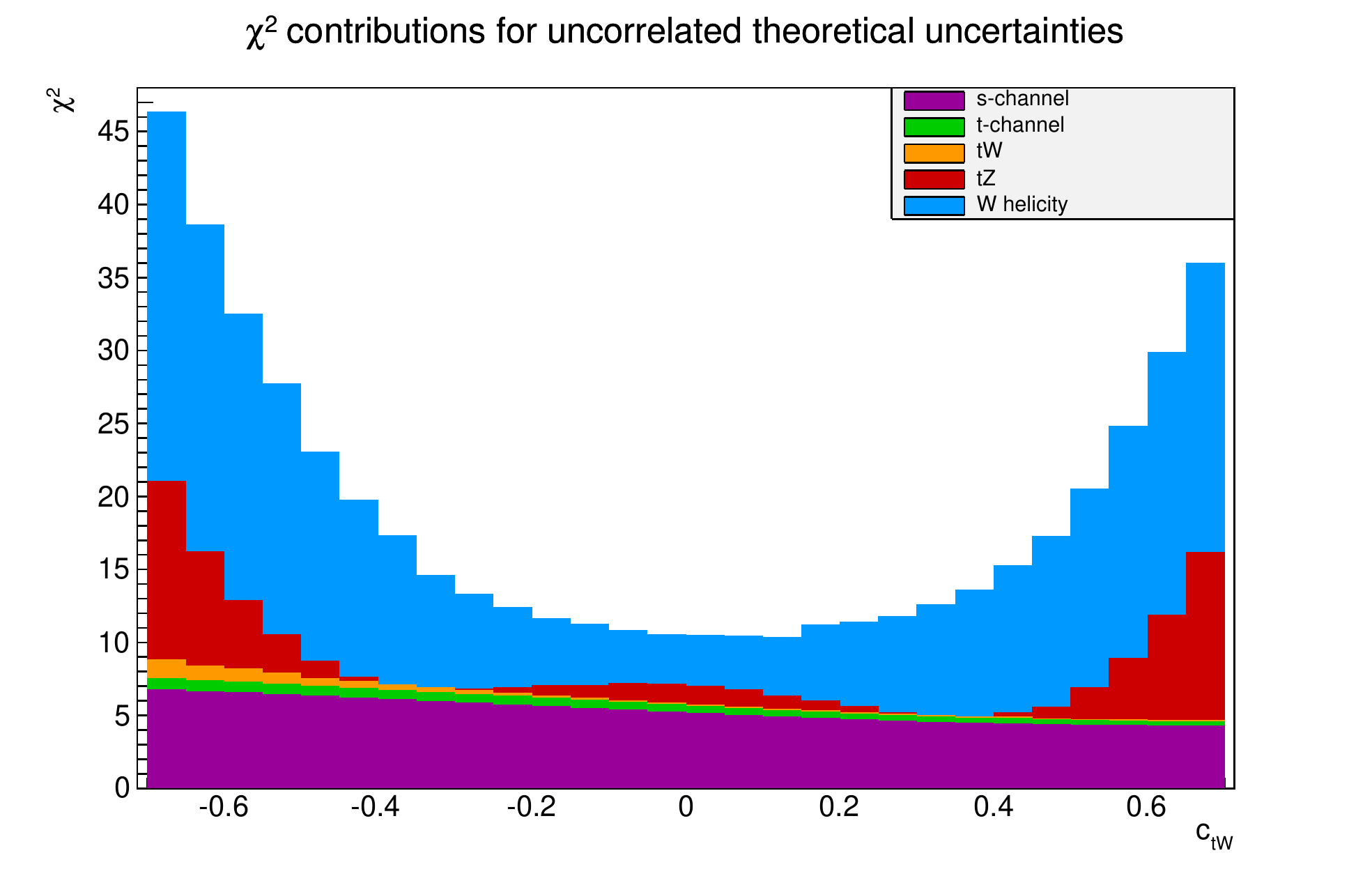}
  \vspace{-3mm}
  \caption{Contributions of measurements of table~\ref{tab:data:measurements} (except $t$-channel distributions) to the log-likelihood $\chi^2$ as a function of the Wilson coefficient $c_{tW}$.
  All measurements of table apart from the $t$-channel distributions are used.
  Unlike figure~\ref{img:meth:chi2}, systematic uncertainties are taken into account but theoretical uncertainties are not.
  }
  \label{img:meth:chi2_hatuncorr}
\end{figure}

Figure~\ref{img:meth:chi2_hatuncorr} shows the log-likelihood over $c_{tW}$ where no averaging has taken place and thus the correlations among theoretical uncertainties have not been taken into account.
While the overall shape of the distribution does not change much, one can clearly see that the contributions of all processes to the log-likelihood get bigger.
In particular, there is a nonzero contribution of the $t$-channel measurements and the contribution of the $s$-channel measurements have an offset of about six, compared to figure~\ref{img:meth:chi2}.
The best-fit point according to this figure is $c_{tW}\simeq +0.15$, as opposed to $c_{tW}\simeq -0.1$ with the correct implementation.

If the theoretical uncertainties were uncorrelated, one would have expected the two figures to be the same.
From the comparison of the two figures one can see that this is not the case.
Not taking correlations among theoretical uncertainties into account, through averaging or otherwise, modifies the covariance matrix in equation~\ref{eq:meth:chi3}.
This leads to an artificial distortion of the log-likelihood. 

In the following, the process of averaging measurements is described in more detail.
Consider a set of $m$ measurements and $n>1$ statistical and systematic uncertainties. 
The theoretical uncertainties do not play a role in the average because they arise from the predictions and not the measurements.
The uncertainty matrix $D \in M_{m\times n}$ can be written as~\cite{Avery1996} 
\begin{equation}
D =
\begin{bmatrix}
    \sigma_{11} & \sigma_{12} & \dots  & \sigma_{1m} \\
    \sigma_{21} & \sigma_{22} & \dots  & \sigma_{2m} \\
    \vdots & \vdots & \ddots & \vdots \\
    \sigma_{n1} & \sigma_{n2} & \dots  & \sigma_{nm}.
\end{bmatrix},
\end{equation}
Then the elements of the covariance matrix $V \in M_{m\times m}$ are defined as
\begin{equation}
V_{ij} = \sum_{k=1}^n \sigma_{ki} \sigma_{kj}
\label{eq_AvgMeas_cov}
\end{equation}
with $i,j=1,\dots,m$. 
In the case of uncorrelated measurements, this simplifies to $V_{ij}= \sum_{k=1}^n \sigma_{ki}^2 \delta_{ij}$.
In the literature, one often finds a factor of $\frac{1}{n-1}$ in the covariance matrix.
This factor is not needed, however, because of the way the matrix is used in the following.
The aim is to find an estimator
\begin{equation}
\bar{x} = \sum_{i=1}^m w_i x_i,
\label{eq_AvgMeas_x}
\end{equation}
consisting of measurements $x_i$ and weights $w_i$ such that $\sum_i w_i =1$.
The goal is to minimize the standard deviation 
\begin{equation}
\sigma_{\bar{x}}^2 = \sum_{i,j=1}^m w_i w_j V_{ij}
\label{eq_AvgMeas_sigma}
\end{equation}
with respect to $\{w_i\}$. 
For this, the Lagrange multiplier technique is used:
\begin{equation}
\sigma_{\bar{x}}^2 = \sum_{i,j=1}^m w_i w_j V_{ij} + \lambda (\sum_{i=1}^m w_i -1),
\end{equation}
By setting $\frac{\text{d}\sigma_{\bar{x}}^2}{\text{d}w_i}=\frac{\text{d}\sigma_{\bar{x}}^2}{\text{d}\lambda}=0$, one finds the solution
\begin{equation}
w_i= \frac{\sum_{k=1}^m V^{-1}_{ik}}{\sum_{j,l=1}^m V^{-1}_{jl}}.
\label{eq_AvgMeas_weights}
\end{equation}
This expression contains the inverse of the covariance matrix.
Note that the inverse not have to exist for the weights to be determined.
In the case where $V$ is not invertible, one can add a small offset $\epsilon$ to the diagonal elements so the inverse exists.
In the calculation of $w_i$, the offset $\epsilon$ finally drops out.

By inserting equation~\ref{eq_AvgMeas_weights} into equations~\ref{eq_AvgMeas_x} and~\ref{eq_AvgMeas_sigma}, one can simplify the formulae to 
\begin{equation}
\sigma_{\bar{x}}^2 = \frac{1}{\sum_{i,j=1}^m V^{-1}_{ij}},\ \text{and}
\label{eq_AvgMeas_sigmaAB}
\end{equation}
\begin{equation}
\bar{x}= \sigma_{\bar{x}}^2\ \sum_{i,j=1}^m x_i\ V^{-1}_{ij}.
\label{eq_AvgMeas_xAB}
\end{equation}

Via construction of a weighted average, one can average iteratively, i.e. first average two measurements, and then average that with another measurement.
Therefore, it suffices to focus on $m=2$.
With two measurements $A$ and $B$, the elements of the covariance matrix~\ref{eq_AvgMeas_cov} become
\begin{equation}
\begin{aligned}
& V_{AA} = \sum_{k=1}^n \sigma_{kA}^2 \\
& V_{BB} = \sum_{k=1}^n \sigma_{kB}^2 \\
& V_{AB} = V_{BA} = \sum_{k=1}^n \rho_k\ \sigma_{kA}\ \sigma_{kB}. \\
\end{aligned}
\label{eq_AvgMeas_covEl}
\end{equation} 
Here, $\rho_k \in [-1,1]$, where $-1$ means that the uncertainties $\sigma_{kA}$ and $\sigma_{kB}$ are fully anticorrelated, +1 fully correlated, and 0 that they are uncorrelated.
The elements of the inverse matrix are
\begin{equation}
V_{AA}^{-1} = \frac{V_{BB}}{\text{det}\ V},\ \ \ V_{BB}^{-1} = \frac{V_{AA}}{\text{det}\ V}, \ \ \ V_{AB}^{-1} = V_{BA}^{-1} = -\frac{V_{AB}}{\text{det}\ V}, 
\label{eq_AvgMeas_invCovEl}
\end{equation}
where $\text{det}\ V = V_{AA}\ V_{BB} - V_{AB}^2$.
Inserting~\ref{eq_AvgMeas_invCovEl} into~\ref{eq_AvgMeas_sigmaAB} and~\ref{eq_AvgMeas_xAB} gives
\begin{equation}
\sigma_{\bar{x}}^2 = \frac{1}{V_{AA}^{-1} + V_{BB}^{-1} + 2\ V_{AB}^{-1}}
\label{eq:meth:totmean}
\end{equation}
\begin{equation}
\bar{x} = \sigma_{\bar{x}}^2 \left(A\ V_{AA}^{-1} + B\ V_{BB}^{-1} + (A+B)\ V_{AB}^{-1} \right)
\label{eq:meth:totsigma}
\end{equation}

The expression for the total uncertainty of the weighted mean in equation~\ref{eq:meth:totmean} does not suffice because various systematic uncertainties might be correlated with systematics of other measurements.
The individual of uncertainties $\sigma_{\bar{x},i}$ of the weighted mean can be calculated using
\begin{equation}
\sigma_{\bar{x},i}^2 =  (w_A \sigma_{iA})^2 + (w_B \sigma_{iB})^2 +2\ \rho_i\ w_A \sigma_{iA}\ w_B \sigma_{iB}.
\label{eq_AvgMeas_indUnc}
\end{equation}
As $\rho_i$ is equal to zero for statistical uncertainties and one for correlated systematic uncertainties, this is equivalent to using
\begin{equation}
\begin{aligned}
& \sigma_{\bar{x},i} = w_A \sigma_{iA} + w_B \sigma_{iB}\quad \text{for correlated systematic uncertainties, and} \\
& \sigma_{\bar{x},i}^2 =  w_A^2 \sigma_{iA}^2 + w_B^2 \sigma_{iB}^2\quad \text{for statistical uncertainties.} \\
\label{eq:meth:avg}
\end{aligned}
\end{equation}

What is left to be proven is that this method of extracting different uncertainties is indeed correct, i.e. 
\begin{equation}
\sigma_{\bar{x}}^2 = \sum_{i=1}^n \sigma_{\bar{x},i}^2.
\end{equation}
This is shown as follows: As before, assume two measurements $A,B$ with covariance matrix elements $V_{AA}, V_{BB}$, and $V_{AB}$.
Further assume that there are $N_\text{stat}$ statistical and $n - N_\text{stat}$ systematic uncertainties, and the uncertainties are sorted such that all statistical uncertainties are listed first, and the systematic uncertainties after that.
The statistical uncertainties are all not correlated at all, $\rho_k = 0$ for $k=1, \dots, N_\text{stat}$, and the systematics are fully correlated, $\rho_k = 1$ for $k=N_\text{stat}+1, \dots, n$.
Then the expression for $V_{AB}$ can be simplified to 
\begin{equation}
V_{AB} = V_{BA} = \sum_{k=N_\text{stat}+1}^n \sigma_{kA}\ \sigma_{kB}. 
\label{eq:meth:VAB}
\end{equation}

Inverting the covariance matrix, one can simplify equations~\ref{eq:meth:totsigma} and~\ref{eq_AvgMeas_weights} to 
\begin{equation}
\begin{aligned}
& \sigma_{\bar{x}}^2 = \frac{V_{AA} V_{BB} - (V_{AB})^2} {V_{AA} + V_{BB} -2 V_{AB}},\\
& w_A = \frac{V_{BB} - V_{AB}}{V_{AA}+V_{BB}-2 V_{AB}}, \quad
 w_B = \frac{V_{AA} - V_{AB}}{V_{AA}+V_{BB}-2 V_{AB}} .
\label{eq:meth:totsigmasimple}
\end{aligned}
\end{equation}

Now, the expression above can be compared with the quadratic sum of the sources of uncertainty of the weighted mean:
\begin{equation*}
\begin{aligned}
\sum_{i=1}^n \sigma_{\bar{x},i}^2 &= \sum_{i=1}^{n_\text{stat}} \sigma_{\bar{x},i}^2 + \sum_{i=n_\text{stat}+1}^n \sigma_{\bar{x},i}^2 \\
&= \sum_{i=1}^{n_\text{stat}} (w_A^2 \sigma_{iA}^2 + w_B^2 \sigma_{iB}^2 + 2 w_A w_B \sigma_{iA} \sigma_{iB} ) + \sum_{i=n_\text{stat}+1}^n (w_A^2 \sigma_{iA}^2 + w_B^2 \sigma_{iB}^2 ) \\
&= w_A^2 \sum_{i=1}^n \sigma_{iA}^2 + w_B^2 \sum_{i=1}^n \sigma_{iB}^2 + 2 w_A w_B \sum_{i=1}^{n_\text{stat}} \sigma_{iA} \sigma_{iB} \\
&= w_A^2 V_{AA}+ w_B^2 V_{BB} + 2 w_A w_B V_{AB} \\
\end{aligned}
\end{equation*}
\begin{equation*}
\begin{aligned}
\qquad \qquad \quad \ &= \frac{(V_{BB} - V_{AB})^2 V_{AA}+ (V_{AA} - V_{AB})^2 V_{BB} + 2 (V_{BB} - V_{AB}) (V_{AA} - V_{AB}) V_{AB}}{(V_{AA}+V_{BB}-2 V_{AB})^2} \\
&= \frac{V_{AA} V_{BB} - (V_{AB})^2} {V_{AA} + V_{BB} -2 V_{AB}} =\sigma_{\bar{x}}^2.
\end{aligned}
\end{equation*}

In the first line, the uncertainties are split up into statistical and systematic uncertainties.
In the second line, equation~\ref{eq:meth:avg} is inserted, which is the claim.
In the fourth and fifth line, equations~\ref{eq:meth:VAB} and~\ref{eq:meth:totsigmasimple} are inserted, respectively.
The sixth line is obtained by evaluating the braces in the enumerator and performing a polynomial division by $(V_{AA}+V_{BB}-2 V_{AB})$.

The averaging of measurements is one of the features that is automated with \textsc{Data-} \textsc{Prep}, a software developed by the author.
A detailed description of this software can be found in appendix~\ref{app:DataPrep}.


\subsection{Correlated systematic uncertainties}
\label{subsec:meth:statcorr}

\begin{figure}[h]
\vspace{3mm}
\centering
  \includegraphics[width=0.9\textwidth]{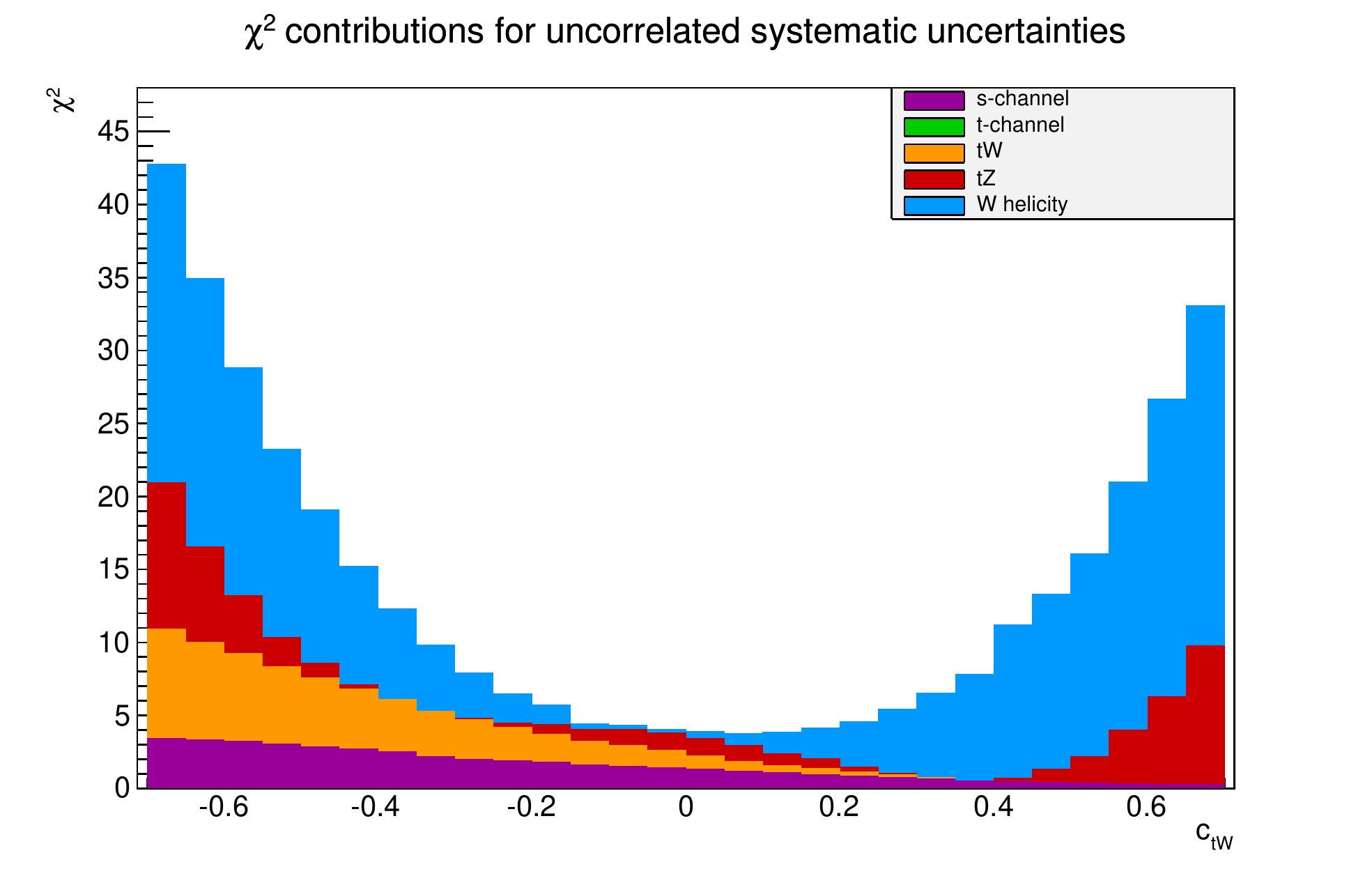}
  \vspace{-3mm}
  \caption{Contributions of measurements of table~\ref{tab:data:measurements} (except $t$-channel distributions) to the log-likelihood $\chi^2$ as a function of the Wilson coefficient $c_{tW}$.
  All measurements of table apart from the $t$-channel distributions are used.
  Unlike figure~\ref{img:meth:chi2}, theoretical uncertainties are taken into account but systematic uncertainties are not.
  }
  \label{img:meth:chi2_statuncorr}
  \vspace{5mm}
\end{figure}

As previously mentioned, systematic uncertainties are often correlated. 
Consider for example the measurements of two cross sections $\sigma_t$ and $\sigma_{\bar{t}}$ of a top and an anti-top quark.
Further assume that these measurements are from the same process, with the same center-of-mass energy and from the same detector.
One would then expect detector-related uncertainties of $\sigma_t$ and $\sigma_{\bar{t}}$ to be correlated.

In the fit from figure~\ref{img:meth:chi2}, correlations between systematic uncertainties are taken into account. 
In contrast, in the fit shown in figure~\ref{img:meth:chi2_statuncorr} correlations between theoretical uncertainties have been taken into account, but not those between systematic uncertainties.
One can see that the distribution of the log-likelihood changes a lot.
The contributions of the $s$-channel production to the log-likelihood increase by about a factor of two, those of the production in association with a $W$ boson by a factor of seven.
The contributions of the $t$-channel production remain zero.

The fact that the production in association with a $Z$ boson remains unchanged is not surprising because there is only one measurement of it and no systematic uncertainties are assumed to be correlated among different processes.
The contributions of the measurements of the helicity fractions increase by roughly a factor of two.
In total, one gets a best-fit point of $c_{tW}\simeq+0.1$ instead of $c_{tW}\simeq-0.1$.

Taking correlations between systematic uncertainties into account is another feature of \textsc{DataPrep}, see appendix~\ref{app:DataPrep}.
This is a huge advantage because of the large number of different systematic uncertainties in the fit.


\subsection{Execution}

\begin{figure}[h]
  \centering
  \includegraphics[width=\textwidth]{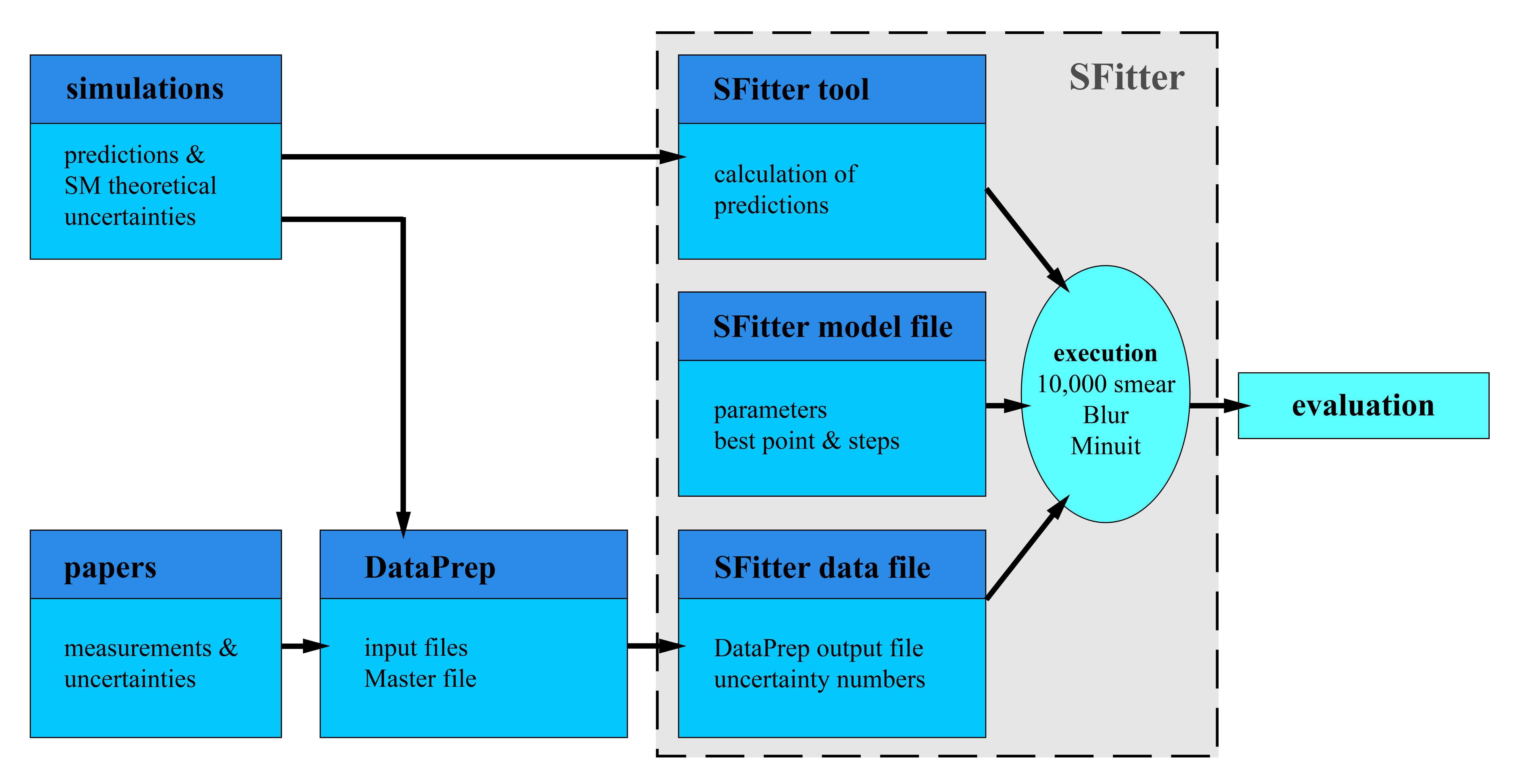}
\caption{Work flow in this thesis. \textsc{DataPrep} is not regarded as part of \textsc{SFitter} as it is still a separate entity at the time of publishing this thesis.}
\label{img:meth:corrhat3}
\end{figure}

In figure~\ref{img:meth:corrhat3}, one can see the work flow adopted in this thesis.
The first step is to extract the predictions of the SM and the dimension-six Wilson coefficients from the \textsc{MadGraph} simulations.
These get used by the \textsc{SFitter} tool to calculate the predictions during the fit.
Also, the measurements and their statistical and systematic uncertainties are extracted from the experimental analyses.
These and the theoretical uncertainties on the SM prediction from the simulations get plugged into the \textsc{DataPrep} input files.
In addition, the so-called Master file of \textsc{DataPrep} needs to edited to specify the different uncertainties and the correlations among them.

The output of \textsc{DataPrep} then is included in the \textsc{SFitter} data file.
The numbers of theoretical, statistical and systematic uncertainties also get added after they have been output by \textsc{DataPrep}.
In the \textsc{SFitter} tool file, the user must specify the way that the SMEFT predictions are calculated.
This is, of course, dependent on which Wilson coefficients are used and on whether the SMEFT order is $\mathcal{O}(\Lambda^{-2})$ or $\mathcal{O}(\Lambda^{-4})$.

The relevant parameters for the fit and their settings are listed in the \textsc{SFitter} model file, and a rough estimate for the best fit point and the step size is determined via $\chi^2$-scans with \textsc{SFitter}.
The step size of a parameter is important so that subprograms of \textsc{SFitter} work properly, and is usually set such that the variation of a parameter by the step size increases the log-likelihood of the fit by one.
When the step sizes are determined, \textsc{SFitter} is executed using $N_{\text{toy}}=10,000$ MC toys.

If not stated otherwise, the \textsc{SFitter} subprograms \textsc{Blur} and \textsc{Minuit} are used.
\textsc{Blur} takes care of an initial blurring of the operators so that the fit can converge better into a global minimum and does not get stuck in a local minimum of the log-likelihood.
It is in principle equivalent to a smear of an observable (but for a fit parameter), where the width of the Gaussian distribution that is used to smear is equal to the step size.
\textsc{Minuit} is a tool to find the best fit point with high precision.
Finally, the \textsc{SFitter} output is evaluated and bounds are derived for the dimension-six operators.

\section{Results}
\label{sec_results}

In this section, the final results are presented. 
In addition, the stability of the fit is assessed regarding variations of the dataset, such as leaving away kinematic distributions or measurements at 7 TeV, and variations of the theory settings, such as using only predictions at LO or leaving away SMEFT terms at order $\mathcal{O}(\Lambda^{-4})$.

\subsection{Standard dataset and theory settings}
\label{sec:results:S}


\subsubsection{Log-likelihood}

\begin{figure}[t!]
\centering
  \hspace{-5mm}
  \includegraphics[width=0.9\textwidth]{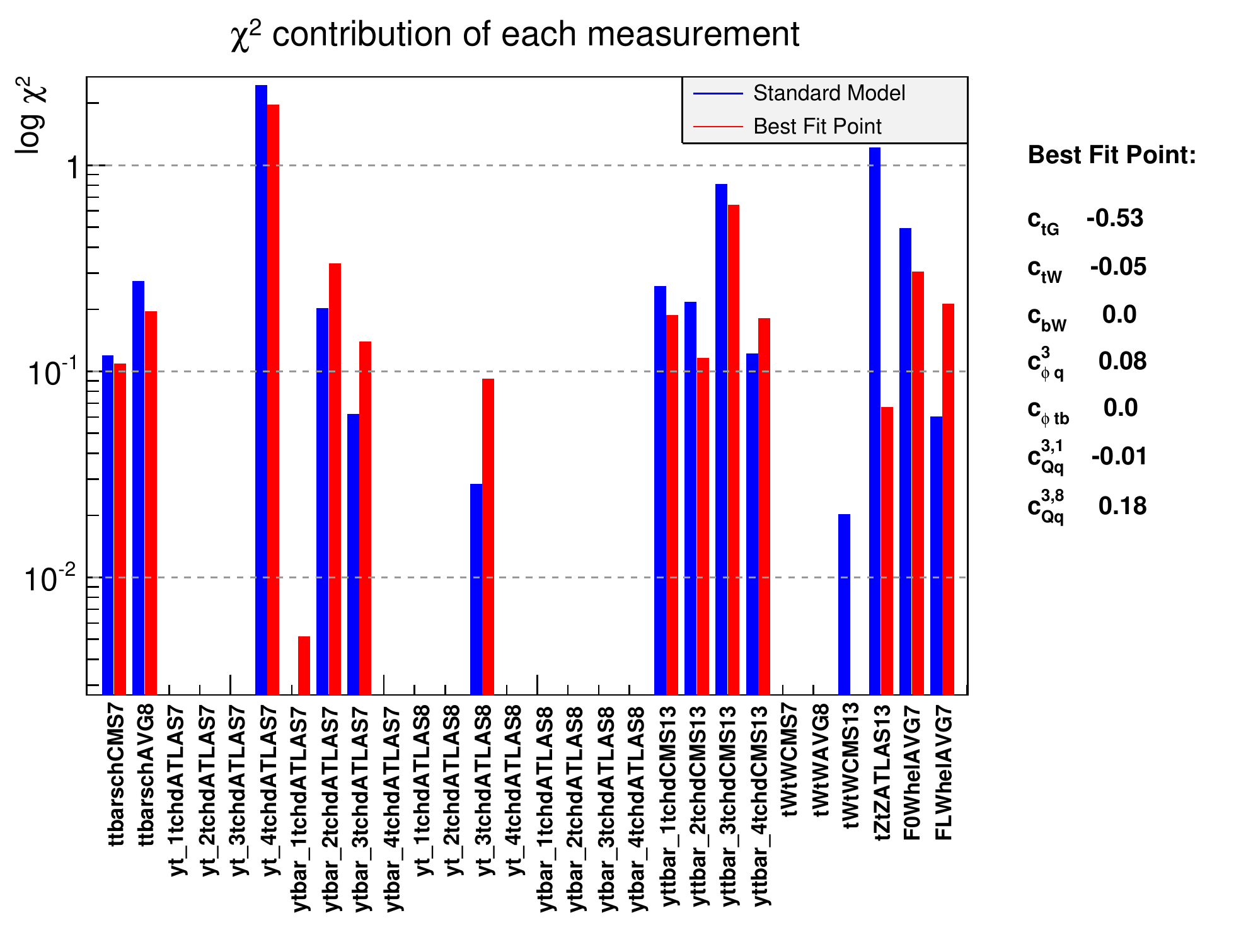}
  \hspace{-2mm}
  \caption{Contribution of each measurement to the log-likelihood, where the name of each measurement is stated in \textsc{SFitter} code. 
  The blue bars indicate the $\chi^2$-values of measurements in the Standard Model, the red bars those at the best fit values, as listed on the right hand side. 
  The $y$-axis is logarithmic to better visualize small contributions. }
  \label{img:results:Schi2}
\end{figure}

Figure~\ref{img:results:Schi2} shows the contributions of each measurement to the log-likelihood at the SM values and at the best fit points.
The best fit points were found by running \textsc{SFitter} with \textsc{Blur} and \textsc{Minuit} using only the nominal values of the measurements.
The measurements denoted in \textsc{SFitter} code, which works the following way:

\begin{itemize}
\item The prefixes \texttt{y} and \texttt{p} indicate distributions of the rapidity $\eta$ and the transverse momentum $p_T$ of the observed top quark, respectively.
\item The observable is indicated as \texttt{t}, \texttt{tbar}, \texttt{ttbar}, \texttt{tW}, \texttt{tZ}, \texttt{F0} or \texttt{FL}, which refers to the cross section of a single top quark, single top antiquark, top quark or antiquark, top quark in association with a $W$ boson or with a $Z$ boson, and the longitudinal helicity fraction and left-handed transversely polarized helicity fraction, respectively.
In the case where the name refers to a distribution, the suffix \texttt{\_x} refers to the x-th bin.
\item The tags \texttt{sch}, \texttt{tch}, \texttt{tchd}, \texttt{tW}, \texttt{tZ} and \texttt{Whel} indicate the physical process, which are single top quark $s$-channel production, $t$-channel production inclusive cross section and kinematic distribution, production in association with a $W$ and a $Z$ boson, and top quark decay, respectively.
\item \texttt{ATLAS}, \texttt{CMS} and \texttt{AVG} indicate the experiment that the measurement was performed by, where \texttt{AVG} means an average of ATLAS and CMS.
\item The last digit(s) indicate(s) the center-of-mass energy $\sqrt{s}$ in TeV. As the helicity fractions are energy-independent and thus all energies are averaged, the smallest energy is added to the name.
\end{itemize}

A log-likelihood of zero indicates that the measurement lies within the band of the theoretical uncertainties.
If the log-likelihood of a measurement is below one, the measurement agrees with the SM prediction within the one-$\sigma$ range.
One can see that all measurements agree with the SM predictions within the two-$\sigma$ range, both in the SM scenario and the best fit point.

Although not all contributions to the log-likelihood are smaller at the best fit point than at the SM, the overall log-likelihood is reduced from about six to about four with SMEFT.
The largest contribution to this reduction comes by far from the single top quark production in association with a $Z$ boson.
Also, the last bin of the $y_t$ distribution at 7 TeV has a relatively high contribution to the log-likelihood.
This is counter-intuitive as one would expect bins that correspond to higher energies to deviate more due to energy-growing SMEFT operators.
However, it is not impossible for the sensitivity of a SMEFT operator to be higher at lower energies.
While this deviation of a low-energy bin is notable, it is not significant.
In addition, the best fit point does not happen to change the contribution of the last bin to the log-likelihood very much, indicating that it does not hugely affect the overall fit.

While most bins of the $t$-channel distributions at 7 and 8 TeV lie within the bounds of the theoretical uncertainties, all bins of the distribution at 13 TeV lie slightly outside of them.
With exception of the of the $t$-channel distributions at 7 TeV, one can see a tendency to higher deviations from the SM at higher energies in all measurements.
This could indicate some potential to constrain energy-growing SMEFT operators when the LHC reaches even higher energies.


\subsubsection{Distributions}

As explained in section~\ref{sec_methodology}, \textsc{SFitter} is run with 10,000 MC toys of the current dataset, using \textsc{Blur} and \textsc{Minuit}. 
Then the distributions of each Wilson coefficient are used to determine their bounds.
These distributions are histograms with the values of the Wilson coefficient on the $x$-axis, and the number of times a value has been the best fit point on the $y$-axis.

\begin{figure}[b!]
\centering
\begin{subfigure}{0.48\textwidth}
  \includegraphics[width=\textwidth]{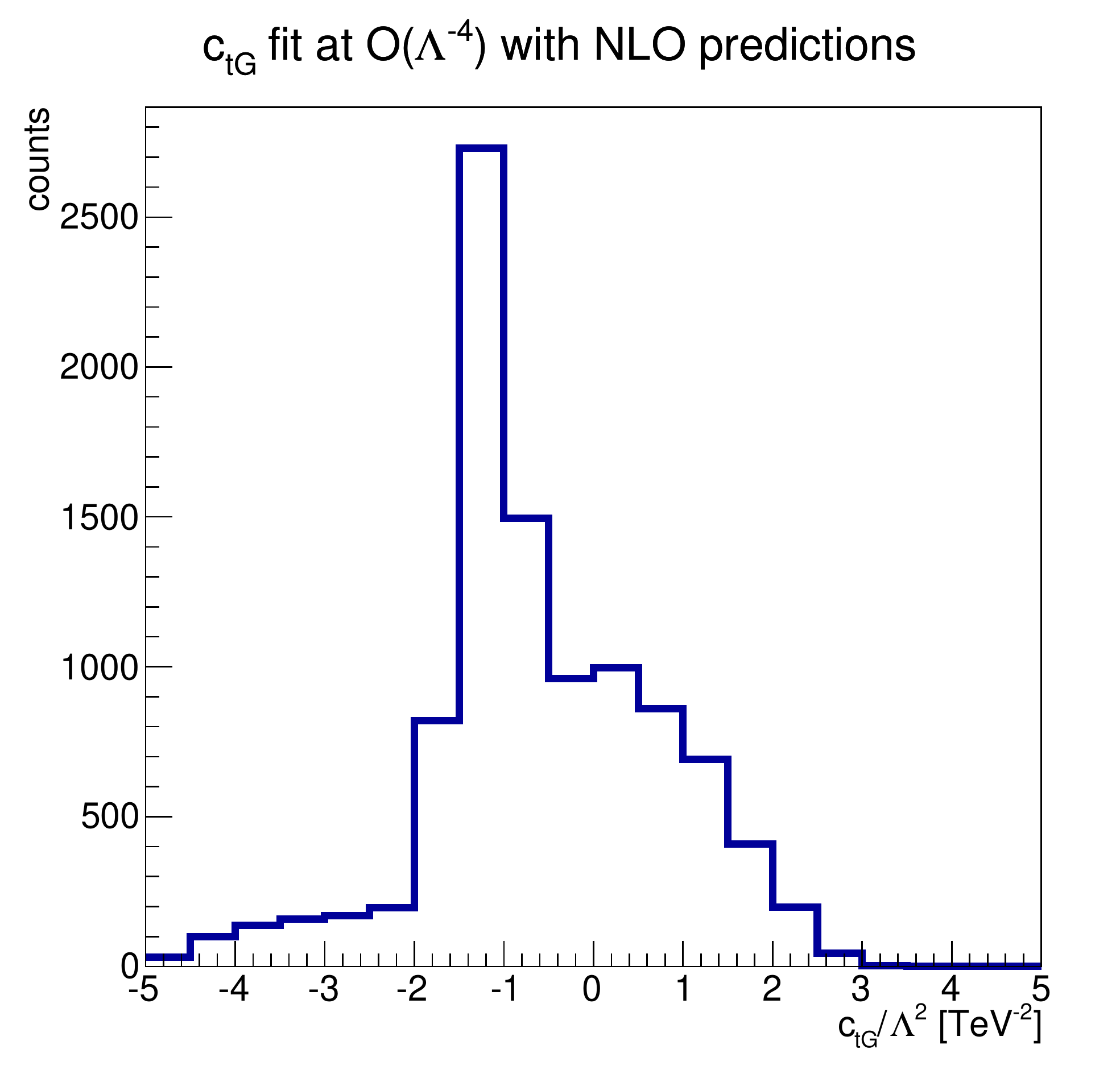}
  \caption{}
\end{subfigure}%
\hspace{1mm}%
\begin{subfigure}{0.48\textwidth}
  \includegraphics[width=\textwidth]{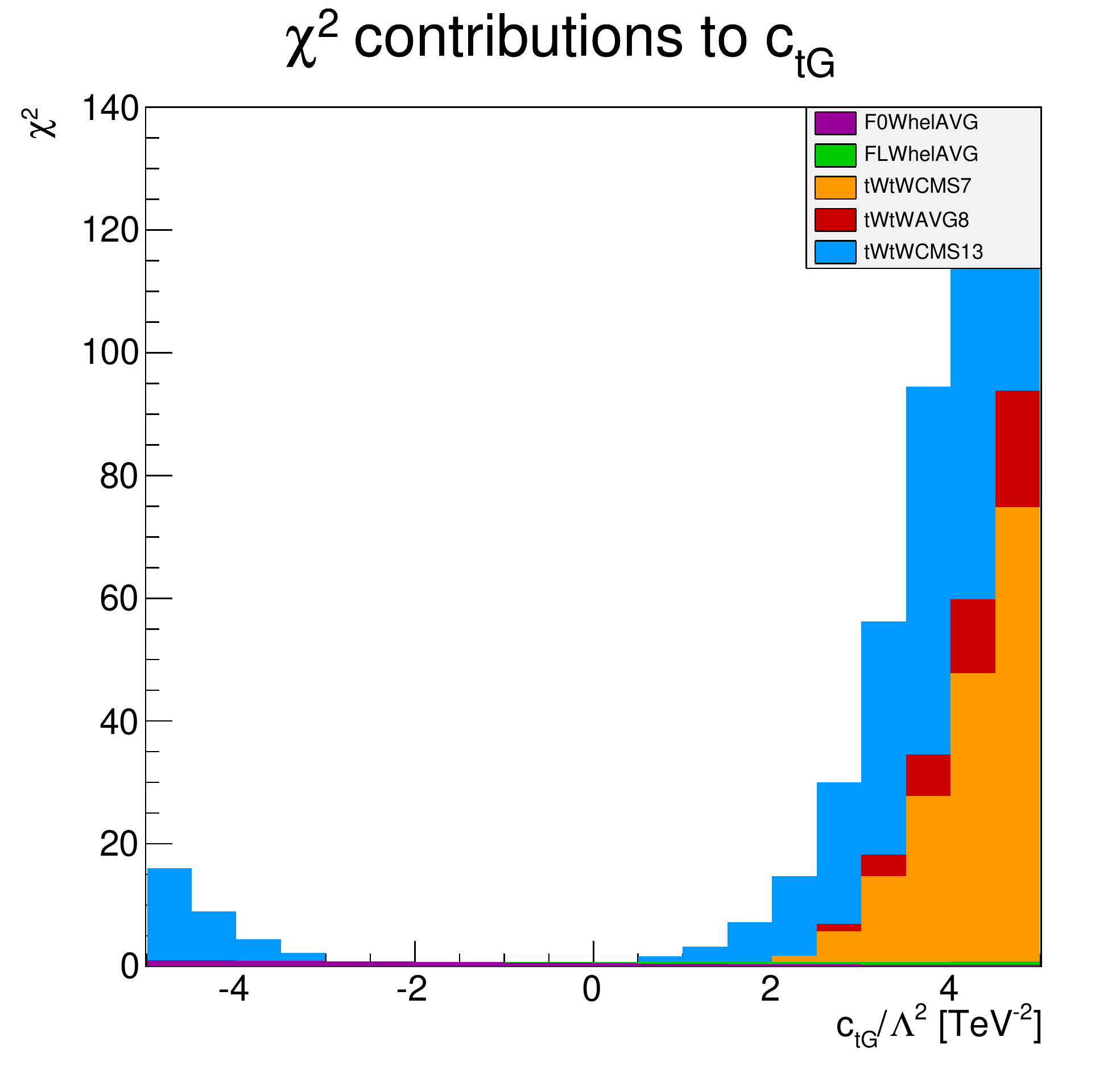}
  \caption{}
\end{subfigure}
\caption{Panel (a): Distribution of the counts of $c_{tG}$ after a global fit of all coefficients with 10,000 MC toys. Panel (b): Contributions of various measurements to the log-likelihood as a function of $c_{tG}$. Only the contributions of measurements which constrain $c_{tG}$ are shown. The contributions of the helicity fractions are in the range of zero to about one. }
\label{img:results:Stg}
\end{figure}

\begin{figure}[t]
\centering
  \includegraphics[width=0.9\textwidth]{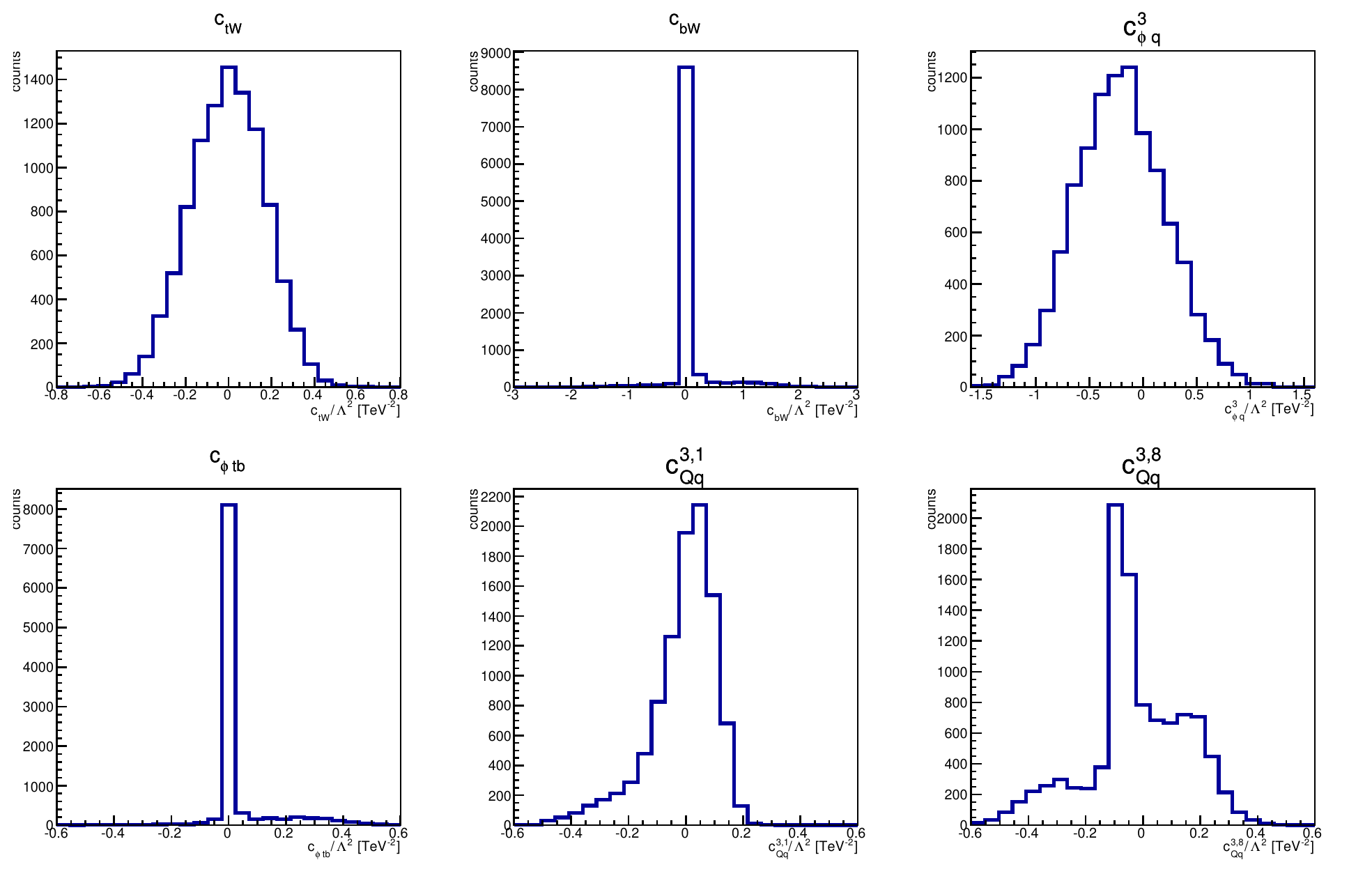}
\caption{ Distributions of all Wilson coefficients except $c_{tG}$, which is shown in figure~\ref{img:results:Stg}(a). Top left: $c_{tW}$, middle: $c_{bW}$, right: $c_{\varphi q}^3$; bottom left: $c_{\varphi tb}$, middle: $c_{Qq}^{3,1}$, right: $c_{Qq}^{3,8}$. Details see text. }
\label{img:results:Sother}
\end{figure}

The left panel of figure~\ref{img:results:Stg} shows an example of such a distribution. 
The right panel shows the contributions to the log-likelihood of the nominal value of those measurements which constrain $c_{tG}$.
One can see that the distribution is zero at a log-likelihood of around 20-30, taking only into account the measurements that are shown.
While one can see a slight bump around zero in the left panel, corresponding to the SM expectation, there is also a prominent peak at about $c_{tG}\simeq-1.3$, which corresponds to the most probable value.
Note that the most probable value is not necessarily equal to the best fit value because the latter is extracted using the nominal value of each measurement, while the former arises from toys.
Effects that cause a range of the toys to heap on one point in the $c_{tG}$-distribution can then cause the most probable value of the toys to differ from the best fit value of the nominal values.

The cause of the peak in the $c_{tG}$-distribution is the following:
The pre-factors $\sigma_i$ and $\tilde{\sigma}_{ij}$ from equation~\ref{eq_SMEFT_sigma} for the single top quark production in association with a $W$ boson are positive and similar in magnitude.
Therefore, the SMEFT contribution of $c_{tG}$ is smallest at around -1.3, and increases parabolically around this point.
This means that if the measurement is below the minimum attainable SMEFT prediction through variation, \textsc{SFitter} converges towards that value of $c_{tG}$ at which the prediction is smallest.
This, as one can see, happens for about a quarter of all toys.
In general, values higher than the peak are more probable than values lower than it because the nominal values of the $tW$-measurements at 7 and 8 TeV are higher than the SM, and only that at 13 TeV is lower.
This explains why the distribution of $c_{tG}$ falls off on the left hand side of the peak.

Figure~\ref{img:results:Sother} show the distributions of the remaining Wilson coefficients.
One can see that $c_{tW}$ and $c_{\varphi q}^3$ resemble a Gaussian centered around zero, as is expected given the excellent agreement of all measurements with the SM.
The Wilson coefficients which only contribute quadratically have a prominent peak around zero.
The reason is that a lot of measurements are slightly below their SM prediction, making an improvement of the log-likelihood impossible given the positive signs of the pre-factors of these coefficients.
The reverse is true where measurements of helicity fractions are concerned.

While $c_{Qq}^{3.1}$ is centered around zero, the falloff in positive direction is a lot more prominent than that in negative direction.
This is largely due to the measurements in $s$-channel, which have the largest contributions to the log-likelihood.
The reason behind the falloff is not only the high sensitivity to $c_{Qq}^{3,1}$ to the $t$-channel measurements, but also the fact that their nominal values lie above the SM predictions.
Combined with the positive pre-factors of this coefficient, this leads to a suppression of higher values and a favoring of lower values.

Finally, the distribution of $c_{Qq}^{3,8}$ has a peak at about -0.1 for similar reasons as the peak in the distribution of $c_{tG}$, as previously discussed.
The difference is that the peak is not caused by the production in association with a $W$ boson, but of that in association with a $Z$ boson.
The bumps at around -0.3 and +0.2 largely arise from the fact that the contributions of the $s$-channel measurements are smallest at $\pm 0.6$ and $\pm 0.3$ at 7 and 8 TeV, respectively.
The fact that the distribution falls off faster on the positive than the negative side arises from the fact that the single top quark in association with a $Z$ boson has higher contributions to the log-likelihood at higher values of $c_{Qq}^{3,8}$.
The fact that the bump on the positive side is a lot more pronounced than that on the negative side arises from interference terms of $c_{Qq}^{3,8}$ with other Wilson coefficients.

\subsubsection{Constraints}

\begin{figure}[p]
\centering
  \includegraphics[width=1.05\textwidth]{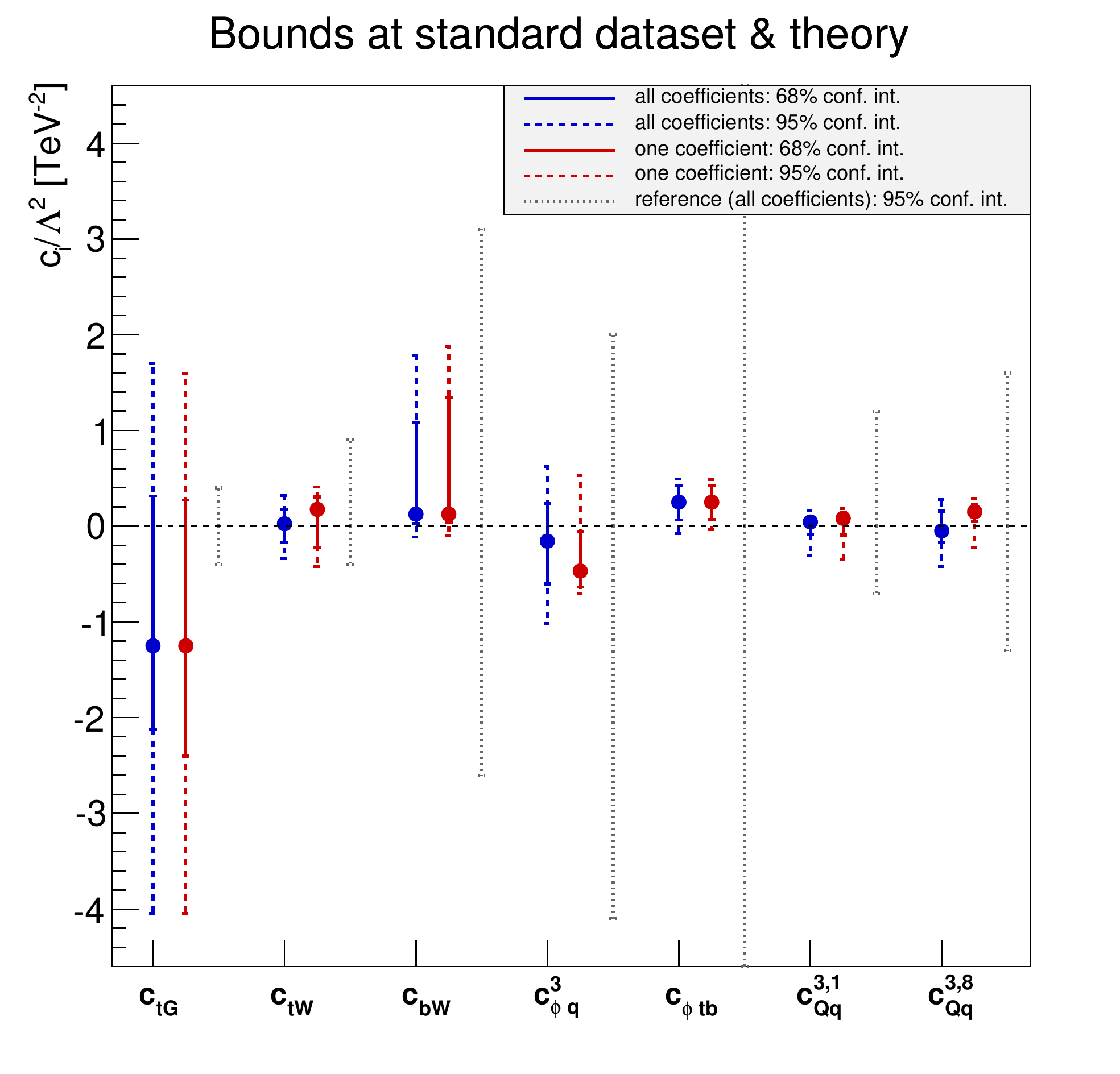}
  \caption{Visualization of the constraints on the Wilson coefficients of table~\ref{tab:results:bounds}. Solid lines mark the 68\%-, dashed lines the 95\% confidence intervals. 
  The blobs mark the most probable value of each coefficient.
  The blue lines indicate the constraints obtained from an fit with all coefficients, the red lines those obtained from fits with the individual coefficients.
  The finely dashed gray line marks the current constraints from the literature, as obtained from fits with all coefficients in question.
  The dashed line at $c_i/\Lambda^2=0$ marks the SM expectation.
  }
  \label{img:results:Sbounds}
\end{figure}

The constraints are obtained by cutting each distribution of a Wilson coefficient in two parts at its most probable value.
In each of the two parts, the 68\% (95\%)-range is extracted using percentiles. 
The constraints are shown in table~\ref{tab:results:bounds} and figure~\ref{img:results:Sbounds}.

All constraints obtained in this study with exception of $c_{tG}$ are more stringent than those obtained in the literature, even though a smaller subset of measurements is used.
Here, the stringency of a constraint is quantified by the difference of the upper and the lower bound - the smaller that value, the more stringent the constraint.
There could be a small effect from not respecting some correlations among theoretical uncertainties, for example when predictions of the same process but different energies were used.
This could lead to a slight artificial improvement of the bounds.
However, based on the findings in sections~\ref{subsec:meth:loglik} to~\ref{subsec:meth:theocorr}, this effect should be rather small.

One can see that the SM expectation lies within all constraints at 95\% confidence level, and almost all  at 68\% confidence level.
The only occurrences where the SM expectation does not lie within the 68\% confidence level are the bounds on $c_{\varphi q}^3$ in an individual fit, and those on $c_{\varphi tb}$ both in a fit with all coefficients and with only the coefficient in question.

In general, the bounds obtained from individual fits do not differ much from those obtained from the fit with all coefficients.
This indicates that the influence of different coefficients on one another is pretty small.
Also, the similarity between the results demonstrates that there are enough different observables in the current dataset such that the correlations among the Wilson coefficients are relatively small.

\begin{table}[t]
  \centering
  \vspace{5mm}
  \begin{tabular}{|llllll|}
  \hline
  coeff. & 68\% all & 95\% all & 68\% indiv. & 95\% indiv. & 95\% reference \\
  \hline
  $c_{tG}$ & [-2.1, 0.3] & [-4.1, 1.7] & [-2.4, 0.3] & [-4.0, 1.6] & [-0.4, 0.4]~\cite{Hartland:2019bjb} \\
  $c_{tW}$ & [-0.2, 0.2] & [-0.3, 0.3] & [-0.2, 0.3] & [-0.4, 0.4] & [-1.8, 0.9]~\cite{Hartland:2019bjb} \\
  $c_{bW}$ & [0.0, 1.1] & [-0.1, 1.8] & [0.0, 1.3] & [-0.1, 1.9] & [-2.6, 3.1]~\cite{Hartland:2019bjb} \\
  $c_{\varphi Q}^3$ & [-0.7, 0.2] & [-1.0, 0.6] & [-0.6, -0.1] & [-0.7, 0.5] & [-4.1, 2.0]~\cite{Buckley:2015lku}\\ 
  $c_{\varphi tb}$ & [0.1, 0.4] & [-0.1, 0.5] & [0.1, 0.4] & [0.0, 0.5] & [-27, 8.7]~~\cite{Hartland:2019bjb} \\
  $c_{Qq}^{3,1}$ & [-0.1, 0.1] & [-0.3, 0.2] & [-0.1, 0.1] & [-0.3, 0.2] & [-1.1, 1.3]~\cite{Hartland:2019bjb} \\
  $c_{Qq}^{3,8}$ & [-0.2, 0.2] & [-0.4, 0.3] & [0.0, 0.2] & [-0.2, 0.3] & [-1.3, 1.6]~\cite{Hartland:2019bjb} \\
  \hline
  \end{tabular}
  \caption{Constraints on the Wilson coefficients (first column) at 68\% and 95\% confidence level, as obtained from a fit with all coefficients (second and third column) and from fits with each coefficient individually (fourth and fifth column). 
  All results are given in units of $\text{TeV}^{-2}$, assuming $\Lambda = 1$ TeV. 
  The last column lists the most stringent bounds to date, found by global analyses of the entire top quark sector.
  }
  \label{tab:results:bounds}
\end{table}

\subsubsection{Correlations}

To quantify the correlations between two Wilson coefficients $c_i$ and $c_j$, the correlation coefficient
\begin{equation}
\rho\ (c_i, c_j) = \frac{ \frac{1}{N_\text{toy}} \sum_{k=1}^{N_\text{rep}} c_i^{(k)} c_j^{(k)} - \langle c_i \rangle \langle c_j \rangle } { \delta c_i\ \delta c_j } 
\label{eq:results:corr}
\end{equation}
is used, where $c_i^{(k)},\ c_j^{(k)}$ are the $k$-th MC toys, $\langle c_i \rangle,\ \langle c_j \rangle$ are the means and $\delta c_i,\ \delta c_j$ the standard deviations of the distributions of the coefficients.
If the two coefficients were totally uncorrelated, one would expect $\rho\ (c_i, c_j) = 0$.
If $i=j$, then the two coefficients are fully correlated by construction, i.e. $\rho\ (c_i, c_i) = 1$.
For $i\neq j$, $\rho$ can reach values from -1 (anti-correlated) to 1 (fully correlated).

\begin{figure}[t!]
\centering
  \includegraphics[width=0.75\textwidth]{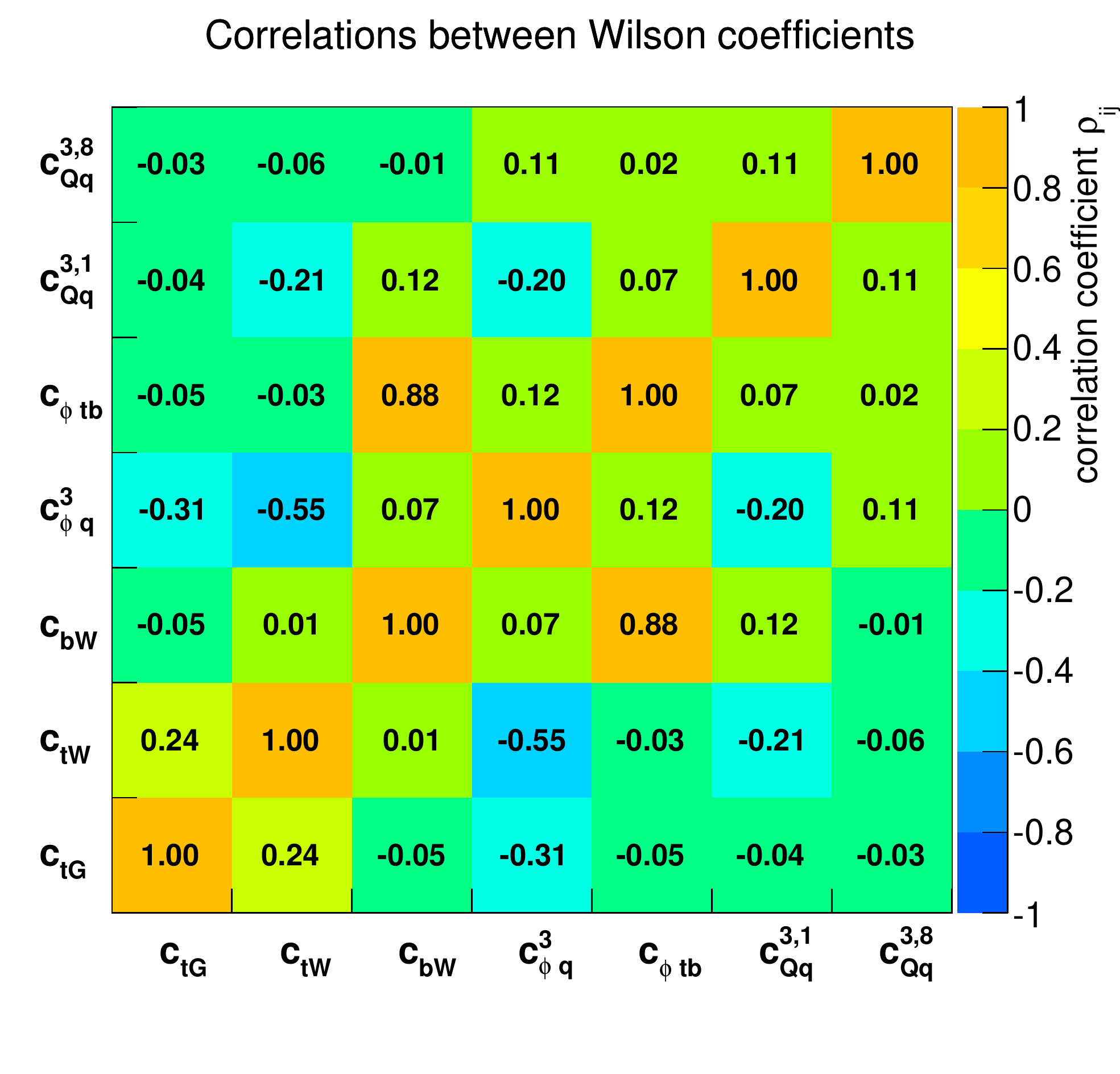}
  \vspace{-8mm}
  \caption{Correlation matrix of the Wilson coefficients. 
  Orange scales indicate that two coefficients are highly correlated, green scales that they are uncorrelated, and blue that they are anticorrelated. 
  }
  \label{img:results:Scorr}
\end{figure}

\begin{figure}[h]
\centering
\begin{subfigure}{0.32\textwidth}
  \includegraphics[width=\textwidth]{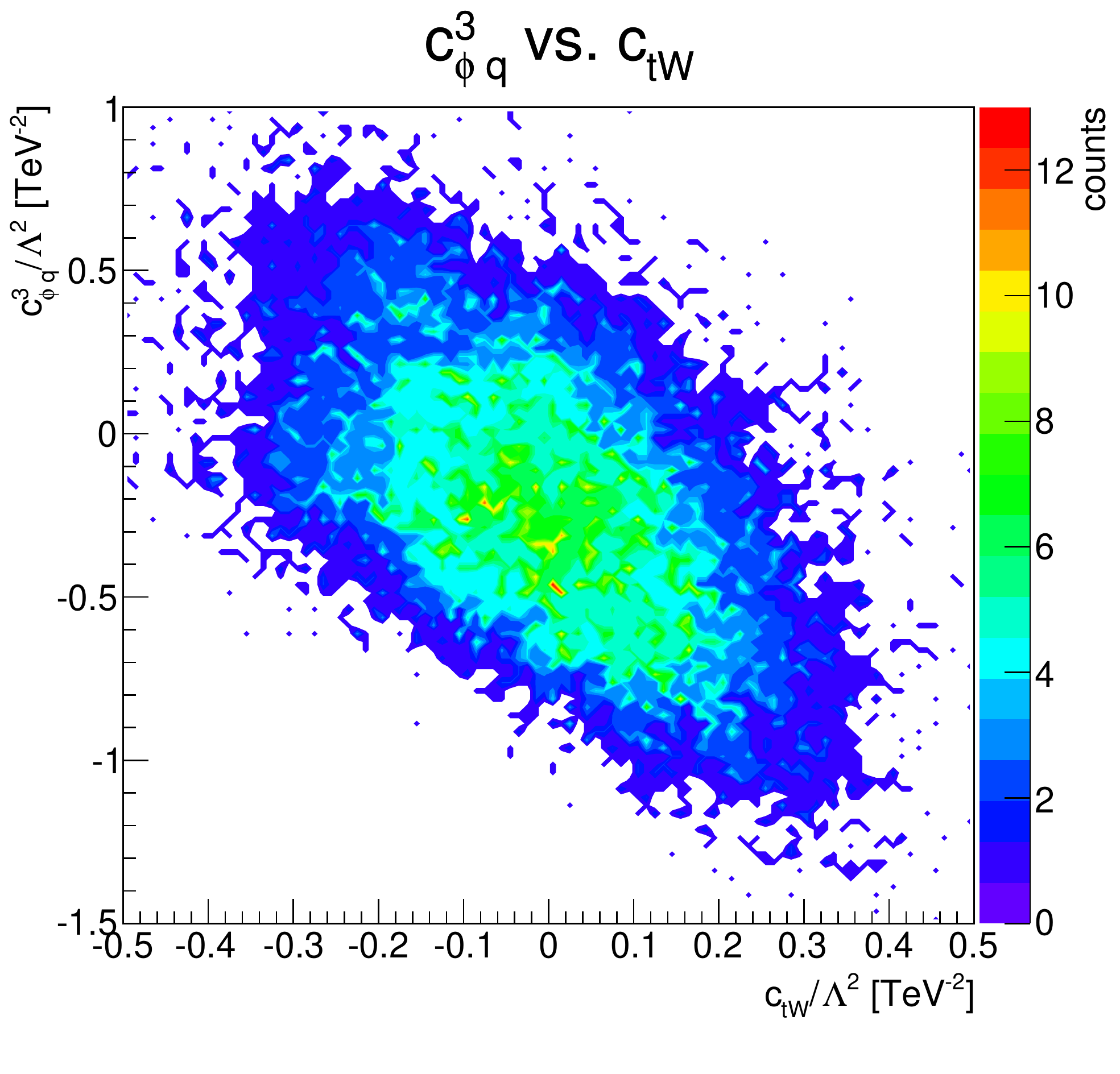}
  \caption{}
\end{subfigure}%
\hspace{1mm}%
\begin{subfigure}{0.32\textwidth}
  \includegraphics[width=\textwidth]{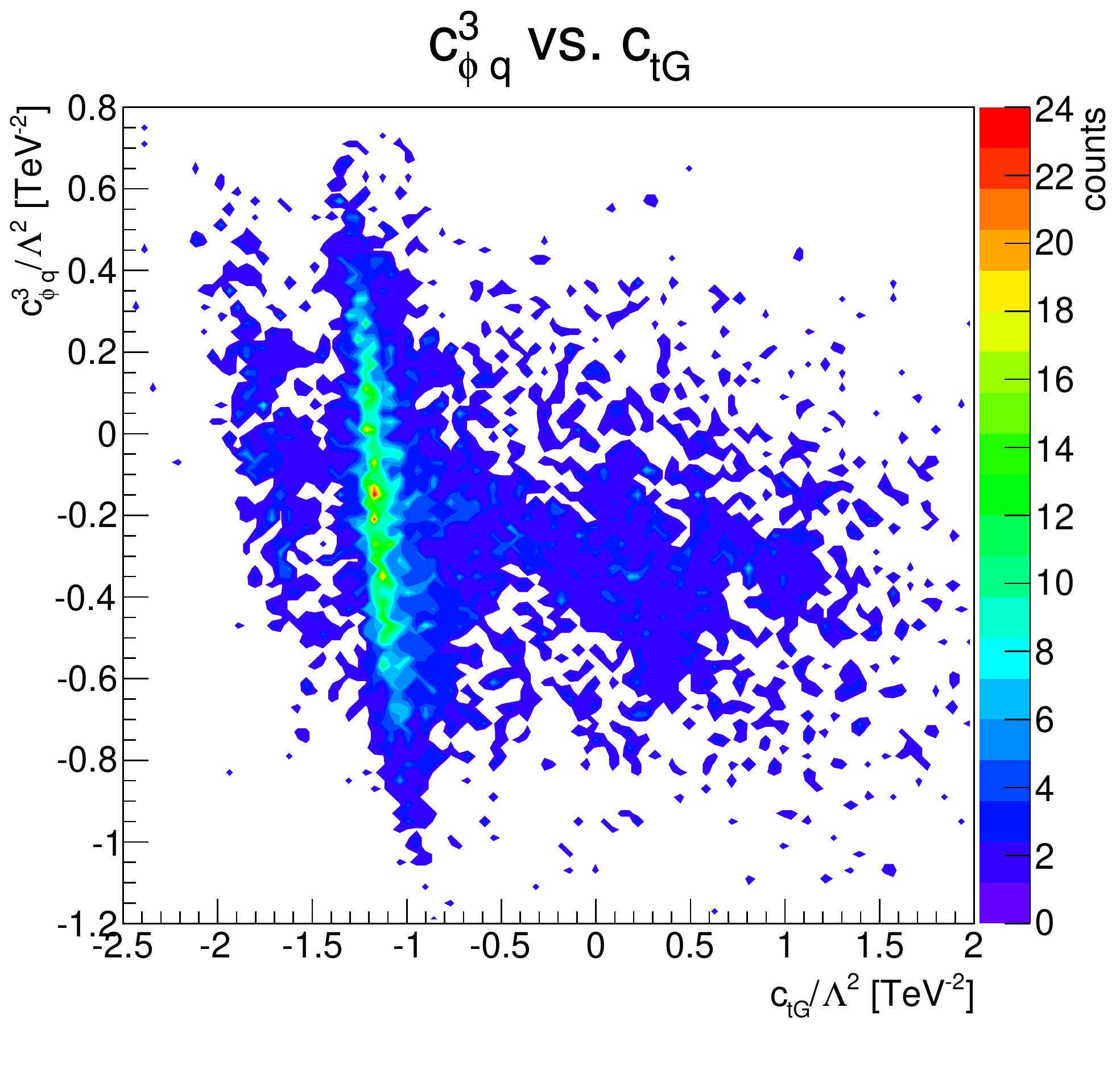}
  \caption{}
\end{subfigure}%
\hspace{1mm}%
\begin{subfigure}{0.32\textwidth}
  \includegraphics[width=\textwidth]{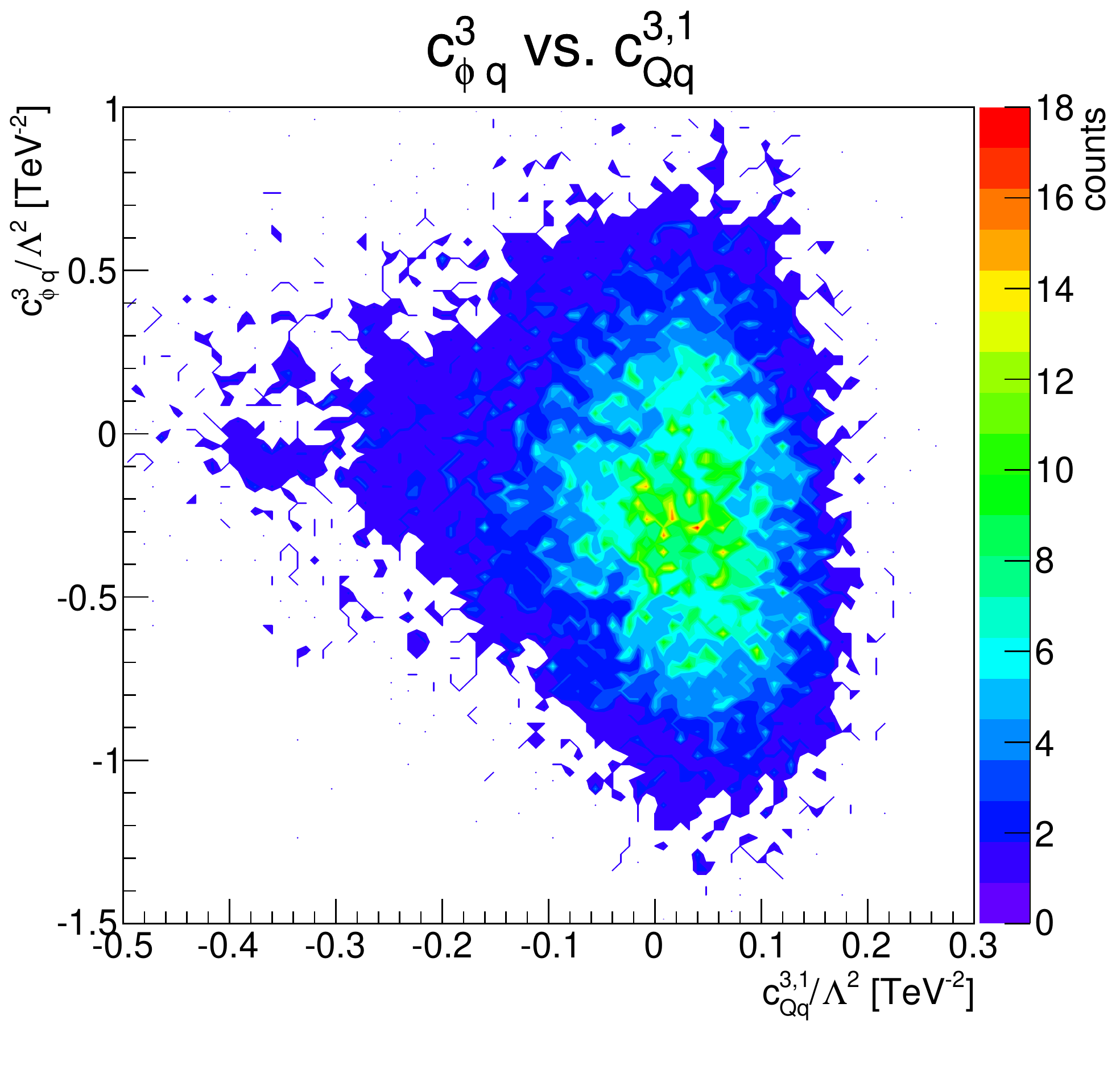}
  \caption{}
\end{subfigure}
\caption{ Two-dimensional distributions of various highly (anti-)correlated fit Wilson coefficients against the number of counts. Panel (a): $c_{\varphi q}^3$ versus $c_{tW}$; (b): $c_{\varphi q}^3$ versus $c_{tG}$; (c): $c_{\varphi q}^3$ versus $c_{Qq}^{3,1}$. Details see text.  }
\label{img:results:Scorr2d}
\end{figure}

Figure~\ref{img:results:Scorr} shows the correlation matrix, which has been extracted from the fit including all coefficients. 
One can see that most coefficients are practically uncorrelated.
The most prominent exceptions are $c_{\varphi tg}$ and $c_{bW}$, which are highly correlated, and $c_{\varphi q}^3$ and $c_{tW}$, which have a sizable anti-correlation.

Figure~\ref{img:results:Scorr2d} shows two-dimensional distributions of various Wilson coefficients.
The $z$-axis counts the number of times that a fit point was obtained, analogously to the one-dimensional distributions previously discussed.
One can clearly see the anti-correlation between $c_{\varphi q}^3$ and $c_{tW}$ from squeezed elliptical shape in panel (a). 
Letting aside the the low counts in the range where $c_{tG}\gtrsim -0.5$, one can also see a squeezed elliptical shape, showing the anti-correlation.
This elliptical shape arises from the peak in the distribution of $c_{tG}$, as can be seen in figure~\ref{img:results:Stg}.
Where $c_{tG}$ has higher values, the two coefficients are rather uncorrelated, which in the relatively small correlation coefficient $\rho_{ij}=-0.31$.
Finally, $c_{\varphi q}^3$ and $c_{Qq}^{3,1}$ have a relatively small correlation coefficient of -0.20.
Therefore, one cannot see a strong correlation between the two, see panel (c), as the shape of the distribution resembles an ellipse that is not squeezed onto a diagonal like in the other two panels.


\subsection{Variations of the dataset}
\label{sec:results:dataset}

\subsubsection{Kinematic distributions}
\label{subsubsec:results:Snd}

As the kinematic distributions have only recently become available, the question is how much they change the outcome of the fit.
In the following, the results from subsection~\ref{sec:results:S} are compared with those where the $t$-channel distributions are replaced by the measurements of the $t$-channel inclusive cross sections.
These measurements are listed in table~\ref{tab:data:measurements}.

\begin{figure}[h]
\vspace*{5mm}
\centering
  \includegraphics[width=0.83\textwidth]{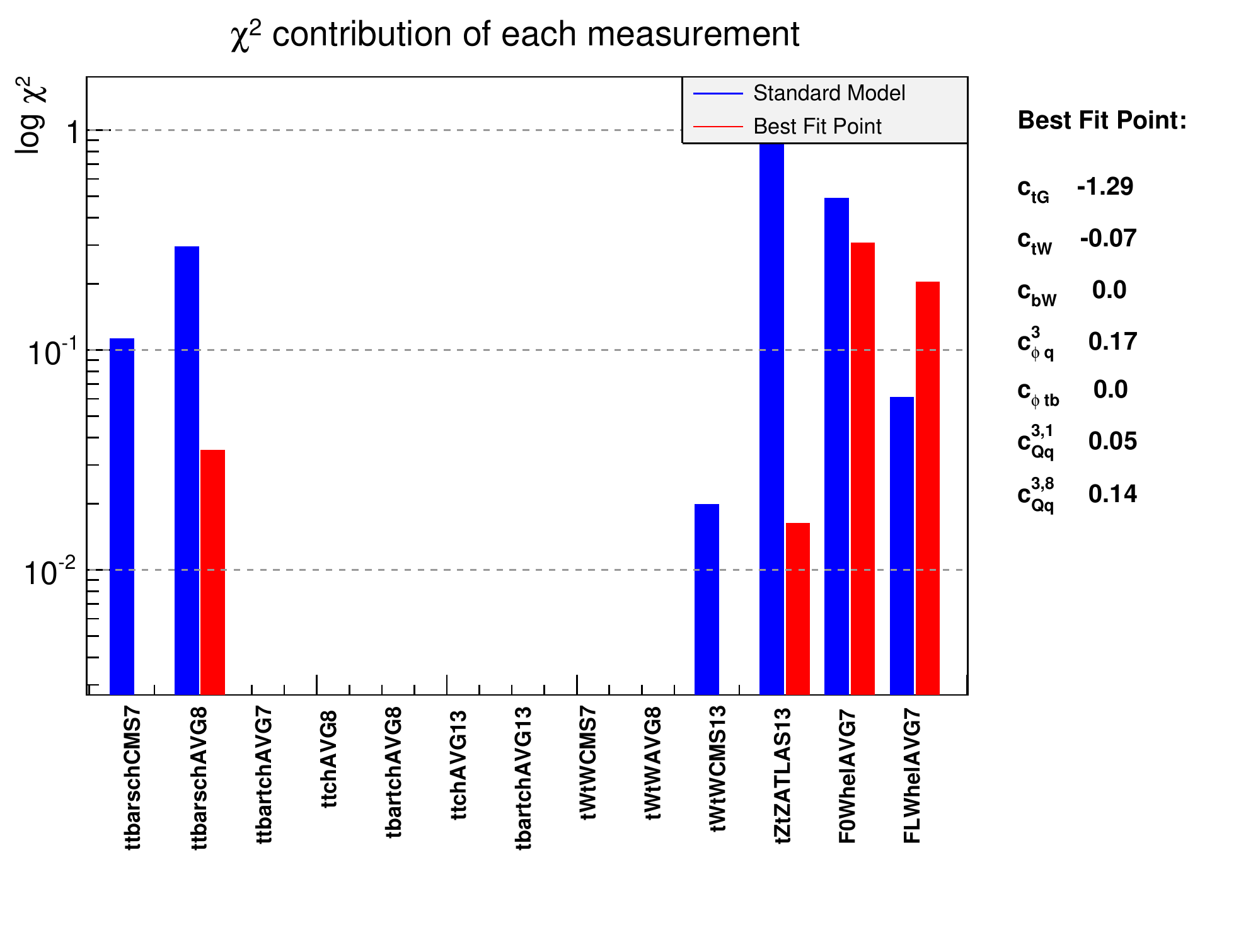}
  \vspace{-8mm}
  \caption{Contribution of each measurement to the log-likelihood, where the name of each measurement is stated in \textsc{SFitter} code, analogous to figure~\ref{img:results:Schi2}. 
 }
  \label{img:results:SndChi2}
\end{figure}

Similarly to the previous subsection, figure~\ref{img:results:SndChi2} shows the contribution of each measurement to the log-likelihood at the SM and at the best fit point.
The SM values apart from the $t$-channel are of course identical with those shown in figure~\ref{img:results:Schi2}.
One can also see that, contrarily to the bins of the $t$-channel distributions, the corresponding inclusive cross sections do not contribute anything at all.

This leads to different contributions to the log-likelihood at the best fit point than those in figure~\ref{img:results:Schi2}. 
The contributions of the measurements of helicity fractions at the best fit point are practically the same with and without $t$-channel distributions.
In contrast, the contributions of measurements of $s$-channel production and of production in association with a $Z$ boson are drastically reduced when no $t$-channel-distributions are included.

\begin{table}[h!]
  \centering
  \vspace{10mm}
  \begin{tabular}{|llllll|}
  \hline
  coeff. & 68\% all (standard) & 68\% all & 95\% all & 68\% indiv. & 95\% indiv.  \\
  \hline
  $c_{tG}$ & [-2.1, 0.3] & [-2.4, 0.0] & [-4.0, 1.5] & [-2.5, 0.3] & [-4.1, 1.6] \\
  $c_{tW}$ & [-0.2, 0.2] & [-0.2, 0.1] & [-0.4, 0.3] & [-0.2, 0.3] & [-0.4, 0.4] \\
  $c_{bW}$ & [0.0, 1.1] & [0.0, 1.4] & [-0.1, 2.1] & [0.0, 1.5] & [-0.7, 2.1] \\
  $c_{\varphi Q}^3$ & [-0.7, 0.2] & [-0.5, 0.2] & [-0.8, 0.6] & [-0.6, 0.1] & [-0.7, 0.5] \\ 
  $c_{\varphi tb}$ & [0.1, 0.4] & [0.1, 0.4] & [-0.1, 0.5] & [0.1, 0.4] & [0.0, 0.5] \\
  $c_{Qq}^{3,1}$ & [-0.1, 0.1] & [-0.1, 0.1] & [-0.1, 0.1] & [0.0, 0.1] & [-0.1, 0.2] \\
  $c_{Qq}^{3,8}$ & [-0.2, 0.2] & [-0.4, -0.1] & [-0.5, 0.2] & [-0.4, 0.1] & [-0.5, 0.2] \\
  \hline
  \end{tabular}
  \caption{Constraints on the Wilson coefficients (first column) at 68\% confidence level, as obtained from a fit at standard settings with all coefficients (second column).
  For the constraints at 95\% confidence level and the comparison with the literature, see table~\ref{tab:results:bounds}.
  In addition, the constraints where kinematic distributions are omitted are listed, as obtained from a fit with all coefficients (third and fourth column) and from fits with each coefficient individually (last two columns). 
  All results are given in units of $\text{TeV}^{-2}$, assuming $\Lambda = 1$ TeV. 
  }
  \label{tab:results:SndBounds}
\end{table}

\begin{figure}[h!]
  \vspace{-3mm}
\centering
  \includegraphics[width=0.8\textwidth]{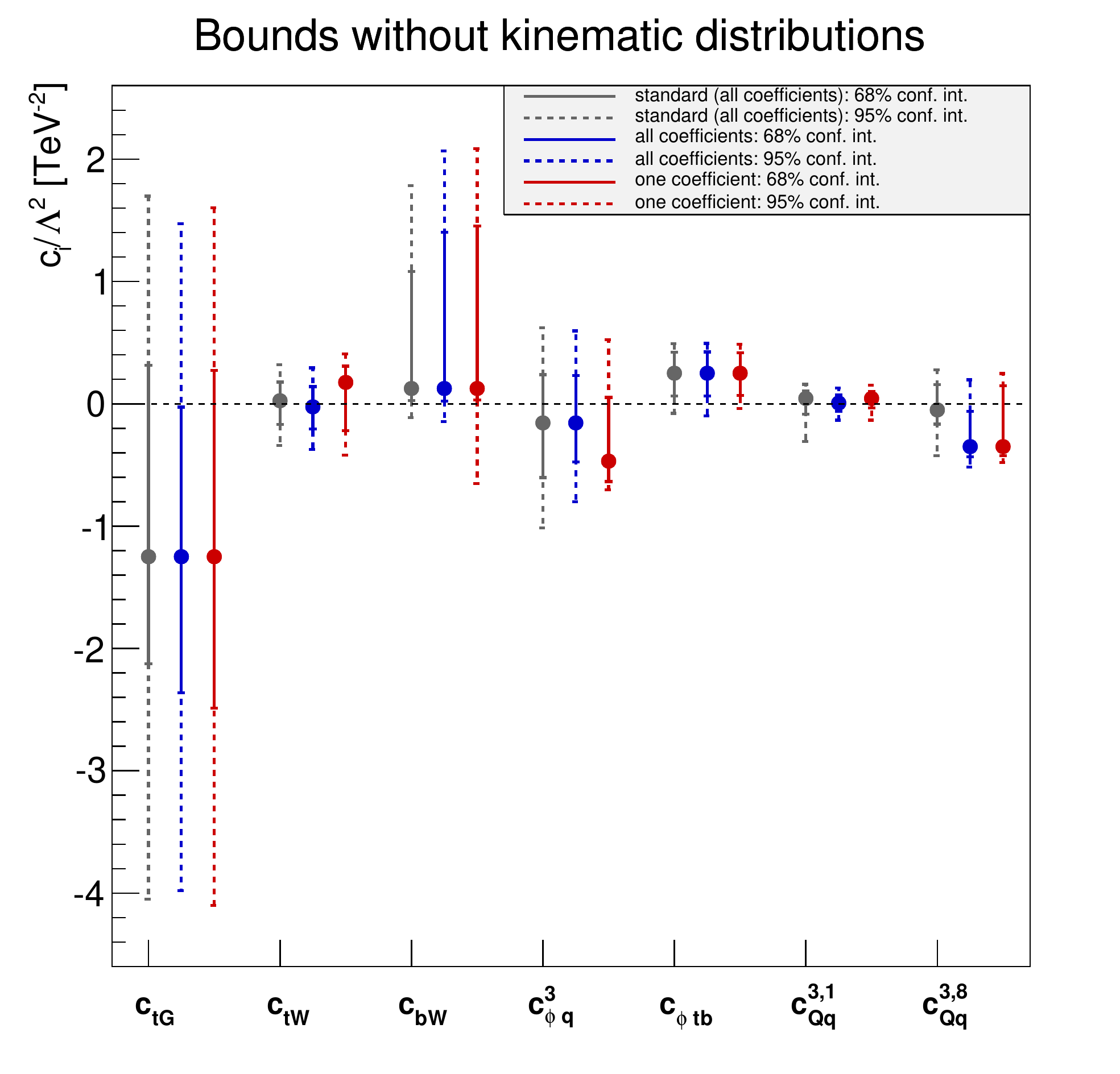}
  \vspace{-3mm}
  \caption{Visualization of the constraints on the Wilson coefficients of table~\ref{tab:results:SndBounds}. Solid lines mark the 68\%-, dashed lines the 95\% confidence intervals. 
  The blobs mark the most probable value of each coefficient.
  The gray lines indicate the constraints from the standard dataset and theory settings.
  The blue lines indicate the constraints obtained from an fit with all coefficients, the red lines those obtained from fits with the individual coefficients, where measurements of $t$-channel distributions have been replaced by the corresponding inclusive cross sections.
  The dashed line at $c_i/\Lambda^2=0$ marks the SM expectation.
  }
  \label{img:results:SndBounds}
\end{figure}

The fact that the contributions of the measurements to the log-likelihood are different to those from the previous subsection undermines the fact that the best fit point is different to the one before.
As a consequence of interference terms between $c_{tG}$ and other Wilson coefficients, the coefficient $c_{tG}$ changes from -0.53 in the previous section to -1.29 here.
The values of other coefficients at the best fit point also change, for example $c_{\varphi q}^3$ and $c_{Qq}^{3,1}$, which are particularly sensitive to $t$-channel production, see section~\ref{subsubsec:data:ts}.

Table~\ref{tab:results:SndBounds} and figure~\ref{img:results:SndBounds} show the constraints derived from fits on the dataset without kinematic distributions.
As before, one can see that all constraints are in excellent agreement with the SM expectation.
The constraints from individual fits also do not change very much in comparison those from a fit with all Wilson coefficients.

One can see that the constraints are not affected much by the omission of kinematic distributions.
In the case of $c_{tW}$ and $c_{\varphi q}^3$, they are slightly less stringent at 68\% confidence level than those where kinematic distributions are included.
This is to be expected as both coefficients are rather sensitive to $t$-channel production.
The constraint on $c_{bW}$ is slightly more stringent at 68\% when kinematic distributions are included.
The other constraints are slightly shifted or remain the same with or without the distributions.
All these changes are not large, however.

From this, one can conclude that kinematic distributions do not have a large impact on the outcome of the fit.
This cannot be generalized to other datasets, for example if one would include measurements of top quark pair production.
However, it shows that in the current dataset, there is no evidence that the impact of operators that grow with the center-of-mass energy of a process is more visible through kinematic distributions.


\subsubsection{Measurements at 7 TeV} 
\label{subsubsec:results:Sn7}

\begin{figure}[h!]
\vspace{10mm}
\centering
  \includegraphics[width=0.83\textwidth]{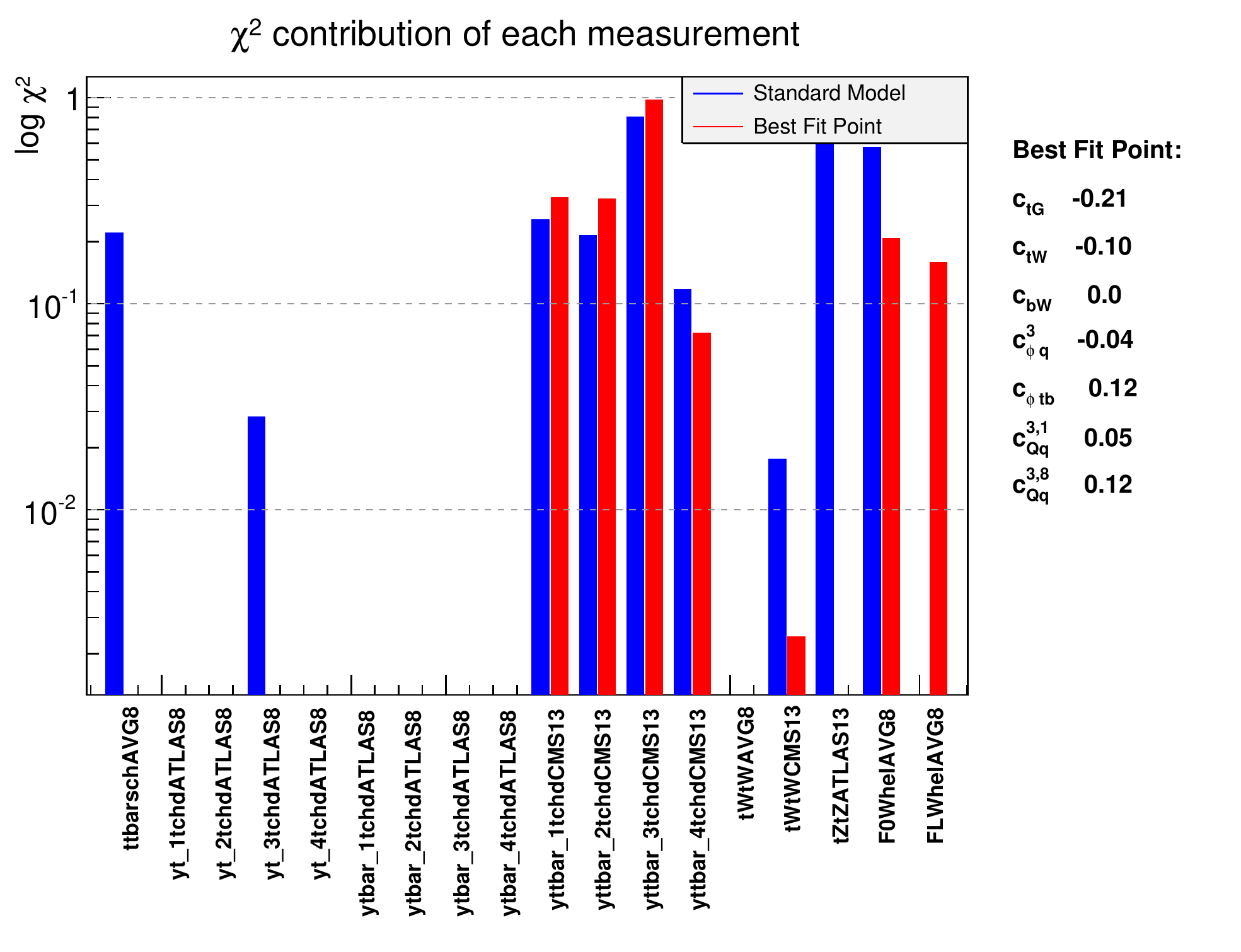}
  \caption{Contribution of each measurement to the log-likelihood, where the name of each measurement is stated in \textsc{SFitter} code, analogous to figure~\ref{img:results:Schi2}. 
 }
  \label{img:results:Sn7Chi2}
\end{figure}
\vspace{15mm}

\begin{table}[h]
  \centering
  \vspace*{6mm}
  \begin{tabular}{|llllll|}
  \hline
  coeff. & 68\% all (standard) & 68\% all & 95\% all & 68\% indiv. & 95\% indiv.  \\
  \hline
  $c_{tG}$ & [-2.1, 0.3] & [-2.2, 0.7] & [-4.2, 2.2] & [-2.7, 0.7] & [-4.4, 2.0] \\
  $c_{tW}$ & [-0.2, 0.2] & [-0.2, 0.1] & [-0.4, 0.2] & [-0.3, 0.1] & [-0.4, 0.3] \\
  $c_{bW}$ & [0.0, 1.1] & [0.0, 1.2] & [-0.7, 2.0] & [-1.3, 0.0] & [-2.2, 1.4] \\
  $c_{\varphi Q}^3$ & [-0.7, 0.2] & [-0.6, 0.3] & [-1.1, 0.8] & [-0.6, 0.1] & [-0.7, 0.6] \\ 
  $c_{\varphi tb}$ & [0.1, 0.4] & [0.1, 0.4] & [-0.2, 0.7] & [0.1, 0.4] & [0.0, 0.5] \\
  $c_{Qq}^{3,1}$ & [-0.1, 0.1] & [-0.1, 0.2] & [-0.4, 0.2] & [0.0, 0.2] & [-0.3, 0.2] \\
  $c_{Qq}^{3,8}$ & [-0.2, 0.2] & [-0.2, 0.1] & [-0.4, 0.3] & [-0.4, 0.1] & [-0.5, 0.2] \\
  \hline
  \end{tabular}
  \caption{Constraints on the Wilson coefficients, analogous to table~\ref{tab:results:SndBounds},
  where now the constraints are listed where measurements at 7 TeV are omitted. 
  }
  \label{tab:results:Sn7Bounds}
\end{table}

\begin{figure}[t!]
\vspace{10mm}
\centering
  \includegraphics[width=0.8\textwidth]{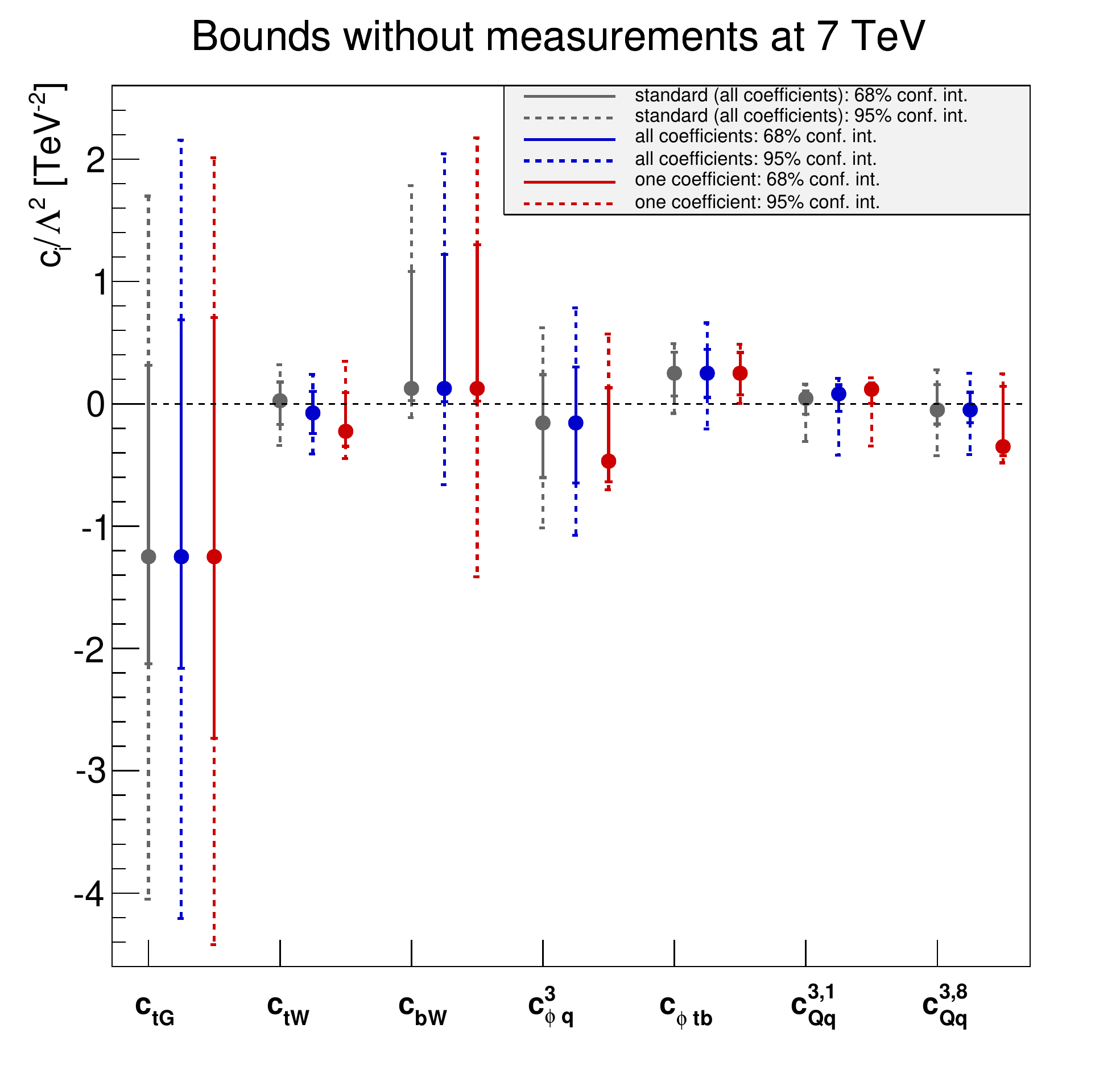}
  \caption{Visualization of the constraints on the Wilson coefficients of table~\ref{tab:results:Sn7bounds}, analogous to figure~\ref{img:results:SndBounds}.
  }
  \label{img:results:Sn7Bounds}
\end{figure}

In this study, the measurements concerning a single top quark at a center-of-mass energy of 7 TeV have been included.
In this section, the impact of these measurements on the constraints on the Wilson coefficients is investigated. 
The constraints of subsection~\ref{sec:results:S} are compared with those arising from fits without measurements at 7 TeV.

Figure~\ref{img:results:Sn7Chi2} shows the contributions of each measurement to the log-likelihood.
The contributions of the helicity fractions at SM differ from those in figure~\ref{img:results:Schi2} because of the measurements at 7 TeV are no more included.
While contribution of the fraction $F_0$ to the log-likelihood remains similar, that of $F_L$ drops to zero when only measurements at 8 TeV are included.

The contributions to the log-likelihood at the best fit point are practically zero for the measurements of the $s$-channel cross section and the $t$-channel distributions at 8 TeV.
The contributions of the first three bins of the distribution at 13 TeV and the helicity fraction $F_0$ are slightly bigger than at the SM values, while all others are smaller.
Most notably, the contribution of the single top quark production in association with a $W$ and $Z$ boson at 13 TeV are a lot smaller at the best fit point than at the SM values.

The smaller contribution to the log-likelihood of the aforementioned measurements can be explained through the change of the values of $c_{tW}$, $c_{\varphi q}^3$, $c_{\varphi tb}$ and the coefficients from the four-quark operators in comparison to the findings of subsection~\ref{sec:results:S}.
While the value of $c_{tG}$ at the best fit point changes from -0.53 to -0.21, one can see practically no change in the contributions of the measurements of the helicity fractions to the log-likelihood.
This once more shows the low sensitivity of this coefficient to the measurements of helicity fractions.

The constraints on the Wilson coefficients without measurements at 7 TeV are shown in table~\ref{tab:results:Sn7Bounds} and figure~\ref{img:results:Sn7Bounds}.
One can see that they become slightly less stringent when measurements at 7 TeV are not included.
This shows that even though the effect is pretty small, these measurements are not entirely superseded by those at 8 and 13 TeV.

\clearpage


\subsection{Variations of theory settings}
\label{sec:results:theory}

\subsubsection{NLO effects}
\label{subsubsec:results:SLO}

Taking into account the accuracy of the included measurements, it is important to use predictions at NLO QCD level, see subsection~\ref{subsec:SMEFT:NLO}.
The impact of NLO effects is studied by comparing the findings in subsection~\ref{sec:results:S} with those obtained from fits where only predictions at LO are used.

\begin{figure}[b!]
\vspace{5mm}
\centering
  \includegraphics[width=0.83\textwidth]{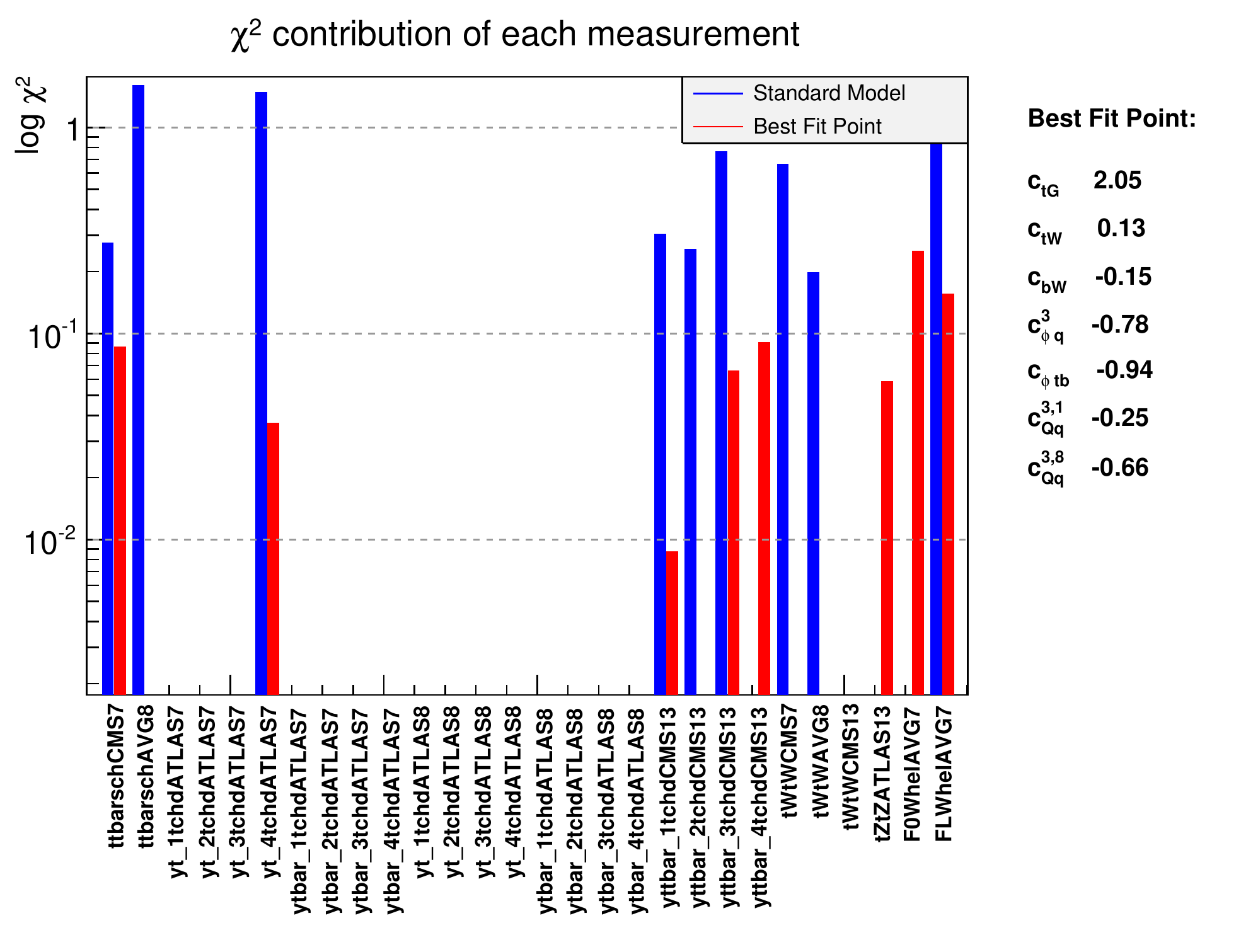}
  \vspace{-2mm}
  \caption{Contribution of each measurement to the log-likelihood, analogous to figure~\ref{img:results:Schi2}. 
}
  \label{img:results:SLOchi2}
\end{figure}

\begin{table}[h]
  \centering
  \vspace{6mm}
  \begin{tabular}{|llllll|}
  \hline
  coeff. & 68\% all (standard) & 68\% all & 95\% all & 68\% indiv. & 95\% indiv.  \\
  \hline
  $c_{tG}$ & [-2.1, 0.3] & [0.7, 3.8] & [-4.5, 4.6] & [0.8, 3.6] & [-3.3, 4.4] \\
  $c_{tW}$ & [-0.2, 0.2] & [0.0, 0.4] & [-0.1, 0.5] & [0.0, 0.3] & [-0.2, 0.4] \\
  $c_{bW}$ & [0.0, 1.1] & [-1.1, 0.3] & [-2.0, 1.6] & [-1.6, 2.3] & [-2.8, 3.0] \\
  $c_{\varphi Q}^3$ & [-0.7, 0.2] & [-0.7, 0.9] & [-1.7, 1.6] & [0.1, 1.5] & [-0.5, 2.1] \\ 
  $c_{\varphi tb}$ & [0.1, 0.4] & [-0.4, 3.8] & [-4.8, 6.4] & [0.1, 4.1] & [-2.4, 5.7] \\
  $c_{Qq}^{3,1}$ & [-0.1, 0.1] & [-0.2, 0.1] & [-0.4, 0.2] & [-0.2, 0.1] & [-0.4, 0.2] \\
  $c_{Qq}^{3,8}$ & [-0.2, 0.2] & [-0.4, 0.7] & [-0.8, 0.9] & [0.0, 0.7] & [-0.1, 0.9] \\
  \hline
  \end{tabular}
  \caption{Constraints on the Wilson coefficients, analogous to table~\ref{tab:results:SndBounds},
  where now the constraints are listed where only predictions at LO are used.
  }
  \label{tab:results:SLObounds}
\end{table}

\begin{figure}[t!]
\vspace{8mm}
\centering
  \includegraphics[width=0.8\textwidth]{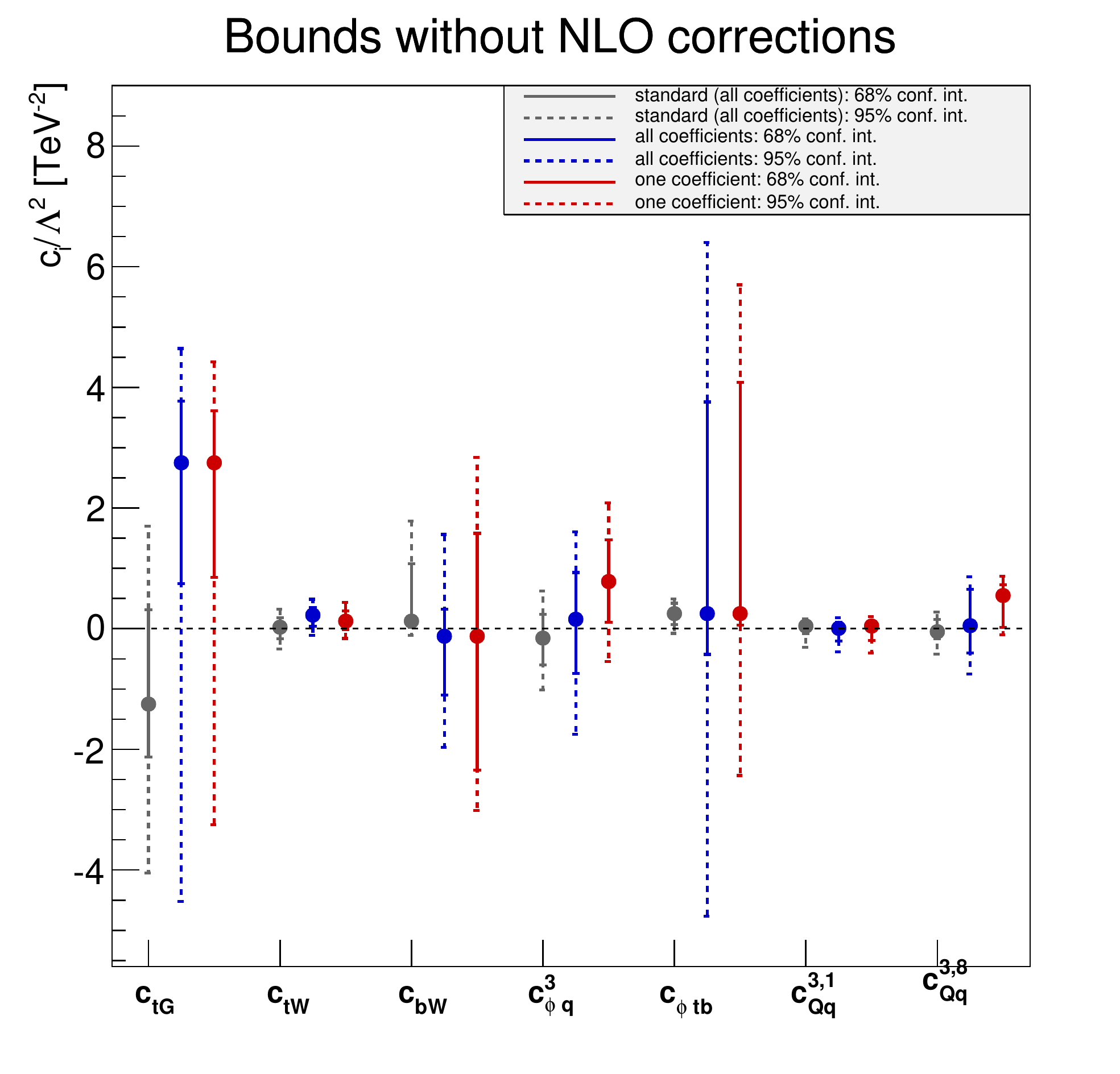}
  \caption{Visualization of the constraints on the Wilson coefficients of table~\ref{tab:results:SLObounds}, analogous to figure~\ref{img:results:SndBounds}.
  }
  \label{img:results:SLObounds}
\end{figure}

Figure~\ref{img:results:SLOchi2} shows the contributions of the measurements to the log-likelihood at SM values and the best fit point.
Comparing this with figure~\ref{img:results:Schi2}, one can immediately see that most contributions at SM values are rather different at LO due to the altered predictions and theoretical uncertainties.

The contributions of the $s$-channel measurements are roughly a factor two to three higher at LO, while those of $t$-channel production slightly decrease in most cases. 
The measurements of single top quark production in association with a $W$ quark at 7 and 8 TeV have nonzero contributions, while those of the associated production processes at 13 TeV have none.
This is the opposite of the findings from section~\ref{sec:results:S}.
Finally, the measurement of the helicity fraction $F_0$ does not contribute to the overall log-likelihood, while that of $F_L$ is about ten times bigger than previously found.

With such a difference in the initial setting, it is not surprising that the best fit point and the contributions of the measurements at this point are very different from the previous findings.
This is reflected in the constraints, as can be seen in table~\ref{tab:results:SLObounds} and figure~\ref{img:results:SLObounds}.

Almost all constraints are a lot less stringent without NLO corrections.
The only exception are the constraints on $c_{tW}$, which appear to be slightly shifted in comparison to those from standard settings.
One can also see that there are stronger differences than in figure~\ref{img:results:Sbounds} between the constraints obtained from a fit with all coefficients and those from individual fits.
This indicates that the terms where two different operators occur play a larger role in this fit.

Even though all constraints agree with the SM expectation and with one one another at 95\% confidence level, the difference between using NLO QCD corrections or not is arguably profound.
This shows that using predictions at NLO level is in principle indispensable.


\subsubsection{Order $\mathcal{O}(\Lambda^{-4})$ effects}
\label{subsubsec:results:SL2}

In this study, order $\mathcal{O}(\Lambda^{-4})$ effects are included in the fits.
The reasons for this have been explained in section~\ref{subsec:SMEFT:intro}.
The impact of these effects are studied by comparing the results from fits that only include terms up to order $\mathcal{O}(\Lambda^{-2})$ with the findings of section~\ref{sec:results:S}.

\begin{table}[b!]
  \centering
  \vspace*{12mm}
  \begin{tabular}{|llllll|}
  \hline
  coeff. & 68\% all (standard) & 68\% all & 95\% all & 68\% indiv. & 95\% indiv.  \\
  \hline
  $c_{tG}$ & [-2.1, 0.3] & [-3.2, 1.0] & [-5.6, 3.1] & [-2.8, 0.9] & [-4.9, 2.9] \\
  $c_{tW}$ & [-0.2, 0.2] & [-0.2, 0.1] & [-0.4, 0.3] & [-0.1, 0.2] & [-0.3, 0.3] \\
  $c_{\varphi Q}^3$ & [-0.7, 0.2] & [-0.8, 0.4] & [-1.4, 0.9] & [-0.9, 0.1] & [-1.4, 0.6] \\ 
  $c_{Qq}^{3,1}$ & [-0.1, 0.1] & [-0.1, 0.1] & [-0.2, 0.2] & [0.0, 0.1] & [-0.1, 0.2] \\
  $c_{Qq}^{3,8}$ & [-0.2, 0.2] & [0.1, 0.5] & [-0.1, 0.7] & [0.1, 0.5] & [-0.1, 0.7] \\
  \hline
  \end{tabular}
  \caption{Constraints on the Wilson coefficients, analogous to table~\ref{tab:results:SndBounds},
  where now the constraints are listed where only effects up to order $\mathcal{O}(\Lambda^{-2})$ are included.
  }
  \label{tab:results:SL2bounds}
\end{table}

\begin{figure}[h]
\vspace{10mm}
\centering
  \includegraphics[width=0.83\textwidth]{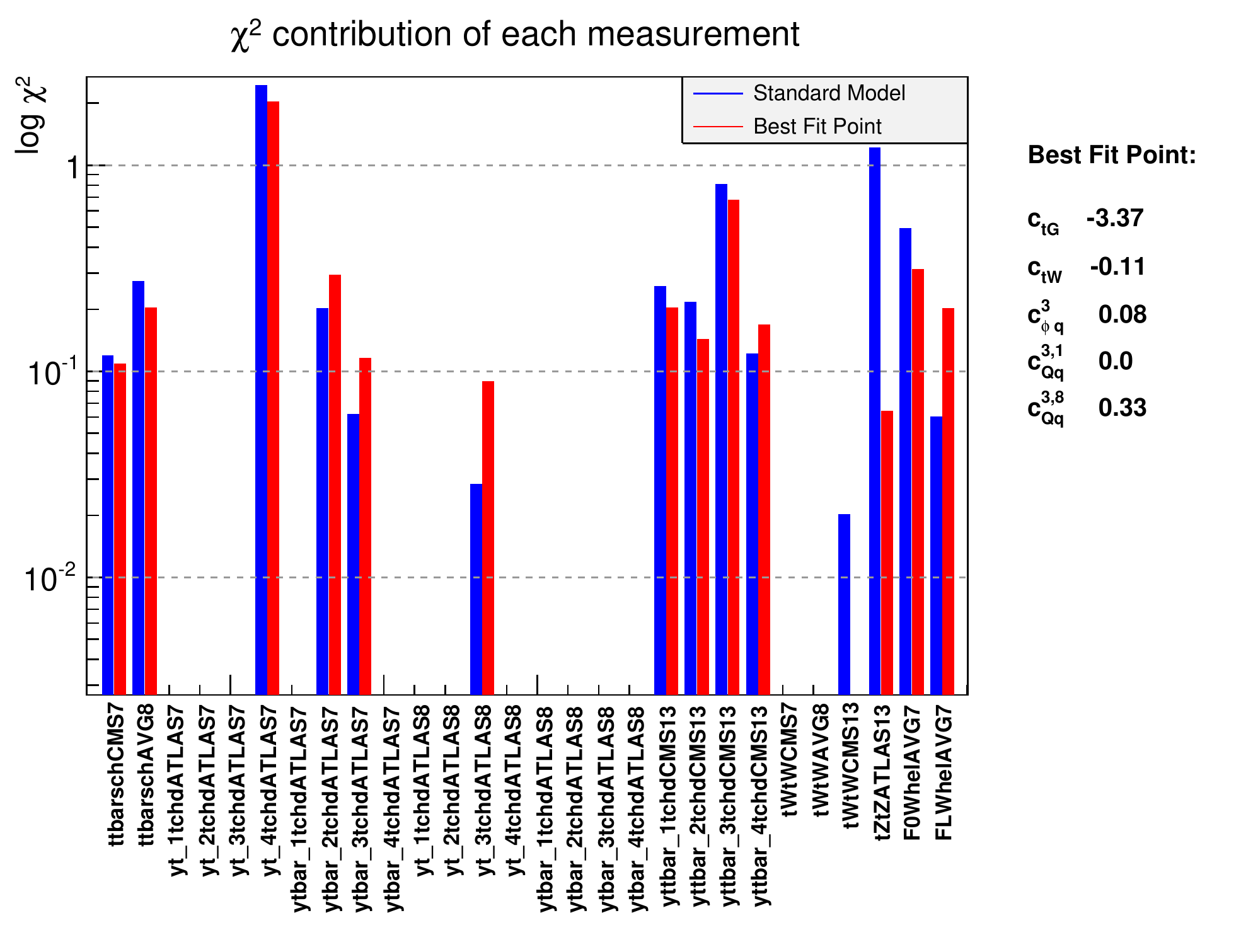}
  \vspace{-2mm}
  \caption{Contribution of each measurement to the log-likelihood, analogous to figure~\ref{img:results:Schi2}.  }
  \label{img:results:SL2chi2}
\end{figure}

\begin{figure}[h]
\vspace{10mm}
\centering
  \includegraphics[width=0.8\textwidth]{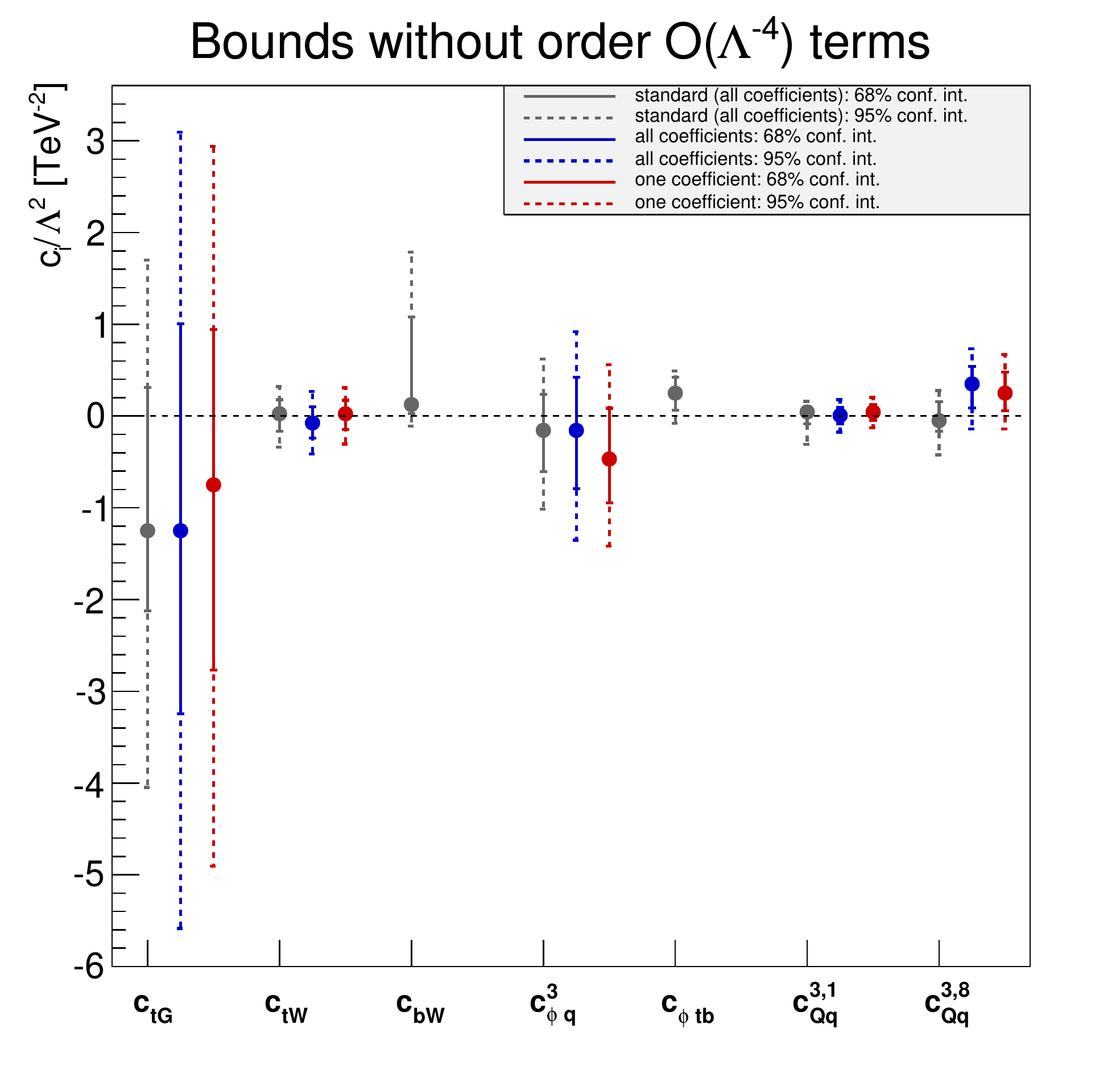}
  \caption{Visualization of the constraints on the Wilson coefficients of table~\ref{tab:results:SL2bounds}, analogous to figure~\ref{img:results:SndBounds}.
  }
  \label{img:results:SL2bounds}
\end{figure}

As can be seen in figures~\ref{img:results:Schi2} and~\ref{img:results:SL2chi2}, the contributions of the measurements to the log-likelihood are very similar with or without $\mathcal{O}(\Lambda^{-4})$.
The best fit points of the fits with and without quadratic terms differ where $c_{Qq}^{3,8}$ and $c_{tG}$ are concerned, but this does not affect the log-likelihood at this point very much.
The fact that $c_{bW}$ and $c_{\varphi tb}$ cannot be constrained at order $\mathcal{O}(\Lambda^{-2})$ does not change anything here because in the best fit point in the higher-order fit, these coefficients are zero.

From table~\ref{tab:results:SL2bounds} and figure~\ref{img:results:SL2bounds} one can see that most bounds are slightly less stringent but still similar when quadratic terms are not included.
The only exception is $c_{Qq}^{3,8}$, which has relatively large quadratic terms and small linear terms.
In summary, one can see that including terms up to order $\mathcal{O}(\Lambda^{-4})$ does lead to more stringent constraints, more reliable constraints in the case of $c_{Qq}^{3,8}$, and makes it possible to constrain Wilson coefficients like $c_{bW}$ and $c_{\varphi tb}$.

\clearpage



\section{Conclusion and Outlook}
\label{sec_conclusion}

In this thesis, a global SMEFT fit has been performed with \textsc{SFitter}, using measurements of the production and decay of a single top quark.
The fit constrains seven Wilson coefficients from dimension-six operators that involve a top quark and are sensitive to the processes in question.
The Wilson coefficients $c_{tZ}$, $c_{\varphi Q}^-$ and $c_{\varphi t}$ are not included in the fit.
These only have a slight sensitivity to the production of a top quark in association with a $Z$ boson, to which only one measurement is available.

The fit includes some methodological improvements in comparison to other studies.
These have been implemented in \textsc{DataPrep}, a software developed by the author.
\textsc{DataPrep} can handle large numbers of correlated and uncorrelated systematic uncertainties in an efficient and user-friendly way.
In addition, it takes correlations between theoretical uncertainties into account by averaging measurements of identical processes and observables at the same center-of-mass energy.

With the improvements on the methodology, the constraints obtained in this study represent a major improvement of those in the literature.
This is quantified by the difference between the upper and the lower constraint on a Wilson coefficient.
The smaller the value, the more stringent a constraint is called.
Even though constraints in the literature concern the entire top quark sector, almost all constraints in this study are three to five times more stringent at 95\% confidence level.
Only the constraints of $c_{tG}$ are six times more stringent in the literature.
In the case of $c_{\varphi tb}$, the improvement in comparison to the literature is even more than 50-fold.
All constraints are in agreement with the SM expectation at 95\% confidence level.

When the measurements of kinematic distributions of the $t$-channel single top quark production are replaced by the corresponding measurements of inclusive cross sections, the constraints on the Wilson coefficients do not change much.
This shows that where single top quark production and decay up to a center-of-mass energy of 13 TeV are concerned, kinematic distributions do not show any more evidence for SMEFT effects than inclusive cross sections.

Measurements at a center-of-mass energy of 7 TeV have been included as this does make the constraints slightly more stringent than without them.
As the effect is small, it might be compensated for when measurements of other processes in the top quark sector, for example top quark pair production, are included in the fit.

All theoretical predictions have been carried out at NLO level.
This is crucial as the constraints are altered when only predictions at LO are used.
Not only are most constraints less stringent, but they also are often shifted in comparison to the constraints obtained using predictions at NLO.
Even though all constraints at 95\% confidence level overlap, the change is still too profound to perform the fit with predictions at LO.

The fit includes terms of order $\mathcal{O}(\Lambda^{-4})$.
This makes sense as the Wilson coefficients $c_{bW}$ and $c_{\varphi tb}$ do not have any effect at lower orders and thus could not be constrained otherwise.
Similarly, $c_{Qq}^{3,8}$ has a relatively big effect at order $\mathcal{O}(\Lambda^{-4})$ in this study but a small contribution at lower orders. 
Reliable constraints can thus not be obtained from a fit at lower orders.
In addition, higher-order terms make the constraints on the other operators slightly more stringent.
This shows that these terms are important for the fit.

For the future of the project, there are plans to make \textsc{DataPrep} a part of \textsc{SFitter} as it is currently still a separate entity.
One feature that is left to be implemented is the averaging of more than two measurements, as is needed for example for the measurements of the helicity fractions because their predictions are independent of the center-of-mass energy.

While this study does not comprise a global fit of the entire top quark sector, it is part of a larger project which includes measurements on top quark pair production and Higgs boson production.
As a global fit of the top quark and Higgs boson sectors has never been performed before, this is another milestone towards a truly global fit of the SMEFT parameter space.


\section*{Acknowledgements}

First and foremost, I thank Tilman Plehn for giving me the opportunity to work on this project and supporting me through the entire process.
A big thank you also goes to Dirk Zerwas for always having an open door and answering all my questions, be it physics or software.
I want to thank Susanne Westhoff for being a great coordinator of the project, and Sebastian Bruggisser and Anke Biek\"otter for the excellent collaboration both on the technical and the physics side.
Thank you for the many ideas you gave me for \textsc{DataPrep} and for answering all my questions concerning the underlying theory.
I am looking forward to merging the $t\bar{t}$ pair and Higgs boson sectors, which you did, with the single top quark results presented in this thesis.
I also thank Remi Lafaye for passing on some deep expertise on the workings of \textsc{SFitter}.
A big thank you goes to Eleni Vryonidou and Cen Zhang, who provided the theoretical predictions and uncertainties for the fit.
Last but not least, I thank all researchers at the Laboratoire de l'Acc\'el\'erateur Lin\'eaire for being a lovely company in lunch breaks and every other occasion.


\clearpage


\begin{appendices} 


\section{Dimension-six Operators}
\label{app:ope}

This section is a summary of the relevant dimension-six operators in single top quark production and decay.
The notation is adopted from~\cite{Grzadkowski:2010es}, where
flavor indices are denoted by $i,j,k$ and $l$;
left-handed fermion SU(2) doublets by $q,l$;
right-handed fermion singlets by $u,d,e$;
the Higgs doublet by $\varphi$;
the antisymmetric SU(2) tensor by $\epsilon = i\tau^2$.
Further, 
\begin{equation*}
\begin{aligned}
& \tilde{\varphi} = \epsilon \varphi^*, 
\qquad (\varphi^\dagger \stackrel{\longleftrightarrow}{iD}_\mu \varphi ) = \varphi^\dagger (iD_\mu \varphi) - (iD_\mu \varphi^\dagger) \varphi, \\
& (\varphi^\dagger \stackrel{\longleftrightarrow}{iD}_\mu^I \varphi ) = \varphi^\dagger \tau^I(iD_\mu \varphi) - (iD_\mu \varphi^\dagger) \tau^I \varphi, 
\quad \text{and} \ \ T^A = \lambda^A/2, 
\end{aligned}
\end{equation*}
where $\tau^I$ are the Pauli matrices and $\lambda^A$ are Gell-Mann matrices.
Non-Hermitian operators are marked with a double dagger. 
For Hermitian operators involving vector Lorentz bilinears, complex conjugation equals transposition of indices: $O^{(ij)*} = O^{(ji)}$ and $O^{(ijkl)*} = O^{(jilk)}$.
The implicit sum over flavor indices only includes independent combinations.
The field strength tensors of the electroweak interaction are denoted as $W_{\mu\nu}^I$ and $B_{\mu\nu}^I$, and the QCD one as $G_{\mu\nu}^A$.
The two groups of relevant SMEFT operators are: the four-quark operators, as listed in equation~\ref{eq-SMEFT-ops1}, and those that contain two quarks coupled to Higgs fields or gauge boson fields, as in equation~\ref{eq-SMEFT-ops2}.


\vspace{5mm}
\begin{equation}
\begin{aligned}
& \mathcal{O}_{qq}^{1(ijkl)} = (\bar{q}_i \gamma^\mu q_j)(\bar{q}_k \gamma_\mu q_l) \\
& \mathcal{O}_{qq}^{3(ijkl)} = (\bar{q}_i \gamma^\mu \tau^I q_j)(\bar{q}_k \gamma_\mu \tau^I q_l) \\
\end{aligned}
\label{eq-SMEFT-ops1}
\end{equation}

\begin{equation}
\begin{aligned}
& ^\ddagger \mathcal{O}_{u\varphi}^{(ij)} = \bar{q}_i u_j \tilde{\varphi} ( \varphi^\dagger \varphi ) \\
& \mathcal{O}_{\varphi q}^{1(ij)} = (\varphi^\dagger \stackrel{\longleftrightarrow}{iD}_\mu \varphi ) ( \bar{q}_i \gamma^\mu q_j ) \\
& \mathcal{O}_{\varphi q}^{3(ij)} = (\varphi^\dagger \stackrel{\longleftrightarrow}{iD}^I_\mu \varphi ) ( \bar{q}_i \gamma^\mu \tau^I q_j ) \\
& ^\ddagger \mathcal{O}_{\varphi ud}^{1(ij)} = (\tilde{\varphi}^\dagger {iD}_\mu \varphi ) ( \bar{u}_i \gamma^\mu d_j ) \\
& ^\ddagger \mathcal{O}_{uW}^{(ij)} = ( \bar{q}_i \sigma^{\mu\nu} \tau_I u_j ) \tilde{\varphi} W_{\mu\nu}^I \\
& ^\ddagger \mathcal{O}_{dW}^{(ij)} = ( \bar{q}_i \sigma^{\mu\nu} \tau_I d_j ) \varphi W_{\mu\nu}^I \\
& ^\ddagger \mathcal{O}_{uB}^{(ij)} = ( \bar{q}_i \sigma^{\mu\nu} u_j ) \tilde{\varphi} B_{\mu\nu}^I \\
& ^\ddagger \mathcal{O}_{uG}^{(ij)} = ( \bar{q}_i \sigma^{\mu\nu} T^A u_j ) \tilde{\varphi} G_{\mu\nu}^A 
\end{aligned}
\label{eq-SMEFT-ops2}
\end{equation}

\clearpage


\section{DataPrep}
\label{app:DataPrep}

\textsc{DataPrep} is a software developed by the author which automates the correlations of theoretical and systematic uncertainties.
This is apt for global fits where a large number of different uncertainties arises from measurements of different processes at different center-of-mass energies and different detectors.
At the time of publishing this thesis, \textsc{DataPrep} is a separate entity that is run before the execution of \textsc{SFitter}.
There are plans, however, to incorporate it in the core of \textsc{SFitter}.

The two main features of \textsc{DataPrep} is handling fully correlated theoretical uncertainties by averaging the corresponding measurements, and handling correlations among systematic uncertainties.
The mathematical process of averaging measurements is described in section~\ref{subsec:meth:theocorr}.
\textsc{DataPrep} automatically averages two measurements if they are of the same final state of the same process and the same center-of-mass energy. 

At this point in time, the averaging of more than two measurements is carried out through a workaround where first the average of two measurements is calculated and the output is re-input for another average.
While this iterative approach is reasonable, it will be automated in the future so the user does not have to run \textsc{DataPrep} multiple times.
In addition, a feature will be implemented so that the user can specify which measurements should be averaged independently of their center-of-mass energies.
This is the case for example with the measurements of the helicity fractions.
So far, this has been carried out with a workaround, but this will be improved in the future.

One interesting aspect is that the weight with which a measurement contributes to the average can be negative, see equation~\ref{eq_AvgMeas_weights}.
This can lead to effects where the average is smaller than each measurement.
While this is mathematically correct, it does not necessarily make sense from a physics perspective.
Therefore, \textsc{DataPrep} is equipped with an option so the user can decide whether to accept measurements with negative weights or to reject these.

One problem with the averaging of measurements is that one cannot usually deal with multiple kinematic distributions of different experiments but of the same process at the same center-of-mass energy.
As the luminosity at the LHC increases, more and more kinematic distributions are becoming available which usually have different binnings and can thus not be averaged.
The theoretical uncertainties of the bins of two distributions are not uncorrelated if they are from the same process at the same energy.
At this point in time, one therefore needs to leave away one of the distributions to avoid over-counting, even though an additional distribution would add new information to the fit.
A possible solution would be to implement correlations among theoretical uncertainties not in \textsc{DataPrep}, but directly in \textsc{SFitter}.
This already exists for systematic uncertainties, but so far has not been necessary for theoretical uncertainties.

Systematic uncertainties are correlated if they are from measurements of the same process at the same center-of-mass energy and the same detector, and carry the same name, e.g. ``Luminosity" or ``Jet-energy-resolution".
The user can also specify if uncertainties with the same name should be correlated across different processes, center-of-mass energies and detectors.
If, for example, the user assumes detector-related uncertainties to be correlated among different processes, he can specify that.

One difficulty is that different analyses often split the uncertainties differently.
For example, one measurement might contain an uncertainty ``jet-related uncertainties", while another contains ``jet energy resolution" and ``jet energy scale".
These measurements might be of the same final state of the same process at the same center-of-mass energy, but at different detectors.
It is then important to get the correlations between the systematic uncertainties right as that impacts the calculation of an average of the to measurements.
In those cases, the user can specify that the uncertainties ``jet energy resolution" and ``jet energy scale" are bundled together (i.e. quadratically added) and correlated with the "jet-related uncertainties" from the other measurement.

More features of \textsc{DataPrep} are that the user can specify uncertainties that should not be used for the output.
This is convenient to study the impact of different uncertainties.
Also, \textsc{DataPrep} subtracts the signal from the background where these are given, and calculate the corresponding uncertainties.
This is especially practical for measurements that contain a Higgs boson as in such analyses the signal and background rates are often given separately.

\clearpage

\end{appendices}


\end{fmffile}
\end{document}